\magnification=\magstep1   \voffset .4in \input epsf                              
\font\huge=cmr10 scaled \magstep2
\def\la{\lambda}   \overfullrule=0pt
\input amssym.def
\def\Z{{\Bbb Z}}  \def\Q{{\Bbb Q}}  \def\N{{\cal N}}
  
 \def\la{\lambda}  \def\ga{\gamma} 
\def\J{{\cal J}}       \def\i{{\rm i}} \def\B{{\cal B}}
\def\si{\sigma} \def\eps{\epsilon}    
    \def\k{\kappa}     
  \def\L{{\Lambda}}  
\def\E{{\cal E}}    \def\D{{\cal D}} 
\def\N{{\cal N}}   

\font\smcap=cmcsc10     \font\little=cmr7
\def\z{{\cal Z}}  

\def\adots{\mathinner{\mkern2mu\raise1pt\hbox{.}\mkern2mu\raise4pt\hbox{.}
    \mkern2mu\raise7pt\hbox{.}\mkern1mu}}
\def\boxit#1{\vbox{\hrule\hbox{\vrule{#1}\vrule}\hrule}}
\def\splus{\,\,{\boxit{$+$}}\,\,}
\def\stimes{\,\,{\boxit{$\times$}}\,\,}

\centerline{{\bf \huge Boundary Conformal Field Theory}}\bigskip

\centerline{{\bf \huge and Fusion Ring Representations}}\bigskip\bigskip

\centerline{{\smcap Terry Gannon}}\bigskip

\centerline{{\it Department of Mathematical Sciences, University of Alberta}}

\centerline{{\it Edmonton, Canada, T6G 2G1}}

\medskip\centerline{{e-mail: tgannon@math.ualberta.ca}}\bigskip\bigskip

{\bf Abstract}

To an RCFT corresponds two combinatorial structures: the amplitude of a torus
(the 1-loop {partition function} of a closed string, sometimes called
a {\it modular invariant}), and a representation of the fusion ring (called a {\it  NIM-rep}
or equivalently a {\it fusion graph}, and closely related to the 1-loop partition function of an
open string).
In this paper we develop some basic theory of NIM-reps, obtain several new
NIM-rep classifications, and compare them with the corresponding
modular invariant classifications. Among other things, we make the following
fairly disturbing observation: there are infinitely many (WZW) modular
invariants which do not correspond to any NIM-rep. The resolution
could be that those modular invariants are physically sick.
Is classifying modular
invariants really the right thing to do? For current algebras, the answer
seems to be: Usually but not always. For finite groups \`a la Dijkgraaf-Vafa-Verlinde-Verlinde,
the answer seems to be: Rarely.               \bigskip\bigskip

\centerline{{\bf 1. Introduction}}\medskip

For many reasons, not the least of which is open string theory, we are interested
in boundary conformal field theory. Although it has apparently never been
established that bulk RCFT necessarily requires for its consistency
that there be a compatible and complete system of boundary conditions, the conventional
wisdom seems to be that otherwise the RCFT would have sick operator product
expansions. In any case, a boundary CFT would seem to be a relatively accessible
halfway-point between  constructing the full CFT from a chiral
CFT($=$vertex operator algebra). This paper explores the relation
between boundary and bulk CFT, by comparing the classification of {\it
modular invariants} with {\it NIM-reps}, and in so doing it probes
this `conventional wisdom'.

Cardy [1] was the first to explain how conformally invariant boundary conditions in CFT
are related to fusion coefficients. In particular, given a bulk CFT and
a choice of (not necessarily maximal) chiral algebra, consider the set of (conformally
invariant) boundary conditions which don't break the chiral symmetry.
These should be  spanned by the appropriate Ishibashi states $|\mu\rangle\!\rangle$,
 labelled
by the spin-zero primary fields $\phi(\mu,\overline{\mu})$ in the theory. We know [2] these boundary
states need not be linearly independent, but we should be able to find
a (unique) $\Z_\ge$-basis $|x\rangle\in{\cal B}$ for them, equal in
number to the Ishibashi states. Then the 1-loop vacuum amplitude ${\cal
Z}_{x|y}$, where the two edges of the cylinder are decorated with
boundary conditions $|x\rangle,|y\rangle\in{\cal B}$, 
can be expanded in terms of the chiral characters $\chi_\la$:
$$\z_{x|y}=\sum_{\la}\,\N_{\la x}^y\,\chi_\la$$
Cardy explained that these coefficients $\N_{\la x}^y$ define a
representation of the chiral fusion ring with nonnegative integer matrices.
Strictly speaking, Cardy only considered the diagonal theory given by the
modular invariant partition function $\z=\sum_\mu|\chi_\mu|^2$.
 The more general
theory has been developed by e.g.\ Pradisi et al [3] (see e.g.\ [4] for a good review),
Fuchs--Schweigert (see e.g.\ [5]), and Behrend et al (see e.g.\ [6]). We will review and axiomatise
the combinatorial essence of this theory below in Section 3, under the name {\it NIM-reps}.

In a remarkable paper, Di Francesco--Zuber [7] sought a generalisation of the
A-D-E pattern of $\widehat{{\rm sl}}(2)$ modular invariants, by assigning
graphs to RCFT. They were (largely empirically) led to introduce what
we now will call {\it fusion graphs}. Over the years the definition was refined,
and their relations to the lattice models of statistical mechanics, structure
constants in CFT, etc were clarified (see the enchanting review in [8]).
In particular, their connection with NIM-reps is now fully understood (see e.g.\ [6]).

The torus partition function ($=$modular invariant) and the cylinder partition
function ($=$NIM-rep) of an RCFT should be compatible. Roughly, the eigenvalues
of the NIM-rep matrices $(\N_\la)_{xy}=\N_{\la x}^y$ should
be labeled by the Ishibashi states, and the Ishibashi states should be labeled
by the spin-0 primaries, i.e.\ the diagonal ($\la=\mu$) terms in the
modular invariant $\z=\sum_{\la,\mu}M_{\la\mu}\,\chi_\la\,{\chi}_\mu^*$.
(Strictly speaking, all this assumes a choice of `pairing' or `gluing
automorphism' $\omega$ --- see \S2.2 below.)

We call a modular invariant {\it NIMmed} if it has a compatible
NIM-rep; otherwise we call it {\it NIM-less}.
In this way, we can use NIM-reps to probe lists of modular invariants.
After all, the definition (see \S2.2) of modular invariants isolates only certain features of
RCFT, and it is not at all obvious that classifying them is really the right thing to do.

NIM-reps and modular invariants, and their compatibility condition,
 also appear very naturally in the context of braided
subfactor theory (see e.g.\ [9,10] for reviews of this remarkable picture,
due originally to Ocneanu).
The term `NIM-rep' [10] is short for `nonnegative integer matrix representation'.

Even if we restrict attention to the current(=affine Kac-Moody) algebras, i.e.\ WZW theories, very
little is known about NIM-rep classifications. The $\widehat{{\rm sl}}(2)$
theories at all levels $k$,
and all $\widehat{\rm sl}(n)$ at level 1, are all that have been done [7,6], 
although Ocneanu [11] has announced a classification
of the subset of $\widehat{{\rm sl}}(3)$ and $\widehat{{\rm sl}}(4)$
NIM-reps (any level) of relevance to subfactors. Although there
isn't  a perfect match with the corresponding modular invariant classifications,
the relation between what superficially seem to be distinct mathematical problems is
remarkable.

Our two main results are:

\smallskip\item{$\bullet$}  We classify the NIM-reps for all $\widehat{{\rm sl}}(n)$
and $\widehat{{\rm so}}(n)$ at level 2. Those of $\widehat{{\rm sl}}(n)$
match up well with the corresponding modular invariant classification;
those of $\widehat{{\rm so}}(n)$ dramatically do not, and in fact most
$\widehat{{\rm so}}(n)$ level 2 modular invariants are NIM-less.\smallskip

\item{$\bullet$} We develop the basic theory of NIM-reps (see especially
Theorem 3 below), and find striking similarities with modular invariants
(compare Theorem 1). \smallskip

In \S3.4 we discuss the rationality and nonnegativity of
the coefficients ${\cal M}_{\la\mu}^\nu$ of the Pasquier algebra  and of the
dual Pasquier algebra $\widehat{\N}_{xy}^z$. In \S4 we give the level
1 NIM-rep classifications for all current algebras.
We relegate the (unpleasant) proofs of the level 2 classifications and Theorem 3
to the Appendix. 

The reader less interested in the details may
wish to jump to \S6, where we find two simple {\it NIM-less} modular invariants,
then to \S7 where we explain using the example of $\widehat{{\rm sl}}(3)$ level 8
how the results of \S3.3  come together to yield NIM-rep classifications, and
finally move to the conclusion, \S8, where
we give some concluding thoughts and speculations.

What do all these NIM-less modular invariants tell us? One possibility
is that this 
picture of the relation of boundary and bulk CFT is too naive ---
e.g.\ it is related to the assumption of the completeness of
boundaries first raised in [3]. Another possibility is that there
are infinitely many modular invariants which cannot be realised as the
torus partition function of a CFT.

What about NIM-reps for higher-rank algebras and levels? We get good control
over the eigenvalues of the fusion graphs. Much more difficult
is, given these eigenvalues, to draw the possible
 fusion graphs. We know (proved below) that there
will only be finitely many of these, but based on considerations given in
 {\bf (6)} in the concluding section, we expect that number to be typically
 quite large. The classes presently worked out  are atypical, because
the critical Perron-Frobenius eigenvalues involved are $\le 2$. As the eigenvalues rise,
we expect the number of NIM-reps to grow out of control. In other words,
we expect that classifying NIM-reps is probably hopeless (and pointless)
for all but the smallest ranks and levels.

On the other hand, [12] suggests that the modular invariant situation for
$\widehat{{\rm so}}(n)$ level 2 is quite atypical, and that we can expect that
 all modular invariants for most current algebras $X_{r,k}$ are related
 to Dynkin diagram symmetries. The corresponding NIM-reps would then be
  fairly well understood (see e.g.\ [13,14] and references therein); in particular
they are probably all NIMmed.
The situation however will probably be much worse for the modular invariants coming from other
(i.e.\ non-WZW) chiral algebras, e.g.\ the untwisted sector in holomorphic orbifolds by finite groups [15]
--- see e.g.\ \S6.

There is a tendency in the literature to focus only on $\widehat{{\rm
sl}}(n)$ (although [16] briefly discussed NIM-reps for $\widehat{G}_2$ level $k$).
This  perhaps is a mistake --- $\widehat{{\rm sl}}(n)$ is very special,
and this limited perspective leads to incorrect intuitions as to characteristic
WZW or RCFT behaviour. We see that here: for instance the NIM-rep vrs modular invariant
situation for $\widehat{\rm so}(n)$ level 2 is quite remarkable, and considerably
more interesting than that for $\widehat{{\rm sl}}(n)$ level 2.

\bigskip\centerline{{\bf 2. Review: Fusion rings and modular invariants}}\medskip

\noindent{{\it 2.1. Modular data of the RCFT.}} \medskip

The material of this subsection
is discussed in more detail in the reviews [17,18].

As is well-known, the RCFT characters $\chi_\la(q)$ yield
a finite-dimensional unitary representation of the modular group SL$_2(\Z)$,
 given by the natural action of SL$_2(\Z)$
 on $\tau={1\over 2\pi\i}{\rm ln}\,q$.
Denote by $S$ and $T$ the matrices associated to $\left(\matrix{0&-1\cr 1&0}
\right)$ and $\left(\matrix{1&1\cr 0&1}\right)$. Then $T$ is diagonal,
and $S$ is symmetric. The rows and columns of $S$ and $T$ are parametrised
by the primary fields $\la\in P_+$. One of these, the `vacuum' 0, is distinguished.

In this paper we will be primarily interested in the data coming from
current
algebras $\widehat{\frak g}$ (${\frak g}$ simple), i.e.\ associated to Wess-Zumino-Witten models. However, unless
otherwise stated,  everything here holds for arbitrary RCFT.

{We will assume} for convenience that $S_{\la 0}>0$ --- this holds in
particular for unitary RCFTs, such as the WZW models. The changes required
for nonunitary RCFT consist mainly of replacing some appearances of the vacuum
with the unique primary $o\in P_+$ with minimal conformal weight. Then
$S_{\la o}>0$. In a unitary
theory, we have $o=0$.
 The ratio $S_{\la o}/S_{0o}$
is called the {\it quantum-dimension} of $\la$, and plays a central role.

The matrix $S^2$ is a permutation matrix $C$, called {\it charge-conjuation}.
It obeys $C0=0$, $T_{C\la,C\la}=T_{\la\la}$, and corresponds to complex conjugation:
$$S_{C\la,\mu}=S_{\la,C\mu}=S_{\la\mu}^*\eqno(2.1)$$  

The fusion coefficients $N_{\la\mu}^\nu$ of the theory are given by Verlinde's
formula [19]:
$$N_{\la\mu}^\nu=\sum_{\gamma\in P_+}{S_{\la\gamma}\,S_{\mu\gamma}\,S_{\nu\gamma}^*
\over S_{0\gamma}}\in\Z_{\ge}:=\{0,1,2,\ldots\}\eqno(2.2)$$
Let $N_\la$ denote the {\it fusion matrix}, i.e.\ the matrix with
entries $(N_\la)_{\mu\nu}=N_{\la\mu}^\nu$. Then $N_0=I$, $N_{C\la}=N_\la^t$, and
$$N_\la\,N_\mu=\sum_{\nu\in P_+}N_{\la\mu}^\nu N_\nu\eqno(2.3)$$

We use the term {\it modular data} for
any matrices $S$ and $T$ obeying these conditions. The ring with
preferred basis
$P_+$ and structure constants $N_{\la\mu}^\nu$ is called the {\it
fusion ring}.
For example, modular data and a fusion ring exist for every choice of
current algebra $\widehat{\frak{g}}=X_r^{(1)}$ and positive integer $k$ (called the {\it
level}) --- of course this is precisely what arises in the WZW models.
At times we will abbreviate this to $X_{r,k}$. The primaries $\la\in
P_+$ for this WZW modular data consist of the level $k$ integrable highest weights
$\la=\la_1\L_1+\cdots+\la_r\L_r$, where the basis vectors $\L_i$ are called fundamental
weights. See e.g.\ Ch.13 of [20] for more details. Explicit formulas for
$S_{\la\mu}$ are given in [20]; see also [21].

%
%
%

The quantum-dimensions in (unitary) RCFT obey $S_{\la 0}/S_{00}\ge 1$. When it equals 1,
$\la$ is called a {\it simple-current} [22]. Then $N_\la$
will be a permutation matrix, corresponding to a permutation $J$ of $P_+$,
and there will be a phase $Q_J:P_+\rightarrow{\Bbb Q}$
such that
$$S_{J\mu,\nu}=e^{2\pi\i \,Q_J(\nu)}\,S_{\mu\nu}\eqno(2.4)$$
The simple-currents form an abelian group, under composition of permutations.
Note that
$$\eqalignno{N_{J\mu,J'\nu}^{JJ'\gamma}=&\,N_{\mu\nu}^{\gamma}&(2.5a)\cr
N_{C\mu,C\nu}^{C\gamma}=&\,N_{\mu\nu}^\gamma&(2.5b)\cr}$$
for any simple-currents $J,J'$, where $C$ as usual is charge-conjugation.

For example, for  $A_{1,k}$ we may take $P_+=\{0,1,\ldots,k\}$
(the value of the Dynkin label $\la_1$),
and then the $S$ matrix is $S_{ab}=\sqrt{{2\over k+2}}\,\sin(\pi\,{(a+1)\,
(b+1)\over k+2})$. Charge-conjugation $C$ is trivial here, but $j=k$ is a
simple-current corresponding to permutation $Ja=k-a$ and phase
$Q_j(a)=a/2$. The fusion coefficients are given by
$$N_{ab}^c=\left\{\matrix{1&{\rm if}\ c\equiv a+b
\ ({\rm mod\ 2)\ and}\ |a-b|\le c\le{\rm min}\{a+b,
2k-a-b\}\cr0&{\rm otherwise}\cr}\right.$$

Write $\xi_N$ for the root of unity $\exp[2\pi\i/N]$.
A fundamental symmetry of modular data is a certain generalisation of
charge-conjugation. For any RCFT,
the entries $S_{\la\mu}$ are sums of roots of unity $\xi_N^m$, all
divided by a common integer $L$. For example for ${{\rm sl}}(n)_k$ we
can take $N=4n\,(n+k+1)$. 
We say that the entries $S_{\la\mu}$  lie in the {\it cyclotomic number
field} $\Q[\xi_N]$. The automorphisms $\si\in{\rm Gal}(\Q[\xi_N]/\Q)$ of
this field preserve by definition both multiplication and addition and fix the
rational numbers; they are parametrised by
an integer $\ell$ coprime to $N$ (more precisely, the action of $\si_\ell$ is
uniquely determined by the relation $\si_\ell(\xi_N^m)=\xi_N^{m\ell}$, so really
$\ell$ is defined modulo $N$). 
To each such integer $\ell$, i.e.\ each automorphism $\si_\ell$, there is a
permutation $\si_\ell$ of $P_+$ and a choice of signs $\epsilon_\ell(\la)=\pm 1$,
such that [23]
$$\si_\ell(S_{\la\mu})=\epsilon_\ell(\la)\,S_{\si_\ell(\la),\mu}=
\epsilon_\ell(\mu)\,S_{\la,\si_\ell(\mu)}\eqno(2.6)$$

This Galois symmetry may sound complicated, but that could be due only to its
 unfamiliarity. It plays a central role in the theory of modular data, modular
 invariants, and NIM-reps (see e.g.\ \S7 below), and makes accessible problems which have no right
 to be so. Algorithms for this Galois symmetry, for the current algebras, are
explicitly worked out in [21].

An important ingredient of the theory is that of {\it fusion-generators}.
We call $\Gamma=\{\ga^{(1)},\ldots,\ga^{(g)}\}\subset P_+$ a fusion-generator if to any $\la\in P_+$
there is a $g$-variable polynomial $P_\la(x_1,\ldots,x_g)$ such that the fusion matrices obey
$$N_\la=P_\la(N_{\ga^{(1)}},\ldots,N_{\ga^{(g)}})$$
 or equivalently, for any $\la,\mu\in P_+$,
$${S_{\la\mu}\over S_{0\mu}}=P_\la({S_{\ga^{(1)}\mu}\over S_{0\mu}},\ldots,
{S_{\ga^{(g)}\mu}\over S_{0\mu}})\eqno(2.7)$$
This says that $\ga^{(1)},\ldots,\ga^{(g)}$ generate the fusion ring, and also
(we will see) the NIM-reps.

One of the reasons fusion rings for the current algebras are relatively tractable
is the existence of small fusion-generators. In particular, because we know
that any Lie character for $X_r$ can be written as a polynomial in the
fundamental weights ${\rm ch}_{\L_1},\ldots,{\rm ch}_{\L_r}$, it is easy to
show [24] that $\Gamma=
\{\L_1,\ldots,\L_r\}\cap P_+$ is a fusion-generator for any $X_r^{(1)}$
level $k$. Smaller fusion-generators usually exist however.
The question  for ${{\rm sl}}(n)_k$ has been studied quite thoroughly in
[25]. For example, $\{\L_1\}$ is a fusion-generator for ${\rm sl}(n)_{k}$ iff both
\smallskip

(i) each prime divisor $p$ of $k+n$ satisfies $2p>{\rm min}\{n,k\}$, and

(ii) either $n$ divides $k$, or gcd$(n,k)=1$.\smallskip

\noindent More generally, for ${{\rm sl}}(n)_k$ the following are always
fusion-generators:
$$\eqalign{\Gamma_{\div}=&\,\{\L_d\,|\, 2d\le n\ {\rm and}\ d\ {\rm divides}\
k+n\}\cr \Gamma_{\div}^\tau=&\,\left\{\matrix{\{\L_d\,|\, 2d\le k\ {\rm and}\
d\ {\rm divides}\ k+n\}& {\rm when}\ k\ {\rm doesn't\ divide}\ n\cr
\{k\L_1,\L_d\,|\, 2d\le k\ {\rm and}\ d\ {\rm divides}\ k+n\}& {\rm when}\ k\ {\rm divides}\ n\cr}\right.
}$$
(Of course, the weight $k\L_1$ in $\Gamma_{\div}^\tau$ is a simple-current.)
 Examples are:

\noindent{} \item{$\bullet$} $\L_1$ is a fusion-generator for ${{\rm sl}}(2)_k$
and ${{\rm sl}}(3)_k$, for any level $k$;

\noindent{} \item{$\bullet$} for ${{\rm sl}}(4)_k$, $\L_1$ is a fusion-generator 
when  $k$ is odd, while both $\{\L_1,\L_2\}$ are needed when $k$ is even;

\noindent{} \item{$\bullet$} $\L_1$ is a fusion-generator for ${{\rm sl}}(n)_1$
for any $n$; it's also a fusion-generator for ${{\rm sl}}(n)_2$
when $n$ is odd, while both $\{\L_1,2\L_1\}$ are needed when $n$ is even.

\bigskip\noindent{{\it 2.2. Modular invariants and their exponents.}}\medskip

The one-loop vacuum-to-vacuum amplitude of a rational conformal field theory is
the modular invariant partition function
$${\cal Z}(q)=\sum_{\la,\mu\in P_+}M_{\la\mu}\,\chi_\la(q)\,\chi_\mu(q)^*\eqno(2.8)$$

\noindent{\bf Definition 1.} {\it By a} {modular invariant} $M$
{\it we mean a matrix with nonnegative integer entries and $M_{00}=1$, obeying
$MS=SM$ and $MT=TM$.}\medskip

Two examples of modular invariants are $M=I$ and $M=C$ (of course these may
be equal). It is known that for any choice of
modular data, the number of modular invariants will be finite [26,10].
We identify the function $\z$ in (2.8) with its coefficient matrix $M$.

The coefficient matrix $M$ of an RCFT  partition function is a modular invariant,
but the converse need not be true. Also, different RCFTs can conceivably have the same
modular invariant. {\it Is the classification
of modular invariants the right thing to do?} Is there actually a resemblance
between the list of modular invariants, and the corresponding list of RCFTs?
Or are we losing too much information and structure by classifying not
the full RCFTs, but rather the much simpler modular invariants?
We return to these questions in the concluding section, \S8.

We have a good understanding now of the modular invariant lists for the
current algebras, at least for small rank and/or level. See [12,18] and references
therein for the main results and appropriate literature.

The most famous modular invariant list is that of $\widehat{{\rm sl}}(2)$,
due to Cappelli-Itzykson-Zuber [27]. The trivial modular invariant $M=I$
exists for all levels $k$; a simple-current invariant $M[J]$ (see
(2.9) below) exists
for all even $k$; and there are exceptionals at $k=10,16,28$. For instance,
the level $28$ exceptional is
$$\z_{28}(q)=|\chi_0+\chi_{10}+\chi_{18}+\chi_{28}|^2+|\chi_6+\chi_{12}+\chi_{16}
+\chi_{22}|^2$$

Cappelli-Itzykson-Zuber noticed something remarkable about their list: it falls
into the A-D-E pattern. Each of their modular invariants $M$ can be identified
with the Dynkin diagram ${\cal G}(M)$ of a finite-dimensional simply laced Lie algebra
(these are the diagrams $A_n,D_n,E_n$ in Figure 1).
The level of the modular invariant, plus 2, equals the Coxeter
number $h$ of ${\cal G}(M)$. Each number $1\le a\le k+1$ will be an {\it exponent} of ${\cal G}(M)$
with multiplicity given by the diagonal entry $M_{a-1,a-1}$.
The Coxeter number $h$ and exponents $m_i$ of the diagram ${\cal G}$ are listed in
Table 1. For instance,
the modular invariant $\z_{28}$ given above corresponds to the $E_8$ Dynkin
diagram: $28+2=30$, the Coxeter number of $E_8$; and the nonzero diagonal entries $M_{bb}$
of $M$ are at $b=0,6,10,12,16,18,22,28$, compared with the $E_8$ exponents
$1,7,11,13,17,19,23,29$ (all multiplicities being 1).
Likewise, the $D_8$ Dynkin diagram has Coxeter number 14, and exponents
$1,3,5,7,7,9,11,13$, and corresponds to the ${{\rm sl}}(2)_{12}$ modular invariant
$$|\chi_0+\chi_{12}|^2+|\chi_2+\chi_{10}|^2+|\chi_4+\chi_8|^2+2|\chi_6|^2$$

\bigskip\centerline {{\bf Table 1. Eigenvalues of graphs in Figure 1}}
$$\vbox{\tabskip=0pt\offinterlineskip
  \def\tablerule{\noalign{\hrule}}
  \halign to 4in{
    \strut#&\vrule#\tabskip=0em plus1em &    
    \hfil#&\vrule#&\hfil#&\vrule#&    
    \hfil#&\vrule#\tabskip=0pt\cr\tablerule         
&&\omit\hfill  Graph\hfill  &&\omit \hfill Coxeter number $h$\hfill  &&
\omit\hfill  Exponents $m_i$\hfill  &\cr\tablerule
&&\hfill  $A_n$, $n\ge 1$\hfill  && \hfill $n+1$ \hfill && \hfill  $1,2,\ldots,n$\hfill& \cr
&&\hfill  $D_n$, $n\ge 4$\hfill  && \hfill $2n-2$ \hfill && \hfill $1,3,\ldots,2n-3,n-1$\hfill  & \cr
&&\hfill  $E_6$\hfill  &&\hfill  12\hfill  &&\hfill  $1,4,5,7,8,11$\hfill  &\cr
&&\hfill  $E_7$\hfill  && \hfill 18\hfill  &&\hfill  $1,5,7,9,11,13,17$\hfill & \cr
&& \hfill $E_8$\hfill  &&\hfill  30\hfill  &&\hfill  $1,7,11,13,17,19,23,29$\hfill & \cr
&& \hfill $T_n$, $n\ge 1$\hfill  && \hfill $2n+1$\hfill  && \hfill $1,3,5,\ldots,2n-1$\hfill & \cr
\tablerule\noalign{\smallskip}
 }} $$

Because of that observation, [7,16] suggested the following general definition.

\medskip\noindent{\bf Definition 2.}
{\it By the} {exponents} {\it of a modular invariant $M$, we mean the multi-set
$\E_M$ consisting
of $M_{\la\la}$ copies of $\la$ for each $\la\in P_+$.} \medskip

(By a `multi-set',
we mean a set together with multiplicities, so $\E_M\subset P_+\times \Z_{\ge}$.)
In other words, the exponents are precisely the spin-0 primary fields
in the theory (periodic sector).
By analogy with the A-D-E classification for $\widehat{{\rm sl}}(2)$,
we would like to assign graphs to a modular invariant in such a way that the
eigenvalues of the graph (that is to say, the eigenvalues of its adjacency
matrix) can be identified with the exponents of the modular invariant.
We shall see next section that there is a natural way to do this: NIM-reps!

For example, $M=I$ has exponents $\E_I=P_+$, while the exponents $\E_C$ of $M=C$
are the
self-conjugate primaries $\la=C\la$. In both $\E_I$ and $\E_C$, all multiplicities are 1,
but simple-current invariants (see (2.9) below)
can have arbitrarily large multiplicities.

It is merely a matter of convention whether $M_{\la,C\la}\ne 0$ or
$M_{\la\la}\ne 0$ is taken as the definition of exponents --- it has to do
with the arbitrary choice of taking the holomorphic and antiholomorphic
(i.e.\ left-moving and right-moving) chiral algebras to be isomorphic or
anti-isomorphic. In the literature both choices can be found.
We've taken them to be anti-isomorphic, hence our definition
of spin-0 primaries.

Implicit in this discussion is the {\it diagonal} (i.e.\ identity) choice of `gluing automorphism'
$\Omega$ [28] or `pairing' $\omega$ [5]. The pairing can be
any permutation of $P_+$ which preserves fusions and conformal weights,
so it must yield an `automorphism invariant', i.e.\ a
modular invariant $M$ which is a permutation matrix:
$M_{\la\mu}=\delta_{\mu,\omega \la}$. For the current algebras,
all possible pairings can be obtained from [24]. This pairing tells one how to
identify left- and right-moving primaries. Definition 2 can now be generalised
to the multi-set $\E_M^\omega$, where $\la$ occurs with multiplicity $M_{\la,\omega\la}$.

In this paper we will limit ourselves to the trivial ($=$identity) pairing
$\omega$. This is permitted for two reasons. First and most important,
$\E_M^\omega=\E_{M\omega^t}$, where $M\omega^t$ is the modular invariant
obtained by matrix multiplication. Second and quite intriguing, in practice
it appears that the question of whether or not $M$ is  NIM-less (see \S3.2 below)
is independent of $\omega$.

Consider  a simple-current $J$ with order $n$ (so $J^n=id.$).
Then we can find an integer $R$ obeying
$T_{J0,J0}T_{0,0}^*=\exp[2\pi\i\,R\,{n-1\over 2n}]$. Define a matrix
$M[J]$ by [22]
$$M[J]_{\la\mu}=\sum_{\ell=1}^n\delta_{J^\ell\la,\mu}\,\delta(Q_J(\la)+
{\ell\over 2n}R)\eqno(2.9)$$
where $\delta(x)=1$ if $x\in\Z$ and 0 otherwise. For example, $M[id.]=I$.
The matrix $M[J]$ will be a modular invariant iff $T_{J0,J0}T_{0,0}^*$
is an $n$th root of 1 (this is automatic if $n$ is odd; for $n$ even, it's
true iff $R$ is even); it will in addition be a permutation matrix iff
$T_{J0,J0}T_{0,0}^*$ has order exactly $n$. For example, for ${{\rm sl}}(2)_k$,
$R=k$ so $M[J]$ is a modular invariant iff $k$ is even, and when $k/2$ is
odd it will in fact be a permutation matrix. The modular invariant
${M}[J]$ for ${{\rm sl}}(2)_{12}$ is given above.

We call these modular invariants {\it simple-current invariants}. This
construction can be generalised somewhat when the simple-current group
isn't cyclic, but (2.9) is good enough here. For any current algebra,
at any sufficiently large level $k$, it appears that the only modular
invariants are simple-current invariants and their product with $C$ (except
for the algebras ${{\rm so}}(4n)$, whose Dynkin
symmetries permit (2.9) to be slightly generalised, and which have other
`conjugations' $C_i\ne C$).

We'll end this subsection by establishing some of the basic symmetries of
modular invariants and their exponents. First, note that $MC=CM$ (since
$C=S^2$), so $\la$ and $C\la$ always appear in $\E_M$ with equal multiplicity.
More generally, the Galois symmetry (2.6) of modular data yields an
important modular invariant symmetry [23]:
$$\eqalignno{M_{\la\mu}=&\,M_{\si(\la),\si(\mu)}&(2.10a)\cr
M_{\la\mu}\ne 0\qquad\Longrightarrow&\qquad \eps_\si(\la)=\eps_\si(\mu)&(2.10b)}$$
for any Galois automorphism $\si$, i.e.\ any $\ell$ coprime to $N$.
One thing (2.10a) implies is that $\la$ and $\si(\la)$ will always have the
same multiplicity in $\E_M$. This is quite strong  --- for instance,
the primaries $0,6,10,12,16,18,22,28$ for ${{\rm sl}}(2)_{28}$
all lie in the same Galois orbit, and indeed they all have the same
multiplicity as exponents of the $k=28$ exceptional modular invariant
(just as they must for the other two $k=28$ modular invariants).

The other fundamental symmetry of modular data is due to simple-currents.
Let $J,J'$ be simple-currents, and suppose that $M_{J0,J'0}\ne 0$.
Then (see e.g.\ [18]) $\forall \la,\mu\in P_+$
$$\eqalignno{M_{J\la,J'\mu}=&\,M_{\la,\mu}&(2.11a)\cr
M_{\la,\mu}\ne 0\qquad\Longrightarrow&\qquad Q_J(\la)\equiv Q_{J'}(\mu)\
({\rm mod}\ 1)&(2.11b)}$$
Thus by (2.11a), $J\in\E_M$ implies that all powers $J^i$ are in $\E_M$, all with multiplicity
1, and also that $\la$ and $J\la$ have the same multiplicity in $\E_M$ for
any $\la\in P_+$.

It is curious that the selection rules (2.10b) and (2.11b) seem to have
no direct consequences for $\E_M$, although they are profoundly important
in constraining off-diagonal entries of $M$.

For later comparison, let's collect some of the main results on the
exponents of  modular invariants:

\medskip\noindent{{\bf Theorem 1.}} Choose any modular data.
 Let $M$ be any modular invariant, and let ${\cal E}_M$ be
its exponents, with $m_\mu$ being the multiplicity $M_{\mu\mu}$ in ${\cal E}_M$.

\item{(i)} There are only finitely many modular invariants for that choice of
modular data. They obey the bound $M_{\la\mu}\le {S_{\la 0}\over S_{00}}\,
{S_{\mu 0}\over S_{00}}$.

\item{(ii)} $m_0=1$.

\item{(iii)} For any simple-current $J$, $m_J=0$ or 1; if $m_J=1$
then $m_{J\la}=m_\la$ for all $\la\in P_+$.

\item{(iv)} For any Galois automorphism $\si$ and any primary
$\la\in P_+$, $m_{\si(\la)}=m_\la$.

\item{(v)} For any symmetry $\pi$ of the fusion coefficients,
and any $\la\in P_+$, we get 
$$\sum_{\mu\in{\cal E}_{M\pi}}{S_{\la\mu}\over S_{0\mu}}\in\Z_{\ge}$$

\medskip In (v) we sum over $\E_M$ as a multi-set, i.e.\ each $\mu$
appears $m_\mu$ times. The sum in (v) will typically be very
large. By a symmetry of the fusion coefficients, we mean a
permutation $\pi$ of $P_+$ for which
$N_{\la,\mu}^\nu=N_{\pi\la,\pi\mu}^{\pi\nu}$ for all $\la,\mu,\nu\in
P_+$. It is equivalent to the existence of a permutation $\pi'$ for
which $S_{\pi\la,\pi'\mu}=S_{\la,\mu}$ --- all such symmetries for the
current algebras were found in [29]. To prove (v), let $\Pi$ and
$\Pi'$ be the corresponding permutation matrices. Then
$$\sum_{\mu\in\E_{M\pi}}{S_{\la,\mu}\over S_{0,\mu}}={\rm Tr}(M\Pi D_\la)=
{\rm Tr}({S}^*SM\Pi D_\la)
={\rm Tr}(M\Pi'SD_\la{S}^*)={\rm Tr}(M\Pi'N_\la)\in\Z_\ge$$
where $D_\la$ is the diagonal matrix with entries $S_{\la\mu}/S_{0\mu}$.
Thm.1(v) seems to be new.

Thm.1 assumes all $S_{\la 0}>0$. If instead we have a {\it nonunitary} RCFT,
let $o\in P_+$ be as in \S2.1. Then we can show $m_o\ge 1$. However the
known proofs that there are finitely many modular invariants, all break down,
as does the proof of (iii).

In \S3.3 as well as paragraph {\bf (4)} in \S8, we are interested in when simple-currents
are exponents. Consider any matrix  $M$ which  commutes with the $T$ of
sl$(2)_k$. That is,
$$M_{ab}\ne 0\qquad \Rightarrow\qquad (a+1)^2\equiv (b+1)^2\ ({\rm mod}\
4(k+2))$$
Thus $a$ is odd iff $b$ is odd, i.e.\ $Q_J(a)\equiv Q_J(b)$ (mod 1),
provided $M_{ab}\ne 0$.
If $M$ is in addition a modular invariant, we get from this that
$$M_{J,J}=\sum_{a,b=0}^kS_{J,a}\,M_{ab}\,S_{J,b}^*=M_{00}=1$$
Thus it is automatic for sl$(2)_k$ that $J\in \E_M$, for all modular invariants $M$.

This argument generalises considerably. 
The norms of the weights of sl$(n)_k$ satisfy
$$(\la|\la)\equiv Q_J(\la)-Q_J(\la)^2/n\quad ({\rm mod\ 2})\eqno(2.12a)$$
where $Q_J(\la)=\sum_i i\la_i$, for the simple-current $J=k\L_1$. For the basic calculation consider sl$(3)_k$.
Then commutation of $M$ with $T$ implies from (2.12a) the selection rule
$$M_{\la\mu}\ne 0\qquad\Rightarrow\qquad Q_J(\la)^2\equiv Q_J(\mu)^2\
({\rm mod}\ 3)\eqno(2.12b)$$
and hence
$$\eqalignno{M_{J,J}+M_{J,J^{-1}}&=\sum_{\la,\mu\in P_+}(\exp[2\pi\i\,{Q_J(\la)-
Q_J(\mu)\over 3}]\!+\!\exp[2\pi\i\,{Q_J(\la)+Q_J(\mu)\over 3}])\,
S_{0\la}M_{\la\mu}S_{0\mu}&\cr =&
\sum_{\la,\mu\in P_+}\!(\cos[2\pi{Q_J(\la)\!-\!
Q_J(\mu)\over 3}]\!+\!\cos[2\pi{Q_J(\la)\!+\!Q_J(\mu)\over 3}])\,
S_{0\la}M_{\la\mu}S_{0\mu}&(2.12c)}$$
where we use the reality of the LHS of (2.12c). But every term on the RHS of
(2.12c) will be
nonnegative: whenever $M_{\la\mu}\ne 0$,  (2.12b) says that the sum
of cosines in (2.12c) will either be ${1\over 2}$ or 2. Thus
(2.12c) will be positive, so either $M_{J,J}\ne 0$ or $M_{J,J^{-1}}\ne 0$.
In other words, for any sl$(3)_k$ modular invariant $M$, either $J\in
\E_M$ or $J\in\E_{MC}=\E_M^C$.

What we find, in this way, for an arbitrary current algebra, is:

\medskip\noindent{{\bf Proposition 2.}} Let $M$ be a modular invariant for
some current algebra $X_{r,k}$ and let $\E_M$ and $\E_M^C=\E_{MC}$ be the
sets of exponents, where $C$ is charge-conjugation (2.1).

\item{(i)} For any sl$(2)_k$, so$(2n+1)_k=B_{n,k}$, sp$(2n)_k=C_{n,k}$, and
$E_{7,k}$, we  have $J\in\E_M$.

\item{(ii)} For sl$(n)_k=A_{n-1,k}$ when $n<8$, as well as $E_{6,k}$,
either $J\in\E_M$ or $J\in\E_M^C$.

\item{(iii)} More generally, for sl$(n)_k=A_{n-1,k}$, define $n'=n$ or $n/2$
when $n$ is odd or even, resp., and similarly $k'=k$ or $k/2$ when $k$ is
odd or even, resp. Define $a_i$ by the
prime decomposition $n'=\prod_i p_i^{a_i}$,
and let $s=\prod_i p_i^{\lfloor a_i/2\rfloor}$.
Assume gcd$(n',k')$ equals 1 or a power of a single prime. Then there exists
an automorphism invariant (=`pairing') $\omega$ such that the
simple-current $J^s=k\L_s$ lies in $\E_M^\omega=\E_{M\omega^t}$.

\item{(iv)} For so$(2n)_k=D_{n,k}$, when 4 doesn't divide $n$,
the simple-current $J_v=k\L_1$ lies in $\E_M$.\medskip

The simple-current $J$ in (i)-(iii) is any generator of the corresponding
(cyclic) groups of simple-currents. By `$\lfloor a_i/2\rfloor$' here we mean to truncate
to the nearest integer not greater than $a_i/2$. Note that $s$ is the
largest number such that $s^2$ divides $n$ or (if $n$ is even) $n/2$.
For instance $s=1,2,3$ for $n=4,8,18$.
The automorphism invariants $\pi$
for sl$(n)_k$ are explicitly given in [24].
To prove (iii), use (2.12) to obtain $M_{J^s,J^{s\ell}}=1$ for some integer
$\ell$ congruent to $\pm 1$ modulo an appropriate power of each prime $p_i$; the automorphism
invariants $\omega$ (when they exist) can be seen to reverse those signs.

When instead distinct primes
divide gcd$(n',k')$, a multiple $s'$ of $s$ will work in (iii): namely,
choose any prime (say $p_1$) dividing both $n'$ and $k'$, and define
$s'=p_1^{\lfloor a_1\rfloor}\prod_i p_i^{\lfloor a_i/2\rfloor}\prod_j
p_j^{a_j}$ where the $p_i$ don't divide $k'$, and the $p_j$ ($j\ne 1$) do.

More generally, suppose some weight $\kappa$ has the
property that, for any $\la,\mu$, we have
$$T_{\la\la}=T_{\mu\mu}\ \Rightarrow\ {\rm either}\
S_{\la\kappa}\,S_{\mu\kappa}\ge 0\ {\rm or}\
S_{\la\kappa}\,S_{\mu\kappa}^*\ge 0\eqno(2.12d)$$
Then, as in (2.12c), for any modular invariant $M$ we must have either 
$\kappa\in\E_M$ or $\kappa\in\E_M^C$.

\bigskip\noindent{{\it 2.3. Quick review of matrix and graph theory.}}\medskip

We will write $A^t$ for the transpose of $A$.
By a $\Z_\ge$-matrix we mean a matrix whose entries are nonnegative integers.
Two $n\times n$ matrices $A$ and $B$ are called {\it
equivalent} if there is a permutation which, when applied simultaneously
to the rows and columns of $A$, yields $B$ --- i.e.\ $B=\Pi A\Pi^t$.
The direct sum $A\oplus B$ of an $n\times n$
matrix $A$ and $m\times m$ matrix $B$ is the $(n+m)\times(n+m)$ matrix
$\left(\matrix{A&0\cr 0&B}\right)$. A matrix $M$ is called {\it decomposible}
if it can be written in the form (i.e.\ is equivalent to) $A\oplus B$, otherwise it is called
{\it indecomposible}.
A matrix $N$ is called {\it
reducible} if it is equivalent to a matrix of the form $\left(\matrix{
A&B\cr 0&C}\right)$ for submatrices $A,B,C$ where $B\ne 0$. Fortunately,
all of our matrices turn out to be irreducible.

For example, two $n\times n$ permutation matrices $\Pi$ and $\Pi'$ are equivalent
iff the corresponding permutations $\pi$ and $\pi'$ are conjugate in the
symmetric group $S_n$ (i.e.\ have the same numbers of disjoint 1-cycles, 2-cycles,
3-cycles, etc). They will be indecomposible iff they are transitive,
i.e.\ iff they are equivalent to the $n\times n$ matrix
$$\Pi^{(n)}:=\left(\matrix{0&0&\cdots&0&1\cr 1&0&\cdots&0&0\cr 0&1&&0&0\cr
\vdots&&\ddots&\vdots&\vdots\cr 0&0&\cdots&1&0}\right)\eqno(2.13)$$
in which case they will also be irreducible.

The eigentheory (i.e.\ the Perron-Frobenius theory --- see e.g.\ [30]) of nonnegative matrices
is fundamental to the study of NIM-reps. The basic result is that if $A$
is a square matrix with nonnegative entries, then there is an eigenvector
with nonnegative entries whose eigenvalue $r(A)$ is nonnegative. The
eigenvector(resp., -value) is called the {\it Perron-Frobenius eigenvector(-value)}.
This eigenvalue has the additional property that if $s$ is any other eigenvalue of
$A$, then $r(A)\ge|s|$.  There are many other results in Perron-Frobenius
theory that we'll use, but we'll recall them as needed.

The matrices with small $r(A)$ have been classified (see especially [31]
for $r(A)<\sqrt{2+\sqrt{5}}\approx 2.058$), but unfortunately with a weaker notion
of `equivalence' than we would like. The moral of the story seems to be that
these matrix classifications are very difficult, and will be hopeless as
$r(A)$ gets much larger than 2; the only hope is to simultaneously
impose other conditions
on the matrix, e.g.\ some symmetries. Fortunately, we can always find other
conditions obeyed by our matrices, besides the value of $r$.

This is one of the places where `irreducibility' simplifies things. For instance,
$\left(\matrix{1&k\cr 0&1}\right)$, $\forall k$, are some of the indecomposable $\Z_{\ge}$-matrices with
maximum eigenvalue $r(A)=1$, 
but the only $2\times 2$ indecomposable {\it irreducible} $\Z_\ge$-matrix with $r(A)=1$
is $\Pi^{(2)}$.

An irreducible $\Z_\ge$-matrix will have at most $r(A)^2$ nonzero entries
in each row, and so for small $r(A)$ will be quite sparse. A sparse
matrix is usually more conveniently depicted as a {\it (multi-di)graph}.
For example,
in Lie theory a Dynkin diagram replaces the Cartan matrix. The same trick
is used here, and is responsible for the beautiful pictures scattered
throughout the NIM-rep literature (see e.g.\ [7,9]).

By  a {\it graph} we allow loops (i.e.\ an edge starting and ending at
the same vertex), but its edges aren't directed and aren't multiple.
A multi-digraph is the generalisation which allows multiple edges and
directed edges. We assign a multi-digraph to a $\Z_\ge$-matrix $A$ as follows.
For any $i, j$,
draw $A_{ij}$ edges directed from vertex $i$ to vertex $j$. Replace
each pair $i\rightarrow j$,$j\rightarrow i$ of directed edges with an
undirected one connecting $i$ and $j$ (except we never put arrows on loops).

We are very interested in the spectra of (multi-di)graphs, i.e.\ the
list of eigenvalues (with multiplicities) of the associated adjacency matrix.
There has been
a lot of work on this in recent years --- see e.g.\ the readable book
[32]. We will state the results as we need them. A major lesson from the
theory: the eigenvalues usually won't determine the  graph. For example,
the graphs $D_4^{(1)}$ and $A_3^{(1)}\cup A_1$ have identical spectra.

There are no nonzero irreducible $\Z_\ge$-matrices with $r(A)<1$. The only
$n\times n$ irreducible indecomposable
$\Z_\ge$-matrix with $r(A)=1$ is $\Pi^{(n)}$ in (2.13), up to equivalence. 
The connected graphs ${\cal G}$ with $r({\cal G})<2$ or $r({\cal G})=2$ ---
i.e.\ symmetric
indecomposable matrices with $r(A)<2$ or $r(A)=2$ --- are given in Figures 1 and
2, and the loop-less ones (=traceless matrices) fall into the famous A-D-E
pattern (in fact these seem to be the two prototypical A-D-E patterns).
Incidentally, Tables 1 and 2 give all the eigenvalues of the graphs in
Figures 1 and 2, respectively. For Figure 1 these eigenvalues are
$2\cos(\pi m_i/h)$.

\bigskip\centerline {{\bf Table 2. Eigenvalues of graphs in Figure 2}}
$$\vbox{\tabskip=0pt\offinterlineskip
  \def\tablerule{\noalign{\hrule}}
  \halign to 5in{
    \strut#&\vrule#\tabskip=0em plus1em &    
    \hfil#&\vrule#&\hfil#&\vrule#&    
    \hfil#&\vrule#\tabskip=0pt\cr\tablerule         
&&\omit\hidewidth Graph \hidewidth&&\omit\hidewidth eigenvalues \hidewidth&&
\omit\hidewidth range \hidewidth&\cr\tablerule
&& \hfill $A_n^{(1)}$, $n\ge 1$ \hfill && \hfill $2\cos(2\pi k/(n+1))$ \hfill && \hfill  $0\le k\le n$\hfill & \cr
&&\hfill  $D_n^{(1)}$, $n\ge 4$\hfill  && \hfill $0,0,2\cos(\pi k/(n-2))$ \hfill && \hfill $0\le k\le n-2$ \hfill & \cr
&&\hfill  $E_6^{(1)}$\hfill  && \hfill $\pm 2,\pm 1,\pm 1,0$\hfill  && &\cr
&&\hfill  $E_7^{(1)}$\hfill  &&\hfill  $\pm2,\pm\sqrt{2},\pm 1,0,0$\hfill  && & \cr
&&\hfill  $E_8^{(1)}$\hfill  && \hfill $\pm 2,\pm 2\cos(\pi/5),\pm 1,\pm 2\cos(2\pi/5),0$\hfill  && & \cr
&&\hfill  ${}^0\!A_n^0$, $n\ge 1$\hfill  && \hfill $2\cos(k\pi/n)$\hfill  && \hfill $0\le k<n$\hfill & \cr
&&\hfill  $D_n^0$, $n\ge 3$\hfill  && \hfill $0,2\cos(2\pi k/(2n-3))$ \hfill &&\hfill  $0\le k\le n-2$\hfill & \cr
\tablerule\noalign{\smallskip}
 }} $$

Let ${\cal G}$ be any multi-digraph. We'll write $r({\cal G})$ for the
Perron-Frobenius eigenvalue of its adjacency matrix. ${\cal G}$ is called
{\it bipartite} if its vertices can be coloured black and white, in such a
way that the endpoints of any (directed) edge are coloured differently.
For example, any tree is bipartite.
If ${\cal G}$ is connected and its adjacency matrix is irreducible, it will
be bipartite iff the number
$-r({\cal G})$ is also an eigenvalue of ${\cal G}$.

\bigskip\bigskip
\centerline{{\bf 3. NIM-reps}}\bigskip

\noindent{{\it 3.1.\ The physics of NIM-reps.}}\medskip

In this section we introduce the main subject of the paper: NIM-reps.
Recall the discussion in the Introduction. Fix an RCFT and a choice of
chiral algebra. In other words, we are given modular data $S$ and $T$
and a modular invariant $M$.
We are interested here in boundary conditions which are not only
conformally invariant,
but also don't break the given chiral algebra.

Let $x\in{\cal B}$ parametrise the $\Z_{\ge}$-basis for the boundary
states in the RCFT (or Chan-Paton degrees-of-freedom for an open string).  Consider the 1-loop
vacuum-to-vacuum amplitude of an open string, i.e.\ the amplitude
associated to a cylinder whose boundaries are
labelled with states $|x\rangle,|y\rangle$. Then we can write them as
$$\z_{x|y}(q)=\sum_{\la\in P_+}\N_{\la x}^y\,\chi_\la(q)\eqno(3.1a)$$
where $\N_{\la x}^y\in\Z_{\ge}$ and $\chi_\la(q)$ are the usual RCFT (e.g.\
current algebra) characters. The parameter $0<q<1$ parametrises the
conformal equivalence classes of cylinders, just as $|q|<1$ did for tori in
(2.8). Depending on how we choose the time direction, we can interpret the
cylinder either as a 1-loop open string worldsheet, or a 0-loop closed string
worldsheet; using this Cardy [1] derived (at least for $M=I$)
$$\z_{x|y}(q)=\sum_{\la\in P_+} \sum_{\mu\in\E}{U_{x\mu}\,S_{\la\mu}\,
U^*_{y\mu}\over S_{0\mu}}\,\chi_\la(q)\eqno(3.1b)$$
Here $\E$ is the exponents $\E_M$ of the modular invariant $M$ for the RCFT.
The matrix entries $U_{x\mu}$ (appropriately normalised) give the
change-of-coordinates from boundary states $|x\rangle$, $x\in{\cal B}$, to
the Ishibashi states $|\mu\rangle\!\rangle$, $\mu\in\E_M$.
The matrices $\N_\la$
given by
$$(\N_\la)_{xy}=\N_{\la x}^y=\sum_{\mu\in\E}{U_{x\mu}\,S_{\la\mu}\,
U^*_{y\mu}\over S_{0\mu}}\eqno(3.1c)$$
 constitute what we will soon call a NIM-rep. Note by taking complex conjugation of
 (3.1c) that $\N^t_\la=\N_{C\la}$. We will require that $U$ be unitary (although
 the physical reasons for this are not so clear). This gives us (3.2a)
 below.

\vfill\eject\noindent{{\it 3.2. Basic definitions.}}

\medskip\noindent{{\bf Definition 3.}} {\it By a} NIM-rep $\N$ {\it we mean an assignment
of a matrix $\N_\la$, with nonnegative integer entries, to each $\la\in P_+$
such that $\N$ forms a representation of the fusion ring:
$$\N_\la\,\N_\mu=\sum_{\nu\in P_+}N_{\la\mu}^\nu\,\N_\nu\eqno(3.2a)$$
for all primaries $\la,\mu,\nu\in P_+$, and also that}
$$\eqalignno{\N_0=&\,I&(3.2b)\cr \N_{C\la}=&\,\N_\la^t\qquad \la\in P_+&(3.2c)\cr}$$

The NIM-rep `${\cal N}$' should not be confused with the fusion `$N$'.
In (3.2c), `$C$' denotes charge-conjugation (2.1), and `$t$' denotes transpose.
Equation (3.2b) isn't significant, and serves to eliminate from consideration the trivial
$\la\mapsto (0)$. Further refinements of Definition 3 are probably desirable,
and are discussed in paragraphs {\bf (4)},{\bf (5)} in \S8.

 The {\it dimension} $n$ of a NIM-rep is the size $n\times
 n$ of the matrices $\N_\la$. Note that our definition is more general (i.e.\
 fewer conditions) than in older papers by (various subsets of) Di Francesco$\&$Petkova$\&$Zuber.
The {\it fusion graphs} of $\N$
 are the multi-digraphs associated to the matrices $\N_\la$. 

 We call two $n$-dimensional NIM-reps $\N,\N'$
 {\it equivalent}
 if there is an $n\times n$ permutation matrix $P$ (independent of $\la\in P_+$)
 such that $\N'_\la=P
\N_\la P^{-1}$ for all $\la\in P_+$. Obviously we can and should identify NIM-reps
equivalent in this sense. More generally, when that same relation holds
for some arbitrary  invertible (i.e.\ not necessarily permutation) matrix $P$,
we will call $\N$ and $\N'$ {\it linearly equivalent}. At least 3 distinct
NIM-reps for ${{\rm sl}}(3)_9$ have been found with identical exponents [7],
which shows that linear equivalence is strictly
weaker than equivalence (similar examples are known [11] for ${{\rm sl}}(4)_8$).
In fact, linear equivalence isn't important,
and doesn't respect the physics.

One way to build new NIM-reps from old ones $\N',\N''$ is {\it direct sum}
$\N=\N'\oplus\N''$: 
$$\N_\la:=\N'_\la\oplus\N''_\la=\left(\matrix{\N_\la'&0\cr 0&\N''_\la\cr}\right)$$
We call such a representation $\N$ {\it decomposable} (or {\it reducible});
$\N$ is {\it indecomposable} when the
$\N_\la$'s cannot be simultaneously put into block form.
Obviously, an arbitrary NIM-rep can always be written as (i.e.\ is
equivalent to) a direct sum of indecomposable NIM-reps, so it suffices to
consider the indecomposable ones. Physically, decomposable NIM-reps would correspond
to completely decoupled blocks of boundary conditions.
We will show in \S3.3 that for any given choice of modular data, there
are only finitely many indecomposable NIM-reps $\N$.

Two obvious examples of  NIM-reps are given by fusion matrices, namely
$\N_\la=N_\la$, and $\N_\la=N_\la^t$. Both are
indecomposable, but they are equivalent: in fact, $N_\la^t=CN_\la C^{-1}$.
We call this obvious NIM-rep the {\it regular} one.

The matrices $\N_\la$ of \S3.1 define a NIM-rep. Thus to
 any RCFT should correspond a NIM-rep, and it should play as fundamental
 a role as the modular invariant.

Let $\N$ be any NIM-rep.
Equation (3.2a) tells us that the matrices $\N_\la$ pairwise commute; (3.2c)
then tells us that they are normal. Thus they can be simultaneously
diagonalised, by a unitary matrix $U$. Each eigenvalue $e_\la(a)$ defines a
1-dimensional representation $\la\mapsto e_\la(a)$ of the fusion ring,
so $e_\la(a)={S_{\la\mu}\over S_{0\mu}}$ for some $\mu\in P_+$. 
Thus any NIM-rep will necessarily obey the Verlinde-like decomposition (3.1c),
for some multi-set $\E=\E(\N)$. We will call $\E$ the {\it exponents}
of the NIM-rep, by analogy with the A-D-E classification of $\widehat{{\rm sl}}(2)$.
Note that $\N$ and $\N'$
are linearly equivalent iff their exponents ${\cal E}(\N),{\cal E}(\N')$ are equal
(including multiplicities).

At first glance, there doesn't seem to be much connection between NIM-reps
and modular invariants. But the discussion in \S1 tells us that
the NIM-rep $\N$ and modular invariant $M$ of an RCFT should obey the compatibility relation
$$\E(\N)=\E_M\eqno(3.3)$$
Thus we want to pair up the NIM-reps with the modular invariants so that
(3.3) is satisfied; any
NIM-rep (resp.\ modular invariant) without a corresponding modular invariant
(resp.\ NIM-rep) can be regarded as having questionable physical merit
(more precisely, before a modular invariant is so labelled, all possible
pairings $\omega$ should be checked --- see \S2.2).

\medskip\noindent{{\bf Definition 4.}} {\it We call a modular invariant $M$}
NIMmed {\it if there exists a NIM-rep $\N$ compatible with $M$ in the sense
of $(3.3)$. Otherwise we call $M$} NIM-less.\medskip

For instance the regular NIM-rep $\N_\la=N_\la$ has exponents $\E=P_+$, as does
the modular invariant $M=I$. Thus they are paired up. It is easy to verify that
the only modular invariant $M$ with $\E_M\supseteq P_+$ is $M=I$, so
it is the unique modular invariant which can be paired with the regular
NIM-rep.
It would be interesting to find other indecomposable NIM-reps with $\E(\N)\supseteq P_+$.
The Cardy ansatz [1] is essentially the statement that $\E(\N)=P_+$ implies
$\N$ is the regular NIM-rep.

Suppose the RCFT has a discrete symmetry $G$. We can consider fields in the
theory with twisted, nonperiodic boundary conditions induced by the action
of $G$. The resulting partition functions ${\cal Z}_{g,g'}(\tau)$ (one for each
twisted sector of the theory) are {\it sub}modular invariants. The philosophy
of [33] is that what one can do (e.g.\ study NIM-reps) with the modular
invariant ${\cal Z}_{e,e}$, can be done as well for the submodular invariants
${\cal Z}_{g,g'}$ --- indeed this is a way of probing the global structure
of the theory. The material of this paper, e.g.\ the Thm.1$\leftrightarrow$Thm.3
correspondence,  should be generalised to this more general situation.

Let $\Gamma=\{\ga^{(1)},\ldots,\ga^{(g)}\}$ be any fusion-generator of $P_+$.
 From (3.1c) and (2.7) it is easy to see that
$$\N_\la=P_\la(\N_{\ga^{(1)}},\ldots,\N_{\ga^{(g)}})\qquad \forall \la\in P_+
\eqno(3.4)$$
for any NIM-rep $\N$.
Thus for $\widehat{{\rm sl}}(2)$ and $\widehat{{\rm sl}}(3)$ NIM-reps,
the entire $\N$ is uniquely determined by knowing the first-fundamental
$\N_{\L_1}$, or equivalently its fusion graph.
But for most choices of ${{\rm sl}}(n)_k$ (see \S2.1
for the complete answer), knowing $\N_{\L_1}$ 
is not enough to determine all of $\N$.

Several fusion graphs for $\widehat{{\rm sl}}(3)$ are given in [7]. They
make no claims for the completeness of their lists, and in fact it is
not hard to find missing ones. To give the simplest example, the 1-dimensional
 sl$(3)_3$ NIM-rep (given by the quantum-dimension 1, 2 or 3) is missing. Incidentally,
1-dimensional sl$(n)_k$ NIM-reps exist only for sl$(n)_1$, 
sl$(2)_2$, sl$(2)_4$, sl$(3)_3$, and sl$(4)_2$ (for a proof, see p.691 of
[34]).

\bigskip\noindent{{\it 3.3. The basic theory of NIM-reps.}}\medskip

This section is central to the paper. Most of the results here are new.
For simple examples  using them, see \S\S6,7. Although we go much further,
some consequences were already obtained in especially [35],
using more restrictive axioms.

Let $\N$ be any indecomposable NIM-rep. Let $\E(\N)$ be its exponents, and
for any exponent $\mu\in\E(\N)$, let $m_\mu$ denote the multiplicity.
Then the matrix $\sum_{\la\in P_+}\N_\la$ is strictly positive, and
$m_0=1$. More generally, the value of $m_0$ tells you how many indecomposable
summands $\N^i$ there are in a decomposable $\N=\oplus_i\N^i$.

To see this, write `$x\sim y$'  if $\N_{\la x}^y\ne 0$ for some $\la$. Then
this defines an equivalence relation on ${\cal B}$: $x\sim x$ because
$\N_{0}=I$; if $x\sim y$ then $y\sim x$, because $\N_{\la x}^y=\N_{C\la,y}^x$;
if $x\sim y$ (say $\N_{\la x}^y\ne 0$) and $y\sim z$ (say $\N_{\mu y}^z\ne 0$)
then $x\sim z$, because  $(\N_\la\N_\mu)_{xz}\ne 0$. So we get a partition
${\cal B}_i$ of ${\cal B}$ such that $\sum_{\la\in P_+}\N_\la$ restricted
to each ${\cal B}_i$ is strictly positive, but $(\sum_{\la\in P_+}\N_\la)_{xy}
=0$ when $x\in{\cal B}_i,y\in{\cal B}_j$ belong to different classes.
This tells us that $\N$ is the direct sum of the $\N^{(i)}$ (the restriction
of $\N$ to the subset ${\cal B}_i$), so $\N$ being indecomposable requires that
there be only one class ${\cal B}_i$ (i.e.\ that ${\cal B}_i={\cal B}$).
The reason this forces $m_0=1$ is because of Perron-Frobenius theory [30]:
 the multiplicity of the Perron-Frobenius eigenvalue for a strictly positive
matrix (e.g.\ $\sum_{\la\in P_+}\N_\la$) is $1$.

Consider $\N$ indecomposable. The Perron-Frobenius eigenspace of $\sum_\la\N_\la$
will then be one-dimensional, spanned by a strictly positive vector $\vec{v}$.
Now the simultaneous eigenspaces of the matrices $\N_\la$ will necessarily
be a partition of the eigenspaces of e.g.\ $\sum_\la\N_\la$. Thus $\vec{v}$
must be an eigenvector (hence a Perron-Frobenius eigenvector) of all $\N_\la$.
Suppose $\vec{v}$ corresponds to exponent $\mu\in\E(\N)$. Then its eigenvalues
$S_{\la\mu}/S_{0\mu}$ must all be positive. The only primary $\mu\in P_+$
with this property for all $\la\in P_+$, is $\mu=0$. This means that
we know the Perron-Frobenius eigenvalue of any matrix $\N_\la$: it's
simply the quantum-dimension
$$r(\N_\la)={S_{\la 0}\over S_{00}}\eqno(3.5)$$

Let $U$ be a unitary diagonalising matrix of all $\N_\la$, as in (3.1c) (its
existence was proved last subsection).
 $U$ will not be unique, but it can be chosen to have properties
reminiscent of $S$. In particular the column $U_{\updownarrow 0}$ can be
chosen to be the Perron-Frobenius eigenvector $\vec{v}$ (normalised),
so each entry obeys $U_{x0}>0$.
We will discuss $U$ in more detail in \S3.4.

This argument also tells us the important fact that if the matrix ${\cal N}_\la$ is a direct
sum of indecomposable matrices $A_i$, then each $A_i$ (equivalently each component of the
fusion graph of $\la$) must have the same maximal eigenvalue $r(A_i)=
{S_{\la 0}/ S_{00}}$. The reason is that ${\cal N}_\la$ has a strictly
positive eigenvector, namely $\vec{v}$.
Moreover, these matrices $A_i$ will be 
{\it irreducible} (see \S2.3 for the definition). This follows for example from Corollary 3.15 of [30]
--- in particular, the Perron-Frobenius vector for both $\N_\la$ and $\N_\la^t=
\N_{C\la}$ is the vector  $\vec{v}>0$.

Clearly, all ${\cal N}_\la$ are symmetric iff all exponents $\mu\in{\cal E}$
satisfy $\mu=C\mu$. More generally,
$$\N_\la=\N_\nu\qquad {\rm iff}\qquad S_{\la\mu}=S_{\nu\mu}\qquad \forall
\mu\in\E(\N)\eqno(3.6)$$
So for any simple-current $J$, $\N_J=I$ iff $Q_J(\mu)\in\Z$ $\forall\mu\in\E$,
in which case $\N_{J\la}=\N_\la$ $\forall \la\in P_+$. More generally, 
by (3.5) $\N_J$ will be a permutation
matrix. If we let $j$ denote the permutation of the vertices ${\cal B}$,
corresponding to $\N_J$, then the order
of $j$ will be the least common multiple of all the denominators of
$Q_J(\mu)$ as $\mu$ runs over $\E$. Thus the order of $j$ will always divide
the order of $J$. Moreover,
$$(\N_{J\la})_x^y=\N_{\la,jx}^y=\N_{\la,x}^{j^{-1}y}$$

We will prove in Theorem 3 below the very useful and nontrivial facts that
the multiplicity $m_J$ of any simple-current must be 0 or 1, and if it is 1 then $J$
will be a symmetry of $\E$ --- i.e.\ $m_{J\mu}=m_\mu$ for all $\mu\in\E$.
It follows from this that  the simple-currents in $\E$ form a group, which we'll denote $\E_{sc}$.

Fix any vertex $1\in \B$.
By an $\N_1$-grading $g$ we mean a colouring $g(x)\in\Q$ of the vertices
${\cal B}$ and colouring $g_\la\in\Q$ of the primaries $P_+$, such that
$g(1)\in\Z$ and
$$\N_{\la x}^y\ne 0\qquad\Rightarrow\qquad g_\la+g(x)\equiv g(y)\quad
{\rm (mod\ }1)\eqno(3.7)$$
Clearly the $\N_1$-gradings form a group under addition; different choices
of `1' yield isomorphic groups. Thm.3(viii)
says that this group is naturally isomorphic to the group of
simple-currents in $\E$. In particular,  
to any simple-current $J\in\E$ we get an $\N_1$-grading as follows.
Define $g_\la=Q_J(\la)$, and put $g(y)=Q_J(\la)$
if $\N_{\la x}^y\ne 0$ for some $\la\in P_+$. This defines an $\N_1$-grading,
and we learn in Thm.3(viii) that all $\N_1$-gradings arise in this way.

Let $A$ be any matrix, and let $m_s$ be the multiplicity of eigenvalue $s$.
If all entries of $A$ are rational, then each eigenvalue $s$ will be an
algebraic number (since it's the root of a polynomial over $\Q$). If $\si$
is any Galois automorphism (of the splitting field of the characteristic
polynomial of $A$), and $s$ is any eigenvalue, then the image $\si(s)$
will also be an eigenvalue of $A$, and the multiplicities $m_s$ and
$m_{\si(s)}$ will be equal.

Now, $\si{S_{\la\mu}\over S_{0\mu}}={S_{\la,
\si\mu}/ S_{0,\si\mu}}$, by (2.6). So what this means is that the multiplicities
$m_\mu,m_{\si(\mu)}$ of $\mu$ and $\si(\mu)$ in the exponents $\E(\N)$ must be equal --- that is,
the exponents $\E(\N)$ obey the same Galois symmetry as the exponents
$\E_M$ (see Thm.1(iv)).

A special case of this is that $\la$ and $C\la$ have the same multiplicity.
That follows from (is equivalent to) the fact that  the entries $\N_{\la x}^y$ are all {\it real}.
The much more
general Galois symmetry follows from (and together with (3.10a) is equivalent
to) the much stronger statement that each $\N_{\la x}^y$
is {\it rational}.

Note that for any $\la\in P_+$,
$${\rm Tr}(\N_\la)=\sum_{\mu\in\E(\N)}{S_{\la\mu}\over S_{0\mu}}\in \Z_{\ge}
\qquad \forall\la\in P_+\eqno(3.8)$$
This is a strong condition for a multi-set $\E$ to obey --- see e.g.\ \S7. If $\E$
obeys the Galois condition $m_\la=m_{\si(\la)}$, as it must, then the sum in (3.8) will
automatically be integral, so the important thing in (3.8) is nonnegativity.

Suppose $\N$ is indecomposable. Then
$$\sum_{\la\in P_+}S_{0\la}\,{\rm Tr}(\N_\la)=\sum_{\la\in P_+}\sum_{\mu\in\E}
S_{0\la}\,{S_{\la\mu}\over S_{0\mu}}={1\over S_{00}}\eqno(3.9a)$$
All LHS terms are nonnegative. By considering the contribution
to the LHS by $\la=0$, we find that the dimension of an indecomposable NIM-rep
is bounded above by $S_{00}^{-2}$.

Moreover, each entry of $\N_\la$ must be
bounded above by the quantum-dimension $S_{\la 0}/S_{00}$. To see this,
note that the matrix $\N_\la\N_\la^t$
 has largest eigenvalue $r=(S_{\la 0}/S_{00})^2$; by Perron-Frobenius
theory any diagonal entry $A_{ii}$ of a nonnegative matrix $A$ will be bounded
above by $r(A)$. Thus for each $i,j$ we get
$$((\N_{\la})_{ij})^2\le (\N_\la\N_\la^t)_{ii}
\le (S_{\la 0}/S_{00})^2\eqno(3.9b)$$

Together, these two bounds tell us that {\it the number of indecomposable NIM-reps,
for a fixed choice of modular data, must be finite.}

We collect next the main things we've established.

\medskip\noindent{{\bf Theorem 3.}} Choose any modular data, and let
$\N$ be any indecomposable NIM-rep, with exponents $\E$ and multiplicities $m_\la$.

\item{(i)} For the given modular data, there are only finitely many
different indecomposable NIM-reps. We have the
bounds $(\N_\la)_{ij}\le S_{\la 0}/S_{00}$ and dim$(\N)\le S_{00}^{-2}$.

\item{(ii)} $m_0=1$.

\item{(iii)} For any simple-current $J$, either $m_J=0$ or 1; if
$m_J=1$ then $m_{J\la}=m_\la$ for any primaries $\la\in P_+$.

\item{(iv)} For any Galois automorphism $\si$ and primary $\la\in
P_+$, $m_{\si(\la)}=m_\la$.

\item{(v)} For any primary $\la\in P_+$, inequality (3.8) holds.

\item{(vi)} For any primary $\la\in P_+$, each indecomposable submatrix
of $\N_\la$ will be irreducible and have largest eigenvalue equal to
the quantum-dimension $S_{\la 0}/S_{00}$ of $\la$. The number of indecomposable
components will precisely equal the number of $\mu\in{\cal E}$ such that
$S_{\la\mu}/S_{0\mu} =S_{\la 0}/S_{00}$. The number of these components which
have a $\Z_m$-grading  is precisely the number of $\mu\in\E$ with $S_{\la\mu}/
S_{0\mu}=e^{2\pi\i/m}\,S_{\la 0}/S_{00}$.

\item{(vii)} No row or column of any matrix $\N_\la$ can be identically 0.

\item{(viii)} Fix any vertex $1\in{\cal B}$. The $\N_1$-gradings of the NIM-rep are in a natural one-to-one
correspondence with the simple-currents $J\in\E$.

\item{(ix)} Let $\E_{sc}$ denote the set of all simple-currents in $\E$,
${\cal S}_{sc}$ denote all simple-currents in $P_+$,
and ${\cal S}_0$ be the set of all simple-currents $J\in P_+$
such that $Q_J(J')\in\Z$ for all $J'\in\E_{sc}$. Then $\|{\cal S}_{sc}\|$
must divide $\|{\cal S}_0\|\,{\rm dim}(\N)$.

\item{(x)} If a primary $\la\in P_+$ has $Q_J(\la)\not\in\Z$ for some
simple-current $J\in\E$, then $\N_{\la x}^x=0$ for all $x\in\B$. 
\medskip

Note that the grading in (vi) applies to an individual matrix $\N_\la$, whereas
that of (viii) refers to a grading valid simultaneously for all matrices $\N_\la$.
Part (vii) comes from applying nonnegativity to $(\N_\la)(\N_\la)^t=I+\cdots$.
Part (x) comes from (3.8) and Thm.3(iii).
The remainder of the proof of Thm.3 is relegated to the end of the appendix.

Compare Theorems 1 and 3: surprisingly, the general properties
obeyed by the
exponents of a modular invariant, and those of a NIM-rep, match
remarkably well. It would be nice to obtain a simple, general, and effective
test for the NIM-lessness of a modular invariant. One candidate is Thm.3(ix):
this author has managed to show for modular invariants, only the weaker
statement that $\|{\cal E}(M)_{sc}\|$ must divide $\|{\cal E}(M)_{sc}\cap{\cal S}_0\|\,
{\rm Tr}(M)$, where ${\cal E}(M)_{sc}$ equals the number of simple-currents
in ${\cal E}_M$.

Thm.3 assumes all $S_{\la 0}>0$. For nonunitary RCFT, let $o\in P_+$ be as
in \S2.1. Then 3(ii) becomes $m_o=1$, but $m_0$ seems unconstrained.
The bound on dim$(\N)$ is now $S_{0o}^{-2}$. In 3(iii) replace $m_J$ with
$m_{Jo}$.

There are several generic constructions of NIM-reps, and 
a systematic study of these should probably be made. We will only mention one, which seems
to have been over-looked in the literature. It involves the notion of
{\it fusion-homomorphism}, i.e.\ a map $\pi:P_+\rightarrow P_+'$ between
the primaries of two (possibly identical) fusion rings, which defines a
ring homomorphism of the corresponding fusion rings: that is,
$$\pi(\la)\stimes\!'\pi(\mu)=\sum_{\nu\in P_+} N_{\la\mu}^\nu\pi(\nu)$$
where $\stimes\!'$ is the fusion product for $P_+'$.
See Prop.3 of [18] for its basic properties. In particular, there exists a
map $\pi':P_+'\rightarrow P_+$ such that [18]
$${{S}'_{\pi \la,\mu'}\over{S}'_{{0}',\mu'}}
={{S}_{\la,\pi'\mu'}\over{S}_{{0},\pi'\mu'}}$$
Also, $\pi \la=\pi \mu$ iff $\mu=J\la$ for
some simple-current $J$ with $\pi(J)=0$.

Suppose $\pi:P_+\rightarrow P_+'$ is a fusion-homomorphism,
and ${\N}$ is a NIM-rep of $P_+'$. Then ${\N}{}^\pi$
defined by $({\N}{}^\pi)_\la={\N}_{\pi \la}$ is a (usually decomposable)
NIM-rep of $P_+$. For a trivial example,
when $\pi$ is a fusion-isomorphism, and $\la\mapsto N_\la$ is
the regular(=fusion matrix) NIM-rep, 
then $\la\mapsto N_{\pi \la}$ will be equivalent  to the regular NIM-rep
(permute the rows and columns by $\pi$).

The exponents ${\cal E}({\N}{}^\pi)$ of $\N{}^\pi$ is the multi-set
$\pi'({\cal E}(\N))$.
If $\pi$ is onto, then it can be shown using [18] that
${\cal N}^\pi$ will be indecomposable iff ${\cal N}$ is.

%
%
%

\bigskip\noindent{{\it 3.4. The diagonalising matrix $U$ and the Pasquier
algebra.}}\medskip

Consider now the diagonalising matrix $U$ of (3.1c). In the event where
some multiplicities $m_\mu$ are greater than 1, it will be convenient at
times to introduce the following explicit notation for the entries of $U$:
write $U_{x,(\mu,i)}$, where $1\le i\le m_\mu$.

We would expect the diagonalising matrix $U$ to obey essentially
the same properties as $S$, except symmetry $S=S^t$ of course (the columns
and rows are labelled by completely different sets $P_+$ and ${\cal B}$).

However, the unitary matrix $U$ is not uniquely determined by (3.1c): for an exponent
$\mu\in\E$ with multiplicity $m_\mu$, we can choose for the $m_\mu$ columns
corresponding to $\mu$ any orthogonal basis of the corresponding eigenspace --- i.e.\
the freedom is parametrised for each $\mu\in\E$ by an $m_\mu\times m_\mu$ unitary matrix
$A^{(\mu)}\in {\rm U}(m_\mu)$. Explicitly, an alternate matrix $U'$ would
be given by the formula
$$U'_{x,(\mu,i)}=\sum_{j=1}^{m_\mu}A_{ij}^{(\mu)}\,U_{x,(\mu,j)}$$
The question we address in this subsection is, is there a preferred choice
for $U$ which realises most of the symmetries of the $S$ matrix which we saw
in \S2.1?

We claim only that the `preferred' matrix $U$ constructed below, diagonalises
the $\N_\la$ as in (3.1c). Its relation to the change-of-coordinate matrix
$U$, which goes from the boundary condition basis $|x\rangle$ to the
Ishibashi states $|\mu\rangle\!\rangle$, is uncertain, although the following
properties are all natural.

As mentioned in \S3.3, the $\mu=0$ column can (and will) be chosen to be
strictly positive. Fix any $\mu\in\E$. Let ${\Bbb K}_\mu$ be the number field generated by
$\Q$ and all ratios $S_{\la\mu}/S_{0\mu}$, for $\la\in P_+$. Then for each
$1\le i\le m_\mu$, we can require that all entries $U_{x,(\mu,i)}$ lie
in a quadratic extension ${\Bbb K}_\mu^i$ of ${\Bbb K}$. For any Galois
automorphism $\si \in {\rm Gal}({\Bbb K}_\mu^i/\Q)$, we can require
$$\si U_{x,(\mu,i)}=\eps_\si(\mu,i)\,U_{x,(\si \mu,i)}\eqno(3.10a)$$
where $\mu\mapsto \si\mu$ is the permutation of (2.6), and where
$\eps_\si(\mu,i)\in\{\pm 1\}$. We will prove this shortly. Also, fix any vertex
$1\in {\cal B}$; we can require $U$ to satisfy
$$U_{x,(J\mu,i)}=e^{2\pi\i\,g(x)}\,U_{x,(\mu,i)}\qquad \forall J\in\E,\
\mu\in\E,\ x\in\B,\ 1\le i\le m_\mu\eqno(3.10b)$$
where $g$ is the $\N_1$-grading associated to $J$ by Thm.3(viii).
Conversely, let $J\in P_+$ be any simple-current, and write $\N_{J,x}^y=
\delta_{y,jx}$ for the appropriate permutation $j$
of $\B$. Then each column $U_{\updownarrow,(\mu,i)}$ is an eigenvector of
$\N_J$ with eigenvalue $e^{2\pi\i\,Q_J(\mu)}$, that is to say
$$U_{jx,(\mu,i)}=e^{2\pi\i\,Q_J(\mu)}\,U_{x,(\mu,i)}\eqno(3.10c)$$

Incidentally, the relation (3.10a) allows us to prove the rationality of the
coefficients of the so-called {\it dual Pasquier algebra}
(or $\widehat{\N}$-algebra). Assume there is some vertex $1\in\B$ such that
the row $U_{1,(\mu,i)}\ne 0$ for all $\mu,i$. Define  [16]
$$\widehat{\N}_{xy}^z:=\sum_{{\mu\in\E\atop 1\le i\le m_\mu}}{U_{x,(\mu,i)}\,
U_{y,(\mu,i)}\,U_{z,(\mu,i)}^*\over U_{1,(\mu,i)}}$$
Then for {\it any} such choice of $1\in\B$, (3.10a) tells us
$$\si\,\widehat{\N}_{xy}^z=
\sum_{{\mu\in\E\atop 1\le i\le m_\mu}}{\eps_\si(\mu,i)\, U_{x,(\si\mu,i)}\,
\eps_\si(\mu,i)\,U_{y,(\si\mu,i)}\,\eps_\si(\mu,i)\,U_{z,(\si\mu,i)}^*\over
\eps_\si(\mu,i)\,U_{1,(\si\mu,i)}}=\widehat{\N}_{xy}^z$$
for all Galois automorphisms $\si$. This is precisely the statement that
each coefficient $\widehat{\N}_{xy}^z$ is rational. This result is new,
although it had been empirically observed in e.g.\ [16] that the coefficients
$\widehat{\N}_{xy}^z$ for each of the then-known NIM-reps always seemed to
be rational.

Ideally, we would like the coefficients $\widehat{\N}$ to be nonnegative
integers. In this case the Perron-Frobenius eigenvalue of $\widehat{\N}_x$
would be given by $U_{x,(0,1)}/U_{1,(0,1)}$, and hence we would have the inequalities
$$\eqalignno{U_{1,(0,1)}\le&\,U_{x,(0,1)}&(3.11a)\cr
|U_{x,(\mu,i)}/U_{1,(\mu,i)}|\le&\,U_{x,(0,1)}/U_{1,(0,1)}&(3.11b)}$$
In particular, (3.11a) is the statement that a normal $\Z_{\ge}$-matrix $A\ne 0$
must have $r(A)\ge 1$, and (3.11b) says that whenever $A\ge 0$ then $|s|\le r(A)$
for any eigenvalue $s$ of $A$. The inequality (3.11a) justifies the empirical
rule of [16] for choosing the vertex $1\in{\cal B}$.

For example, consider the sl$(2)_{16}$ NIM-rep called $E_7$: its diagonalising
matrix $U$ is given in [6]. We can see by inspection that its dual Pasquier
coefficients cannot all be in $\Z_{\ge}$. In particular, (3.11a) identifies
the vertex 1, and then we find (3.11b) is not satisfied.

On the other hand, the coefficients of the {\it Pasquier algebra} (or ${\cal
M}$ algebra) [16]
$${\cal M}_{(\la,i),(\mu,j)}^{(\nu,k)}:=\sum_{x\in\B} {U_{x,(\la,i)}\,U_{x,(\mu,j)}
\,U^*_{x,(\nu,k)}\over U_{x,(0,1)}} $$
will in general {\it not} be rational --- they will be rational iff the
analogue of (3.10a) holds for rows. The coefficients ${\cal M}$ can be rational,
only when all entries of $U$ lie in a cyclotomic field (the proof in [23] for $S$ works
here).   
We will return to this shortly.

See e.g.\ [6] for a discussion of the (dual) Pasquier algebra.
Note that our matrix $U$ is denoted there by $\psi$, and our $\B$ is
denoted there by ${\cal V}$.
It has appeared in other related contexts --- see e.g.\ the {\it classifying
algebra} in e.g.\ [14] and references therein.
Unlike fusion coefficients, neither the
coefficients ${\cal M}$ nor $\widehat{\N}$ need be integral or nonnegative,
and both depend on the choice of $U$.

To see (3.10a), first note that finding an orthogonal basis of eigenvectors
for the $\mu$-eigenspace amounts
to solving a system of linear equations with coefficients in the
cyclotomic field ${\Bbb K}_\mu$. Find any such basis $\vec{u}_{(\mu,i)}$,
so that each 1-coordinate $(\vec{u}_{(\mu,i)})_1$ is rational.
Then hit these vectors $\vec{u}_{(\mu,i)}$ componentwise
 by $\si$ to yield an orthogonal basis of eigenvectors for the $\si(\mu)$-eigenspace.
Note that
$\si(\mu)=\mu$ iff the automorphism $\si$ is trivial in ${\Bbb K}_\mu$, so these bases will be
well-defined. When $\si(\mu)=J\mu$ for some simple-current $J\in\E$, then (3.10b)
will be automatic; otherwise note from the proof of Thm.3(viii) given in the
appendix that the vectors $(\vec{u}^J_{(\mu,i)})_x:=e^{2\pi\i \,g(x)}(\vec{u}_{(\mu,i)})_x$
are orthogonal eigenvectors
for $J\mu$. Run this construction through a set of representatives $\mu$ of the orbits
in $\E$ of the group $\langle{\rm Gal}(\overline{\Q}/\Q),\E_{sc}\rangle$;
normalising the resulting eigenbases (this is where the quadratic extensions
${\Bbb K}_\mu^i$ and the signs $\epsilon_\si$ arise), gives a unitary
diagonalising matrix $U$ satisfying (3.10).

Unlike the entries of $S$, those of $U$ will {\it not} in general lie in a
cyclotomic field, and there won't in general be a Galois action on the
{\it rows} of $U$. A simple example of this is the sl(2)$_{10}$
exceptional called $E_6$, whose diagonalising matrix is [6]
$$ U={1\over 2}\left(\matrix{a&{1}&b&b&{1}&a\cr
b&{1}&a&-a&-{1}&-b\cr c&0&-d&-d&0&c\cr b&-{1}&a&-a&
{1}&-b\cr a&-{1}&b&b&-{1}&a\cr d&0&-c&c&0&-d}\right)$$
where $a,b$ equal $\sqrt{(3\mp\sqrt{3})/6}$, respectively, and $c,d$ equal
$\sqrt{2}a,\sqrt{2}b$, respectively. Note first that $\sqrt{3+\sqrt{3}}$
does not lie in any cyclotomic field, and so neither do $a,b,c,d$. In fact
the smallest normal extension of $\Q$ containing $\sqrt{3+\sqrt{3}}$ is
$\Q[\sqrt{2},\sqrt{3+\sqrt{3}}]$ (note that $\sqrt{3+\sqrt{3}}\,\sqrt{3-\sqrt{3}}
=\sqrt{3}\,\sqrt{2}$), and the corresponding Galois group is the nonabelian
quaternion group $Q_8=\{\pm 1,\pm {\rm i},\pm {\rm j},\pm {\rm k}\}$.
The Galois automorphism sending $\sqrt{2}$ to itself and $\sqrt{3\pm\sqrt{3}}$
to $\sqrt{3\mp\sqrt{3}}$ interchanges for instance columns 1 and 3 with $\eps=-1$,
but doesn't send the first row anywhere. ($U$ here is unique, up to phases
for each column; no choice of phases however will give us a cyclotomic field.)

Of course, the simplest and most important example of a Galois
automorphism is complex conjugation $z\mapsto z^*$. Eq.(3.10a) becomes
$$U_{x,(\mu,i)}^*=U_{x,(C\mu,i)}\eqno(3.12a)$$
where $\mu\mapsto C\mu$ is charge-conjugation (2.1) --- the parity
$\epsilon_*(\mu,i)$ in (3.10a) will be +1 here because the
normalisation of the columns of $U$ only involves rescaling by a real
number. Using the facts that $U$ is unitary and $C$ is an involution,
we get that $U^tU$ is a permutation matrix:
$$(U^tU)_{(\mu,i),(\nu,j)}=\delta_{\nu,C\mu}\,\delta_{j,i}\eqno(3.12b)$$
In many examples, the analogue of (3.12a) for rows also holds: that
is, there is an invertible involution $\iota$ of ${\cal B}$ such that
$$U_{x,(\mu,i)}^*=U_{\iota x,(\mu,i)}\eqno(3.12c)$$
When this holds, we get $\N_{C\la,\iota x}^{\iota y}=\N_{\la x}^y$ and
$(UU^t)_{xy} =\delta_{y,\iota x}$. Since Tr$(U^tU)={\rm Tr}(UU^t)$, the
number of fixed-points of $\iota$ would equal the number of $\mu\in\E$
with $C\mu=\mu$, counting multiplicities. It is easy to show that
$\iota$ exists iff the NIM-rep $\la\mapsto \N_{C\la}$ is equivalent to
$\la\mapsto \N_\la$ --- even when $\iota$ doesn't exist, they will be {\it linearly equivalent}.
Also, $\iota$ exists iff the corresponding Pasquier algebra ${\cal M}$ has
{\it real} structure constants. The existence of $\iota$ is assumed in
the axioms of [7,16] and it holds in all examples of NIM-reps known to
this author, but probably NIM-reps without an $\iota$ can be
found for sl$(3)_k$ or sl$(4)_k$.

\bigskip\bigskip\centerline{{\bf 4. The current algebras at level 1}}\bigskip 

In the next two sections we obtain several new NIM-rep classifications for the
current algebras, and compare them to the corresponding modular invariant
classifications.

We begin in \S4.1 by finding all NIM-reps for any modular data obeying the 
restrictive property that all primaries are simple-currents. This allows
us immediately to do all simply-laced current algebras at level 1. The
NIM-reps for the ${B}^{(1)}$- and ${C}^{(1)}$-series at level 1 follow from the
$\widehat{{\rm sl}}(2)$ classification,  so we
repeat the $\widehat{{\rm sl}}(2)$ classification  in
\S4.3. 

In all these cases, the NIM-rep and modular invariant classifications match up
fairly well: each modular invariant has a unique NIM-rep, and most NIM-reps are paired
with a unique modular invariant. The only interesting situation here is
${\rm so}(8n)_1$, where different modular invariants correspond to identical
NIM-reps.

Note that NIM-reps (unlike modular invariants) depend only on the fusion
ring. When two fusion rings are isomorphic, their NIM-reps will be identical.
In [29] we found all isomorphisms $X_{r,k}\cong X'_{r',\ell'}$
among the fusion rings of current algebras.
The complete list is: ${{\rm sp}}(2n)_k\cong{{\rm sp}}(2k)
_n$ for all $n,k$; all ${{\rm so}}(2n+1)$ at level
1 are isomorphic to ${{\rm sl}}(2)_2\cong{{\rm sp}}(4)_1\cong
{E}_{8,2}$; ${{\rm sl}}(2)_k\cong{{\rm sp}}(2k)_1$; 
${{\rm so}}(2n)_1\cong{{\rm so}}(2m)_1$ whenever $n\equiv m$ (mod 2),
and in addition odd $m$ are isomorphic to ${{\rm sl}}(2)_2$;
${{\rm sl}}(3)_1\cong {E}_{6,1}$; ${{\rm sl}}(2)_1\cong
{E}_{7,1}$; ${F}_{4,1}\cong{G}_{2,1}$; ${F}_{4,3}\cong
{G}_{2,4}$; and finally ${E}_{8,3}\cong{F}_{4,2}$.

Coincidentally, when the fusion rings of $X_{r,k}$ and $X'_{r',k'}$
 are isomorphic, it turns out that their modular invariant classifications 
will usually be identical. The only exception is 
${{\rm so}}(4n)_1$, which has either 2 or 6 modular invariants, depending on
whether or not $n$ is odd.

\bigskip\noindent{\it 4.1. All primaries are  simple-currents.}\medskip

The simple-currents (i.e.\ the primaries with quantum-dimension 1 ---
see \S2.1) always
form an abelian group, called the {\it centre} of the modular data.
Any NIM-rep, when restricted to the centre, yields a group-representation of the
centre by permutation matrices. In this subsection we consider the
special case where all primaries $\la\in P_+$ are simple-currents (the modular data though
is otherwise general --- it may or may not come from a current algebra).

\medskip\noindent{{\bf Proposition 4.}} Consider any modular
data. Suppose all primaries in $P_+$ are simple-currents. 

\smallskip\item{(a)} The indecomposable NIM-reps are in one-to-one correspondence with the
subgroups ${\cal J}$ of the centre: ${\cal J}\leftrightarrow\N({\cal J})$.
The exponent $\E$ of the NIM-rep $\N({\cal J})$ is ${\cal J}$. (We will
explicitly construct $\N({\cal J})$ below.) The NIM-rep is uniquely specified by
its exponents.

\smallskip\item{(b)} The exponent of any modular invariant is a subgroup of the
centre. Thus any modular invariant is NIMmed. However, some subgroups
(hence NIM-reps) may be realised by none or by several modular invariants.
There may be more/less/the same number of modular invariants as NIM-reps.\medskip

In particular, choose any subgroup ${\cal J}$ of the centre $P_+$, and put
$k=\|{\cal J}\|$. Define a
$k$-dimensional NIM-rep as follows. Let ${\cal J}'$ be the
subset (in fact subgroup) of $P_+$, consisting of all primaries $J'$ for which
$Q_J(J')\in \Z$ for all $J\in{\cal J}$. There will be $\|P_+\|/k$
such $J'$. This is a subgroup because of the relation $Q_J(J'J'')=Q_J(J')+
Q_{J}(J'')$ which holds for any simple-currents $J,J',J''$, and which follows
immediately from (2.4). Now consider the quotient group
$P_+/{\cal J}'=\{[J_0],[J_1],\ldots,[J_{k-1}]\}$. It will in fact be
isomorphic to ${\cal J}$. Define the NIM-rep $\N({\cal J})$ by
$$(\N({\cal J})_J)_{ij}=\delta_{[JJ_i],[J_j]}\qquad \forall J\in P_+$$
So the rows and columns of $\N({\cal J})$ are essentially labelled by
the elements of $P_+/{\cal J}'$. To get that the exponents of $\N({\cal J})$
are ${\cal J}$, use the fact that $J'\in P_+$ is sent to $I$ iff $J'\in{\cal J}'$,
and so $Q_J(J')\in\Z$ for any exponent $J$ and any $J'\in{\cal J}'$.

The two extremes are when the subgroup is all of $P_+$, in which case the NIM-rep
is given by fusion matrices, and when the subgroup is $\{0\}$, in
which case the NIM-rep is the constant $\N_J=1$.


It is clear from Thm.1(iii) and Thm.3(iii) that the exponents of a modular invariant
and a NIM-rep must both form a subgroup of the centre $P_+$. It is not
obvious that there is only one NIM-rep realising that subgroup.
To see the general argument, it is perhaps easiest to consider an example:
$P_+\cong \Z_4\times\Z_3\times\Z_3\cong{\cal J}$.
 Let $J_1,J_2,J_3$ be the corresponding generators.
Let $\N$ be any NIM-rep with exponents $P_+$. We know from 
Thm.3(x) that Tr$(\N_J)=0$
provided $J\ne 0$, so the permutation associated to $\N_J$, for any
$J\ne 0$, can have no fixed-points. Thus the permutation associated to $\N_{J_1}$ must be
a disjoint product of nine 4-cycles. By relabelling the rows/columns appropriately,
we may take it to send $i+4j+12k$ ($i\in\Z_4,
j\in\Z_3,k\in\Z_3$) to $(i+1\,({\rm mod\ 4}))+4j+12k$.
Likewise, $\N_{J_2}$ must be a
disjoint product of 12 3-cycles, and it must commute with $\N_{J_1}$, so
we may take the corresponding permutation to send $i+4j+12k$ to $i+
4(j+1\,({\rm mod \,3}))+12k$.  The matrix $\N_{J_3}$ is handled similarly;
none of its 3-cycles can coincide with those of $\N_{J_2}$ because otherwise
$\N_{J_3}\N_{J_2}^{-1}=\N_{J_3J_2^{-1}}$ would have fixed points and nonzero
trace. So we can likewise fix $\N_{J_3}$. Manifestly, the resulting NIM-rep
is the regular NIM-rep corresponding to the fusion matrices.

\bigskip\noindent{\it 4.2. The simply-laced algebras at level} $1$. \medskip

The algebra $\widehat{\rm sl}(n)=A_{n-1}^{(1)}$, $n\ge 2$,  at level 1 has $n$ primaries,
$P_+=\{0,\L_1,\ldots,\L_{n-1}\}$. Put $\L_0=0$, then $\L_i=J^i$ for the simple-current
$J=\L_1$. The centre of sl$(n)_1$ is the cyclic group $\Z_n$, so there is an indecomposable
NIM-rep corresponding to each divisor $d$ of $n$. In particular, the
exponents will be generated by $J^d$, the subgroup ${\cal J}'$ defined
above will be generated by $J^{n/d}$, and the resulting NIM-rep will be
$n/d$-dimensional. This classification is given in [6].

There is a modular invariant, namely $M[J^d]$ in (2.9), for any divisor $d$ of $n$
for which $(n-1)d$ is even [36]. It has
exponents $\langle J^{n/d}\rangle$ and corresponds to the NIM-rep $\N(\langle
J^{n/d}\rangle)$.

 The algebra $\widehat{{\rm so}}(2r)={D}_r^{(1)}$, $r\ge 4$,
at level $1$ has 4 primaries $P_+=\{0,J_v=\L_1,J_s=\L_r,J_c=\L_{r-1}\}$,
 all of which are simple-currents. For $r$ odd they define the cyclic group
$\langle J_s\rangle\cong\Z_4$, while for $r$ even they define the
group $\langle J_v,J_s\rangle\cong \Z_2\times\Z_2$. Thus there are
precisely three indecomposable NIM-reps for $r$ odd --- one for each choice of exponents $\E=\{0\},
\{0,J_v\},\{0,J_v,J_s,J_c\}$. For $r$ even, there are precisely five indecomposable
NIM-reps --- one for each choice of exponents
$$\E=\{0\},\,\{0,J_v\},\,\{0,J_s\},\,\{0,J_c\},\,\{0,J_v,J_s,J_c\}$$

For $D_{r,1}$, when 4 does not divide $r$, there are only two modular
invariants [26]:
$M=I$ (which has exponents $\{0,J_v,J_s,J_c\}$) and $M=C_1$, the permutation
fixing 0 and $\L_1$ and interchanging $\L_r\leftrightarrow\L_{r-1}$
(which has
exponents $\{0,J_v\}$). When 4 divides $r$, there are six modular
invariants [26]: along with $I$ and $C_1$, these are
$M[J_s]$, $C_1\,M[J_s]$, $M[J_s]\,C_1$, and $C_1\,M[J_s]\,C_1$
(with exponents $\{0,J_s\},\{0\},\{0\},\{0,J_c\}$, resp.). In particular, both
$$C_1\,M[J_s]=\bigl(\chi_0+\chi_{\L_{r-1}}\bigr)\,\bigl({\chi}_0^*+
{\chi}_{\L_{r}}^*\bigr)\ ,\qquad
M[J_s]\,C_1=\bigl(\chi_0+\chi_{\L_{r}}\bigr)\,\bigl({\chi}_0^*+
{\chi}_{\L_{r-1}}^*\bigr)$$
 correspond to the identical NIM-rep (namely $\N_J=1$ $\forall J$).

The algebra $E_{6,1}$ has centre $\{0,\L_1,\L_5\}\cong\Z_3$, two indecomposable
NIM-reps, and two modular invariants ($M=I$ and $M=C$). The algebra
$E_{7,1}$ has centre $\{0,\L_6\}\cong\Z_2$, two indecomposable NIM-reps, and
one modular invariant ($M=I$). The algebra $E_{8,1}$ has trivial centre $\{0\}$,
one indecomposable NIM-rep, and one modular invariant.

\vfill\eject\noindent{\it 4.3. The algebra} $\widehat{{\rm sl}}(2)={A}_1^{(1)}$, 
{\it at level} $k$.\medskip

Because we'll be needing it in the next two subsections, we repeat
here the NIM-rep classification for $\widehat{{\rm sl}}(2)$, which was
 first given in [7].

Let $\N$ be any indecomposable NIM-rep of $A_{1,k}$. Its modular data is given
in \S2.1. A fusion generator for $A_{1,k}$ is $\L_1$, so it suffices to
give $\N_1=\N_{\L_1}$. For $k$ odd, the fusion graph for $\N_1$ is either
$A_{k+1}$ or the tadpole $T_{(k+1)/2}$ (see Figure 1). For $k$ even, the
possible fusion graphs are $A_{k+1}$ and $D_{k/2+2}$, except for $k=10,16$
or 28 where in addition there are $E_6,E_7,E_8$ respectively.

The modular invariants for $A_{1,k}$ were found in [27]. Each corresponds
to a unique NIM-rep, namely one of A-D-E type, as is well-known.

\bigskip\noindent{\it 4.4. The algebra} $\widehat{{\rm
so}}(2r+1)={B}_r^{(1)}$, {\it for  $r\ge 3$
at level} $1$.\medskip

The weights here are $P_+=\{0,\L_1,\L_r\}$.
For $B_{r,1}$ the only modular invariant [26] is the identity matrix
$I$. We learned above that its fusion ring is isomorphic to that of
${{\rm sl}}(2)_2$ (the isomorphism sends $\L_r$ to the fusion generator $\L_1$
of ${{\rm sl}}(2)_2$) and so we can read off its NIM-reps from the classification
of \S4.3: we find that there is only the `regular' one, given by the
fusion matrices, which assigns to the generator $\L_r$ the fusion graph $A_3$.

\bigskip\noindent{\it 4.5. The algebra} $\widehat{{\rm
sp}}(2r)={C}_r^{(1)}$, {\it for $r\ge 2$
at level} $1$.\medskip

Here, $P_+=\{0,\L_1,\ldots,\L_r\}$. Write $\L_0$ for 0.
The fusion-isomorphism between $C_{r,1}$ and $A_{1,r}$ identifies the
primary $\L_i$ of $C_{r,1}$ with the primary $i\L_1$ of $A_{1,r}$.
The NIM-reps for $C_{r,1}$ are thus of A-D-E or tadpole type, exactly as
in \S4.3. 

The modular invariants for $C_{r,1}$ [26] fall into the A-D-E pattern, and
are in a natural one-to-one correspondence with those of $A_{1,r}$ (again
using the identification $\L_i\leftrightarrow i\L_1$).

Thus the NIM-rep $\leftrightarrow$ modular invariant situation for $C_{r,1}$
is identical to that of $A_{1,r}$.

\bigskip\noindent{\it 4.6. The algebras ${G}_2^{(1)}$ and ${F}_4^{(1)}$
 at level $1$}.\medskip

$G_{2,1}$ has $P_+=\{0,\L_2\}$. We compute the quantum-dimension:
${S_{\L_2,0}\over S_{0,0}}={1+\sqrt{5}\over 2}$, the Golden Mean. Thus there's
a Galois automorphism $\sigma$ for which $\si{1+\sqrt{5}\over 2}=
{1-\sqrt{5}\over 2}={-2\over 1+\sqrt{5}}$. Applying that $\si$ to
the quantum-dimension and using (2.6), we see that $\si 0=\L_2$. Thus for any
(indecomposable) NIM-rep of $G_{2,1}$, $m_{\L_2}=m_0=1$, and the NIM-rep must be
2-dimensional. It is now trivial to find it:
$$\N_0=\left(\matrix{1&0\cr 0&1}\right),\qquad \N_{\L_2}=
\left(\matrix{1&1\cr 1&0}\right)$$
and the fusion graph of $\L_2$ is the tadpole $T_2$.

The only modular invariant [26] is $M=I$, which is paired with $T_2$.

The situation is completely identical for $F_{4,1}$: $P_+=\{0,\L_4\}$ here,
and the fusion-isomorphism identifies $\L_4$ with $\L_2$. There is again
only one NIM-rep and one modular invariant, and again the graph is the tadpole
$T_2$.

%

\bigskip\centerline{{\bf 5. The unitary and orthogonal algebras at level 2}}

\bigskip \noindent{\it 5.1. $\widehat{\rm sl}(n)$ at level $2$}.\medskip

\noindent Consider next $\widehat{\rm sl}(n)=A_{n-1}^{(1)}$ at level 2. The
weights $\la$ are all of the form $\la(ab):=\L_a+
\L_b$, for $0\le a,b<n$. Since $\la(ab)=\la(ba)$, we will usually require
$a\le b$. 

The simple-current $J$ and charge-conjugation $C$ act on $P_+$ by:
$$J\la(ab)=\la(a+1,b+1)\ ,\qquad C\la(ab)=\la(n-b,n-a)$$
$J$ has order $n$. For any divisor $d$ of $n$, we get
the modular invariant $M[J^d]$ for ${{\rm sl}}(n)_2$ given in (2.9),
where $Q_{J^d}(\la(ab))=d\,(a+b)/n$ and $R_{J^d}=2d$. For example, 
$M[J^n]=I$ and $M[J]=C$.

The remaining, exceptional, sl$(n)_2$ modular invariants ${\cal E}^{(n,2)}$ are [37]
$$\eqalignno{{\cal E}^{(10,2)}=\sum_{i=0}^9&\,|\chi_{\la(i,i)}+\chi_{\la(i+3,
i+7)}|^2+\sum_{i=0}^4|\chi_{\la(i,i+3)}+\chi_{\la(i+5,i+8)}|^2&\cr
{\cal E}^{(16,2)}=\sum_{i=0}^7&\,\bigl(\,|\chi_{\la(i,i)}+\chi_{\la(i+8,i+8)}|^2+
|\chi_{\la(i,i+4)}+\chi_{\la(i+8,i-4)}|^2+|\chi_{\la(i,i+8)}|^2&\cr
&+|\chi_{\la(i,i+6)}+\chi_{\la(i+8,i-2)}|^2
+(\chi_{\la(i+3,i+5)}+\chi_{\la(i-5,i-3)})\,
\chi_{\la(i,i+8)}^*&\cr
&+\chi_{\la(i,i+8)}\,(\chi_{\la(i+3,
i+5)}+\chi_{\la(i-5,i-3)})^*\,\bigr)&\cr
{\cal E}^{(28,2)}=\sum_{i=0}^{13}&\,\bigl(\,|\chi_{\la(i,i)}
+\chi_{\la(i+14,i+14)}+\chi_{\la(i+5,i-5)}+\chi_{\la(i-9,i+9)}|^2
&\cr&+|\chi_{\la(i+3,i-3)}+\chi_{\la(i-11,i+11)}+\chi_{\la(i+6,i-6)}+
\chi_{\la(i-8,i+8)}|^2\bigr)&\cr}$$
together with the matrix products $C\cdot {\cal E}^{(10,2)}$, $C\cdot {\cal
E}^{(16,2)}$, ${1\over 2}M[J^4]\cdot{\cal E}^{(16,2)}$, and $C\cdot
{\cal E}^{(28,2)}$.

 Note the strong resemblance of
the exceptional modular invariants here to the so-called $E_6,E_7,E_8$
exceptionals of $\widehat{{\rm sl}}(2)$ [27]. This is not a coincidence, and is a 
consequence of  a duality between $\widehat{{\rm sl}}(n)$ level $k$, and 
$\widehat{{\rm sl}}(k)$ level $n$. See also the resemblance between (A.1) and
the $S$ matrix for $\widehat{{\rm sl}}(2)$ level $n$.

We next turn to the NIM-reps. The proof that our list is complete,
is given in \S A.1. Write $n=2^hm$ where $m$ is odd.
We know $J=\la(11)$
and $\L_1=\la(01)$ are fusion-generators, so so are $J^{m}$, $J^{2^h}$
 and $\la:=J^{(m-1)/2}\L_1=\la({m-1\over 2},{m+1\over 2})$. Thus, the NIM-rep
 is uniquely defined once the matrices $A:=\N_\la$, $P':=\N_{J^{m}}$ and
 $P'':=\N_{J^{2^h}}$ are known. The reason it
 is more convenient to use these fusion-generators is Lemma A in
 the appendix --- roughly, the matrix $A$ is nearly symmetric, and
 its failure to be symmetric is governed by the permutation matrix $P'$.

The matrix
$A$ comes from the disjoint union of equivalent diagrams taken from
Figure 3. Each of those diagrams $X_n(k)$ corresponds to a digraph, as
follows. Number the diagram nodes from 1 to $n$, say
from left to right, top to bottom.
The weight ($k$ or $2k$) of each node tells how many
vertices are represented by that node. So each vertex in the digraph
will be labelled
by a pair $(v,i)$, where $v$ is the number of the node in the diagram,
and $i$ runs from 1 to the weight of that node (we take it modulo the weight).
Suppose nodes $v<v'$ are
adjacent to each other in the diagram. If they have identical weight ($k$ say),
put a directed edge from $(v,i)$ to $(v',i)$, and from $(v',i)$ to $(v,i+1)$.
If nodes $v<v'$ have weights $2k$ and $k$,
respectively, then draw directed edges from $(v,i)$ to $(v',i)$, and from
$(v',i)$ to both $(v,i+1)$ and $(v,i+k+1)$. 
If nodes $v<v'$ have weights $k$ and $2k$,
respectively, then draw directed edges from $(v,i)$ to both $(v',i)$ and
$(v',i+k)$, and from $(v',i)$ to $(v,i+1)$. The digraph corresponding to $A_3(4)$ is
given in Figure 4 --- note there the $k=4$ vertical copies of $A_3$.

\bigskip \epsfysize=1.5in \centerline{\epsffile{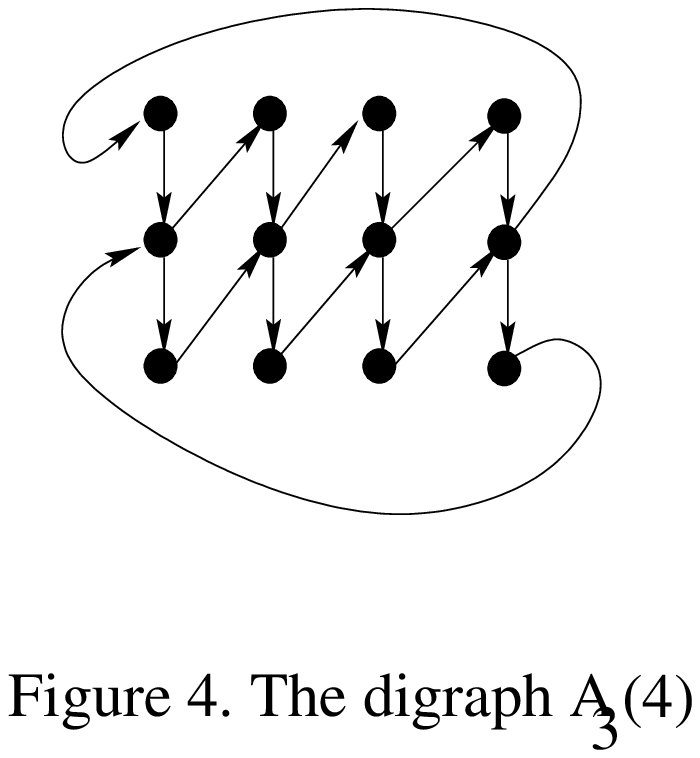}}
\bigskip

The matrix $A$ will consist of $d$ disconnected copies of a digraph
$X_s(2^\ell)$ taken from Figure 3, for some divisor $d$ of $m$ and some $\ell\le h$.
The order-$d$ permutation $P''$
takes vertex $(v,i)$ of the $j$th component digraph, to vertex 
$(v,i)$ of the $(j+1)$-th component digraph (where $j+1$ is taken mod $d$). 
The order $2^\ell$ or $2^{\ell+1}$ permutation $P'$ maps each component to
itself, and takes the vertex $(v,i)$ to the vertex $(v,i+1)$, where $i+1$
is taken modulo the weight $2^{\ell}$ or $2^{\ell+1}$ of the node $v$.

So once we know the matrix $A=\N_\la$, or equivalently the digraph $X_s(2^\ell)$
 from Figure 3 and its multiplicity $d$, then in principle we know the
 entire NIM-rep, using the above prescription and the sl$(n)_{2}$
 fusions in (A.2).

The complete list of indecomposable NIM-reps are:\smallskip

\noindent{{\bf (i)}} {\it Valid whenever $4$ divides $n$:} Choose any
divisor $2^\ell d$ of $n/2$ (where $d$ is odd).
Then the matrix $A$ corresponds
to $d$ copies of the digraph $D_{(n+4)/2}(2^\ell)$. The result is a
$2^{\ell-1} d \,(n+4)$-dimensional NIM-rep which we will denote by
$\N_D({n+4\over 2};2^\ell d)$. It has exponents consisting of all
$\la(i,j)$ for which $n/(2^\ell d)$ divides $i+j$. Each of these
$\la(i,j)$ has multiplicity 1 except for the fixed-points $j-i={n/2}$,
which have multiplicity 2.

For another class, choose any divisor $2^\ell d$ of $n/4$ (again $d$
is odd). Then the matrix $A$ corresponds to $d$ copies of the digraph
$C_{(n+2)/4}(2^\ell)$. The resulting $d 2^{\ell-1}(n+4)$-dimensional
NIM-rep is denoted $\N_C({n+2\over 2};2^\ell d)$. Its exponents consist
of all $\la(i,j)$ for which $n/(2^\ell d)$ divides $i+j$, together with
the fixed-points $\la({m\over 2^{\ell+2}d}i,{m\over 2^{n+2}d}i+{n\over 2})$; all have
multiplicity 1.

\smallskip\noindent{{\bf (ii)}} {\it Valid whenever $n$ is odd:} Choose any
divisor $d$ of $n$. Then the matrix $A$ corresponds to
$d$ copies of the tadpole $T_{(n+1)/2}$. The resulting $d\,{n+1\over 2}$-dimensional
NIM-rep will be denoted $\N_T({n+1\over 2};d)$. It has exponents consisting
of all $\la(i,j)$ for which ${n/d}$ divides $i+j$ --- all with
multiplicity 1.

\smallskip\noindent{{\bf (iii)}} {\it Valid whenever $n$ is even:} Choose any
odd divisor $d$ of $n$. Then the matrix $A$ corresponds to
$d$ copies of the digraph $B_{(n+2)/2}(2^{h-1})$. The resulting $d\,(n+1)\,2^{h-1}$-dimensional
NIM-rep will be denoted $\N_B({n+2\over 2};d)$. It has exponents consisting
of all $\la(i,j)$ for which ${m\over d}$ divides $i+j$ --- all with
multiplicity 1.

\smallskip\noindent{{\bf (iv)}} {\it Valid whenever $n/2$ is odd:} Choose any
odd divisor $d$ of $n$. Then the matrix $A$ corresponds to
$d$ copies of the digraph $C_{(n+2)/2}(1)$. The resulting $d\,{n+4\over 2}$-dimensional
NIM-rep will be denoted $\N_C({n+2\over 2};d)$. It has exponents consisting
of all $\la(i,j)$ for which ${n\over d}$ divides $i+j$, together with
the fixed-points $\la({m\over d}i,{m\over d}i+{n\over 2})$; all have
multiplicity 1.

\smallskip\noindent{{\bf (v)}} {\it Only for} $\widehat{{\rm sl}}(10)$:
Choose either $d=1$ or $d=5$. Then the matrix $A$ corresponds to
$d$ copies of the digraph $F_4(1)$. The resulting $6d$-dimensional
NIM-rep will be denoted $\N_{F4}(d)$. It has exponents
$$\E=\{\la({5\over d}i,{5\over d}i),\la({5\over d}i+1,{5\over d}i+4),
\la({5\over d}i+2,{5\over d}i+8)\,|\,0\le i< 2d\}$$ 
All those primaries have multiplicity 1.

\smallskip\noindent{{\bf (vi)}} {\it Only for} $\widehat{{\rm sl}}(16)$:
Choose any $0\le \ell\le 3$. Then the matrix $A$ corresponds to
the digraph $E_7(2^\ell)$. The resulting $2^\ell 7$-dimensional
NIM-rep will be denoted $\N_{E7}(2^\ell)$. It has exponents
$$\E=\{\la(2^{3-\ell}i,2^{3-\ell}i),\,\la(2^{3-\ell}i+2,2^{3-\ell}i+14),\,
\la(2^{3-\ell}i+3,2^{3-\ell}i+13),\,\la(2^{3-\ell}j+4,2^{3-\ell}j+12)\}$$
where $i,j$ range over $0\le i< 2^{\ell+1}$ and $0\le j<2^\ell$.
All those primaries have multiplicity 1.

\smallskip\noindent{{\bf (vii)}} {\it Only for} $\widehat{{\rm sl}}(28)$:
Choose any divisor $2^\ell d$ of 14 (take $d$ odd). Then the matrix $A$ corresponds to
$d$ copies of the digraph $E_8(2^\ell)$. The resulting $2^{3+\ell}d$-dimensional
NIM-rep will be denoted $\N_{E8}(2^\ell d)$. It has exponents
$$\E=\{\la({14\over 2^\ell d}i,{14\over 2^\ell d}i),\,\la({14\over 2^\ell d}i+3,{14\over 2^\ell d}i+25),\,
\la({14\over 2^\ell d}i+5,{14\over 2^\ell d}i+23),\,\la({14\over 2^\ell d}i+6,
{14\over 2^\ell d}i+22)\}$$
where $i$ ranges over $0\le i< 2^{\ell+1}d$.
All those primaries have multiplicity 1.

\medskip
For example, sl$(10)_2$ has six NIM-reps: $\N_B(6;5)$, $\N_B(6;1)$, $\N_C(6;5)$,
$\N_C(6;1)$, $\N_{F4}(5)$ and $\N_{F4}(1)$. We will find next that these are in precise one-to-one
correspondence with the six sl$(10)_2$ modular invariants.

All sl$(n)_2$ modular invariants are uniquely NIMmed. In particular, for $n$ odd, $M[J^d]$
corresponds to the tadpole NIM-rep $\N_T({n+1\over 2};d)$. When both $n-1$ and
$n/d$ are odd, $M[J^d]$ corresponds to $\N_B({n+2\over 2};{d_o})$.
 Otherwise $d$ divides $n/2$: when $d$ is even or odd, respectively,
$M[J^d]$ corresponds to $\N_D({n+4\over 2};d)$ and $\N_C({n+2\over 2};d)$.
We use here `$d_o$' to denote the odd  part of $d$,
i.e.\ $d/d_o$  is a power of 2.

Note that $\N_D({n+4\over 2};{\rm odd})$ and $\N_C({n+2\over 2};{\rm even})$
aren't paired to any modular invariant, when 4 divides $n$. Our
classification overlaps the A-D-E-T one of sl(2)$_k$, at sl(2)$_2$;
note that in our notation the single NIM-rep there corresponds to diagram
$C_2(1)$ and not $A_3(1)$, because the simple-current $\N_J$ should
have order 2, not 1.

The sl$(10)_2$ exceptional modular invariants $\E^{(10,2)}$ and $C\cdot \E^{(10,2)}$ correspond
to the NIM-reps $\N_{F4}(5)$ and $\N_{F4}(1)$, respectively. The sl$(16)_2$ exceptionals
$\E^{(16,2)},C\cdot \E^{(16,2)}$ and ${1\over 2}M[J^4]\cdot \E^{(16,2)}$
correspond to $\N_{E7}(8),\N_{E7}(4)$ and $\N_{E7}(2)$ respectively.
Finally, the sl$(28)_2$ exceptionals correspond to $\N_{E8}(14)$ and $\N_{E8}(2)$,
respectively.

The only remarkable thing about this NIM-rep $\leftrightarrow$ modular invariant
classification is how well they match: all but the exceptional NIM-reps
$\N_{E8}(1)$ and $\N_{E8}(7)$ are paired to a unique modular
invariant, except when $n$ is a multiple of 4.

\bigskip \noindent{\it 5.2. $\widehat{{\rm so}}(odd)$ at level $2$}.

Consider $\widehat{{\rm so}}(n)=B_r^{(1)}$, where
 $n=2r+1$. The set $P_+$ consists of precisely
$r+4$ weights, which we'll name as follows: $0$, $2\L_1$, $\L_r$, $\L_1+\L_r$, $\ga^i:=
\L_i$ for $i<r$,        and $\ga^r:= 2\L_r$.
  Write $\ga^0$ for the weight 0. The simple-current is $2\L_1=J$; it
   fixes all $\ga^1,\ldots,
\ga^r$. The {\it spinors} are $\L_r$ and $J\L_r=\L_1+\L_r$. For the
 additional modular invariants existing when  $n$ is a perfect square,
 the following notation is convenient:
if $8|r$, write $\la^r:= \L_r$ and $\mu^r:= J\L_r$; otherwise write
$\la^r:= J\L_r$ and $\mu^r:= \L_r$. Also write ${\cal C}=\{\ga^a\ne
0\,|\,\sqrt{n}\ {\rm divides}\ a\}$.

Very atypically for the current algebras, the list of modular invariants for ${{\rm so}}(n)_2$ is messy.
Define matrices $\B(d,\ell)$, $\B(d_1,\ell_1|d_2,\ell_2)$, $\B^i$, $\B^{ii}$,
$\B^{iii}$, $\B^{iv}$ by:
$$\eqalignno{\B(d,\ell)_{J^i\ga^a,J^i\ga^b}=&\left\{\matrix{2&{\rm if}\
d|a,\  d|b, {\rm and\ both}\ a\ne 0,\ b\ne 0\cr
0&{\rm if\ either}\ n\!\not|da\ {\rm or}\ b\not\equiv\pm a\ell\ ({\rm mod}\ 
d)\cr 1& {\rm otherwise}\cr}\right. &\cr
\B(d,\ell)_{J^i\L_r,J^i\L_r}=&\,1&\cr}$$
where $a,b\in\{0,1,\ldots,r\}$ and $i\in\{0,1\}$, and make all other matrix entries 0;
$$\eqalignno{\B(d_1,\ell_1|d_2,\ell_2)=&\,
{1\over 2}(\B(d_1,\ell_1)+\B(d_2,\ell_2))\,M[J]&\cr
\B^{i}_{00}=\B^{i}_{0\ga}=\B^{i}_{\ga 0}=&\,\B^i_{\ga\ga'}=\B^i_{\la^r\ga}=
\B^{i}_{\ga\la^r}=\B^{i}_{\mu^r\mu^r}=\B^{i}_{\la^r,J0}=
\B^{i}_{J0,\la^r}=1&\cr
\B^{ii}_{00}=\B^{ii}_{0\ga}=\B^{ii}_{\ga 0}=&\,\B^{ii}_{\ga\ga'}=
\B^{ii}_{0\la^r}=\B^{ii}_{\la^r 0}=
\B^{ii}_{\la^r\la^r}=\B^{ii}_{\la^r\ga}=\B^{ii}_{\ga\la^r}=1&\cr}$$
and all other entries are 0, where $\ga,\ga'\in{\cal C}$. Finally,
$\B^{iii}:=\B^i\,M[J]$ and $\B^{iv}:=M[J]\,\B^i$.

By `$n\!\not|da$' in the definition of ${\cal B}(d,\ell)$, we mean that
$n$ does not divide $da$. By `$b\not\equiv\pm a\ell$ (mod $d$)' there we
mean that $b$ is congruent mod $d$ to neither $a\ell$ nor $-a\ell$.

In [38] we proved that the modular invariants of $\widehat{{\rm so}}(n)_2=B_{r,2}$
are precisely:\smallskip

\item{(a)} $\B(d,\ell)$ for any divisor $d$ of $n=2r+1$ obeying $n|d^2$,
and for any integer $0\le \ell<{d^2\over 2n}$ obeying $\ell^2\equiv 1$
(mod ${d^2\over n}$);\smallskip

\item{(b)} $\B(d_1,\ell_1|d_2,\ell_2)$ for any divisors $d_i$ of $n$ obeying
$n|d_i^2$, and for any integers $0\le \ell_i< {d_i^2\over 2n}$ obeying
$\ell_i^2\equiv 1$ (mod ${d_i^2\over n}$);\smallskip

\item{(c)} when $n$ is a perfect square, there are four remaining
modular invariants: $\B^i$, $\B^{ii}$, $\B^{iii}$, and $\B^{iv}$.\medskip

The only redundancy here is that $\B(d_1,\ell_1|d_2,\ell_2)=\B(d_2,
\ell_2|d_1,\ell_1)$. The simple-current invariants are $\B(n,1)=I$ and $\B(n,1|n,1)=M[J]$.
For example, when $3\le r\le 10$, respectively, there are precisely
2, 9, 2, 2, 5, 2, 2, and 5 modular invariants for $B_{r,2}$.
The nine
$B_{4,2}$ modular invariants are: ${\cal B}(9,1)=I$, 
$$\eqalignno{{\cal
B}(9,1|9,1)=&\,|\chi_{0000}+\chi_{2000}|^2
+2|\chi_{1000}|^2+ 2|\chi_{0100}|^2+2|\chi_{0010}|^2+2|\chi_{0002}|^2&\cr
{\cal B}(3,1)=&\,|\chi_{0000}+\chi_{0010}|^2+
|\chi_{2000}+\chi_{0010}|^2+|\chi_{0001}|^2+|\chi_{1001}|^2&\cr
{\cal B}(3,1|3,1)=&\,|\chi_{0000}+\chi_{2000}+2\chi_{0010}|^2&\cr
{\cal B}(3,1|9,1)=&\,|\chi_{0000}+\chi_{2000}+\chi_{0010}|^2+2|\chi_{0010}|^2
+|\chi_{1000}|^2+ |\chi_{0100}|^2+|\chi_{0002}|^2&\cr
{\cal B}^i=&\,|\chi_{0000}+\chi_{0010}|^2+(\chi_{2000}\!+\!\chi_{0010})\,{
\chi_{1001}}^*+\chi_{1001}\,({\chi_{2000}\!+\!\chi_{0010}})^*+|\chi_{0001}|^2&\cr{\cal B}^{ii}=&\,|\chi_{0000}+\chi_{0010}+\chi_{1001}|^2&\cr
{\cal B}^{iii}=&\,(\chi_{0000}+\chi_{0010}+\chi_{1001})\,(
\chi_{0000}+\chi_{2000}+2\chi_{0010})^*&\cr
{\cal
B}^{iv}=&\,(\chi_{0000}+\chi_{2000}+2\chi_{0010})\,(\chi_{0000}+
\chi_{0010}+\chi_{1001})^*&}$$

In expressing our NIM-reps as explicitly as possible, we will use the
following matrices. By `$1_{\ell m}$' we mean
the $\ell\times m$ matrix, all of whose entries equal $1$. Write
$0_{\ell m}$ for the $\ell\times m$ zero-matrix, and $I_m$ for the
$m\times m$ identity. Write $B_{\ell m}(a)$ for the $\ell\times m$
matrix defined by 
$$B_{\ell m}(a)=\left(\matrix{2a\cdot 1_{\ell-2,m-2}&a\cdot 1_{\ell-2,1}&
a\cdot 1_{\ell-2,1}\cr
a\cdot 1_{1,m-2}&x&x-1\cr a\cdot 1_{1,m-2}&x-1&x}\right)=\left(\matrix{2a&\cdots&2a&a&a\cr \vdots&&\vdots&\vdots&\vdots\cr
2a&\cdots&2a&a&a\cr a&\cdots&a&x&x-1\cr a&\cdots&a&x-1&x}\right)\eqno(5.1)$$
where $x={a+1\over 2}$. For any integers $m,i$, define $M^{(m|i)}$ to be the $m\times m$
`off-diagonal' matrix with entries $M^{(m|i)}_{ab}=\delta_{b,a+i\ {\rm
mod}\ m}$. Put $M^{(m|i,j)}=M^{(m|i)}+M^{(m|j)}$. So $M^{(m|1,-1)}$ is
the adjacency matrix for the circle, i.e.\ the graph $A^{(1)}_{m-1}$.
Put $\widetilde{M}^{(\ell)}$ for the $\ell\times\ell$
adjacency matrix for the $D_\ell^0$ graph (see Figure 2), where the loop is at $1$,
the branch is at $\ell-2$, the degree-1 vertices
are at $\ell-1$ and $\ell$, and the other vertices are numbered sequentially
in the obvious way. For example,
$$M^{(4|1,-1)}=\left(\matrix{0&1&0&1\cr 1&0&1&0\cr 0&1&0&1\cr 1&0&1&0}\right)
\qquad \widetilde{M}^{(4)}=\left(\matrix{1&1&0&0\cr 1&0&1&1\cr 0&1&0&0\cr
0&1&0&0}\right)$$

Recall the graphs in Figure 2. For later convenience in this subsection,
we will identify $A_0^{(1)}$ with ${}^0\!A_1^0$, i.e.\ the matrix $(2)$,
and $D_2^0$ with ${}^0\!A_2^0$, i.e.\ the matrix $\left(\matrix{1&1\cr 1&1}\right)$.

There are four classes of NIM-reps:

\smallskip\noindent {\bf (i)} {\it Provided $n$ is a perfect square:} choose any integer $m\ge 1$
such that $m$ divides $\sqrt{n}$. Put $\N(m)_{\ga^i}=M^{(m|i,-i)}$,
$\N(m)_J=I_m$, and
$$\N(m)_{\L_r}=\N(m)_{\L_1+\L_r}={\sqrt{n}\over m}\cdot 1_{mm}$$
This defines an $m$-dimensional NIM-rep with exponents ${\cal E}=\{0,2\langle
\ga^{n/m}\rangle\}$, where we use the short-hand $\langle\ga^d\rangle:=
\{\ga^d,\ga^{2d},\ldots,\ga^{(n-d)/2}\}$ for $d$ dividing $n$ (i.e.\ all
$\ga^i$, $1\le i\le r$, where $d$ divides $i$). The
coefficient `2' in this ${\cal E}$ means each of  
these $\ga^{in/m}$'s appear with multiplicity 2.

The fusion graph of $\N(m)_{\L_1}$ is the circle $A_{m-1}^{(1)}$.
The simplest example is the 1-dimensional NIM-rep given by quantum-dimension:
$\ga^i\mapsto 2$, $\L_r\mapsto \sqrt{n}$.

\smallskip\noindent {\bf (ii)} {\it Provided $n$ is a perfect square:}
Choose either spinor $\si=\L_r$ or $\L_1+\L_r$, and choose any integer
$m\ge 2$ such that $2m-3$ divides $\sqrt{n}$. Put $\N'(m,\si)_{\L_1}=
\widetilde{M}^{(m)}$. The other matrices ${\cal N}'(m,\si)_{\ga^i}$
can now be constructed recursively from (A.4a),(A.4b) --- more on this shortly. Put
$\N'(m,\si)_J=I_{m-2}\oplus\left(\matrix{0&1\cr 1& 0}\right)$,
$\N'(m,\si)_\si=B_{mm}(a)$
(see (5.1)), where $a=\sqrt{n}/(2m-3)$. The matrix $\N'(m,\si)_{J\si}$
for the other spinor is the same except with $x$ and $x-1$ interchanged in the bottom
right $2\times 2$ block of (5.1). 

$\N'(m,\si)$ is an $m$-dimensional NIM-rep with exponents $\E=\{0,\si,
\langle\ga^{n/(2m-3)}\rangle\}$. $\L_1$ has fusion graph $D_m^0$.
The simplest example is $\ga^i\mapsto \left(\matrix{1&1\cr 1&1}\right)$,
$J\mapsto \left(\matrix{0&1\cr 1&0}\right)$, and $\L_r\mapsto
\left(\matrix{x&x-1\cr x-1&x}\right)$ where $x=(\sqrt{n}+1)/2$.

\smallskip\noindent {\bf (iii)} {\it Valid for any $n$:} Choose any
$m\ge m'\ge 1$ such that $\sqrt{{n\over mm'}}\in\Z$. Put $\N(m,m')_{\ga^i}
=M^{(m|i,-i)}\oplus M^{(m'|i,-i)}$, $\N(m,m')_J=I_{m+m'}$, and
$$\N(m,m')_{\L_r}=\N(m,m')_{\L_1+\L_r}=\left(\matrix{0_{mm}&b\cdot 1_{mm'}\cr
b\cdot 1_{m'm}&0_{m'm'}}\right)$$
where $b=\sqrt{n/mm'}$.

This is an $(m+m')$-dimensional NIM-rep with exponents $\E=\{0,J0,2\langle
\ga^{n/m}\rangle,2\langle\ga^{n/m'}\rangle\}$. The fusion graph of $\L_1$
consists of two circles: $A_{m-1}^{(1)}\cup A_{m'-1}^{(1)}$. The simplest
possible example of this NIM-rep is $\ga^i\mapsto \left(\matrix{2&0\cr 0&2}\right)$,
$\L_r\mapsto\left(\matrix{0&\sqrt{n}\cr \sqrt{n}&0}\right)$.

\smallskip\noindent {\bf (iv)} {\it Valid for any $n$:} Choose any integers
$m\ge m'\ge 2$ such that $\sqrt{{n\over (2m-3)(2m'-3)}}\in\Z$. The
matrices $\N'(m,m')_{\ga^i}$ are the direct sums $\N'(m,\L_r)_{\ga^i}\oplus
\N'(m',\L_r)_{\ga^i}$ of matrices of NIM-rep (ii). Put
$\N'(m,m')_J=I_{m-2}\oplus\left(\matrix{0&1\cr 1&0}\right)\oplus I_{m'-2}
\oplus \left(\matrix{0&1\cr 1&0}\right)$ and
$\N'(m,m')_{\L_r}=\left(\matrix{0&B\cr B^t&0}\right)$, where $B=B_{mm'}(a)$
for $a=\sqrt{{n\over (2m-3)(2m'-3)}}$. The matrix
for the other spinor, $\L_1+\L_r$, is the same except with $x$ and $x-1$ interchanged
in both $B$ and $B^t$ (see (5.1)).

$\N'(m,m')$ is an $(m+m')$-dimensional NIM-rep, with exponents $\E=\{
0,J,\L_r,\L_1+\L_r,\langle\ga^{n/(2m-3)}\rangle,\langle\ga^{n/(2m'-3)}\rangle\}$.
The fusion graph of $\L_1$ is $D_m^0\cup D_{m'}^0$. It may seem that we've
`broken the symmetry' between $\L_r$ and $\L_1+\L_r$, and so there should be 
another NIM-rep with the images of $\L_r$ and $\L_1+\L_r$ interchanged (as we
did in {\bf (ii)}). However, these two NIM-reps are equivalent here (and they
aren't in {\bf (ii)}).

The simplest example
is $\ga^i\mapsto\left(\matrix{1&1\cr 1&1}\right)\oplus
\left(\matrix{1&1\cr 1&1}\right)$, $J\mapsto \left(\matrix{0&1\cr 1&0}\right)
\oplus\left(\matrix{0&1\cr 1&0}\right)$, and
$$\L_r\mapsto\left(\matrix{0&0&x&x-1\cr 0&0&x-1&x\cr x&x-1&0&0\cr x-1&x&0&0}
\right)$$
for $x={\sqrt{n}+1\over 2}$.\medskip

 The proof of this NIM-rep classification is deferred to  Appendix A.2.
Note that each of these $B_{r,2}$ NIM-reps has a different set $\E$ of exponents.

In {\bf (i)} and {\bf (iii)} we could explicitly write all matrices
$\N_{\ga^i}$. This is much
harder in {\bf (ii)} (though they are constructable recursively by (A.4a),(A.4b)). We'll
make only the following remark: the graph for $\N'(m,\si)_{\ga^i}$ will
consist of one $D_\ell^0$-type component
containing the nodes $m-1,m$, and precisely ${{\rm gcd}(i,2m-3)-1\over  2}$
${}^0\!A_k^0$-type components of equal size.

Which of the modular invariants are NIMmed? We find
that the exponents for ${\cal B}(d,\ell)$ are $\{0,J0,\L_r,J\L_r,\langle
\ga^{m}\rangle,\langle \ga^{n/m}\rangle\}$ where
$m={\rm gcd}(d,(\ell-1){n\over d})$. There is one and only one NIM-rep
corresponding to ${\cal B}(d,\ell)$, namely $\N'({m+3\over 2},{n+3m\over 2m})$.
 Note that the square-root condition is automatically obeyed.
For example, the identity modular
invariant $I={\cal B}(n,1)$ corresponds to $\N'(r+2,2)$.

The exponents of ${\cal B}(d,\ell|d',\ell')$ are $\{0,J0,\langle\ga^{m}\rangle,
\langle\ga^{n/m}\rangle,\langle\ga^{m'}\rangle,\langle\ga^{n/m'}\rangle\}$
where $m={\rm gcd}(d,(\ell- 1){n\over d})$ and $m'={\rm gcd}
(d',(\ell'-1){n\over d'})$. These have a corresponding NIM-rep iff
both $d=d'$ and $\ell=\ell'$, in which case the NIM-rep is given by $\N(m,n/m)$.
The simple-current extension $M[J]=\B(n,1|n,1)$ corresponds to $\N(n,1)$.

When $\sqrt{n}$ is a perfect square, we get the four additional modular
invariants ${\cal B}^i,\ldots,{\cal B}^{iv}$. Note that ${\cal B}^i$ and
${\cal B}^{ii}$ have exponents $\{0,\mu^r,\langle\ga^{\sqrt{n}}\rangle\}$
and $\{0,\la^r,\langle\ga^{\sqrt{n}}\rangle\}$ respectively, and so
correspond to $\N'({\sqrt{n}+3\over 2},\mu^r)$ and
$\N'({\sqrt{n}+3\over 2},\la^r)$, respectively (both $\la^r$ and $\mu^r$ are
defined at the beginning of this subsection). Both ${\cal B}^{iii}$ and its transpose
${\cal B}^{iv}$ have the same exponents, namely $\{0,2\langle\ga^{\sqrt{n}}
\rangle\}$, and so correspond to NIM-rep $\N(\sqrt{n})$.

In other words, only
 the following NIM-reps will have an associated modular invariant:
$\N(\sqrt{n}),\N'(\sqrt{n},\L_r),\N'(\sqrt{n},
\L_1+\L_r),\N(m,{n\over m})$ and $\N'(m,{n\over m})$ (for any divisor $m$
of $n$).

For example, we gave earlier explicitly the nine different modular invariants for
$\widehat{{\rm so}}(9)$. For each of these, in the order given above, the
fundamental weight $\L_1$ corresponds to the fusion graph $D_6^0\cup {}^0\!A_2^0$,
$A_8^{(1)}\cup {}^0\!A_1^0$, $D_3^0\cup D_3^0$, $A_2^{(1)}\cup A_2^{(1)}$, $-$,
$D^0_3$, $D_3^0$, $A_2^{(1)}$ and $A_2^{(1)}$. Only the last two have
identical NIM-reps. Only ${\cal B}(3,1|9,1)$ is NIM-less --- an elementary
proof of this is given in \S6.

For larger rank, the number of NIM-less ${{\rm so}}(n)_2$ modular
invariants will typically far exceed the NIMmed ones: the former grows
like $D^2$ while the latter grows like $D$, where $D$ is the number of
divisors of $n$.
For example, when $n$ is a power $p^{a}$ of a prime, the number
of NIM-reps grows like $a^2/4$, while the NIMmed modular invariants grow
like $a$ and the NIM-less ones grow like $a^2/8$.

All modular invariants for ${{\rm so}}(n)_2$ will be NIMmed, iff
$n$ is a prime. Other small ranks with NIM-less modular invariants are
${{\rm so}}(15)_2$ (for ${\cal B}(15,1|15,4)$), ${{\rm so}}(21)_2$
(for ${\cal B}(21,1|21,8)$), ${{\rm so}}(25)_2$ (for ${\cal B}(5,1|25,1)$),
and ${{\rm so}}(27)_2$ (for ${\cal B}(9,1|27,1)$).

\bigskip\noindent{{\it 5.3. $\widehat{{\rm so}}(even)$ at level $2$.}}\medskip

Consider ${{\rm so}}(n)_2=D_{r,2}$, where $n=2r$. 
There are $r+7$ weights: 0, $2\L_1$, $2\L_{r-1}$, $2\L_r$,
$\L_r$, $\L_1+\L_{r-1}$, $\L_{r-1}$, $\L_1+\L_r$,
$\la^i:= \L_i$ for $1\le i\le r-2$, and $\la^{r-1}:=
\L_{r-1}+\L_r$. Write
$\la^0$ for the weight $0$ and $\la^r$ for $2\L_r$. There are three (nontrivial) simple-currents,
namely $J_v=2\L_1$, $J_s=2\L_r$ and $J_c=2\L_{r-1}$.  The simple-current
$J_v$ fixes the $\la^i$ ($1\le i<r$). The four spinors are
$\L_r,\L_{r-1},J_v\L_r=\L_1+\L_{r-1},J_v\L_{r-1}=\L_1+\L_r$.
For the additional modular invariants occurring when $r$ is a perfect square, 
it is convenient to write ${\cal C}_j=\{\la^b\ne 0\,|\,2{b\over \sqrt{r}}\equiv \pm j\
 ({\rm mod}\ 8)\}$ for $j=0,1,2,3,4$. 

Write $C_0=I$, and let $C_1$ be the permutation of $P_+$ interchanging
$\L_{r-1}\leftrightarrow \L_r$, $2\L_{r-1}\leftrightarrow 2\L_r$, and 
$\L_1+\L_{r-1}\leftrightarrow \L_1+\L_r$, and fixing all other weights.
We call these $C_i$ {\it conjugations} because they correspond to
 symmetries of the unextended Dynkin diagram. Charge-conjugation $C$ for
 $D_{odd,2}$ is $C_1$, and for $D_{even,2}$ is $I$. When $r=4$, there are
 four other conjugations, corresponding to so(8) triality.

Define the matrices $\D(d,\ell)$, $\D(d_1,\ell_1|d_2,\ell_2)$, $\D^i$, $\D^{ii}$,
and $\D^{iii}$, as follows:
$$\eqalignno{\D(d,\ell)_{J_v^i\la^a,J_v^i\la^b}=&\left\{\matrix{2&{\rm if}\
d|a,\ d|b,\ 2d|(a+b),\ {\rm and}\ \{a,b\}\subseteq\{1,\ldots,r-1\}\cr
0&{\rm  if\ either}\ r\!\not| da\ {\rm or}\ 
b\not\equiv \pm a\ell\ ({\rm mod}\ 2d)\cr 1&{\rm otherwise}}\right. &\cr
\D(d,\ell)_{\la_s\la_s}=&\left\{\matrix{1&{\rm if}\ 2d\!\not|
r\cr 2&{\rm if}\ \la_s\in\{\L_r,\L_1+\L_{r-1}\}\ {\rm and}\ 2d|r\cr
0&{\rm otherwise}}\right.&\cr}$$
and all other entries are 0,
where $a,b\in\{0,1,\ldots,r\}$, $i\in\{0,1\}$, and $\la_s$ is any spinor;
$$\eqalignno{\D(d_1,\ell_1|d_2,\ell_2)=&\,
{1\over 2}(\D(d_1,\ell_1)+\D(d_2,\ell_2))\,M[J_v]&\cr
\D^{i}_{J_v^j\L_r,J_v^j\L_r}=&\,\D^{i}_{\L_r,\mu}=\D^{i}_{\mu,\L_r}=
\D^{i}_{J_v\L_r,\mu'}=\D^{i}_{\mu',J_v\L_r}=\D^{i}_{\la\la'}
=\D^{i}_{\ga\ga'}&\cr=&\,\D^{i}_{J',J''}=\D^{i}_{J',\nu}=
\D^{i}_{\nu,J'_v}=\D^{i}_{J'J_v,\nu'}=\D^{i}_{\nu',J'J_v}=1&\cr}$$
 where $\la,\la'\in{\cal C}_0\cup{\cal C}_4$, $\mu\in 
{\cal C}_1$, $\mu'\in{\cal C}_3$, $\ga,\ga'\in{\cal C}_2$, $\nu\in
{\cal C}_0$, $\nu'\in{\cal C}_4$, $J',J''\in\J_s$, and $j\in\{0,1\}$.
All other entries  equal 0.  Finally, $\D^{ii}=\D^i\,M[J_v]$ and $\D^{iii}
=M[J_v]\,\D^{i}$.

In [38] we proved that the modular invariants for ${{\rm so}}(n)_2=
D_{r,2}$ are: (for arbitrary conjugations $C_i,C_j$)\smallskip

\item{(a)} $C_i\,\D(d,\ell)\,C_j$ for  any divisor $d$ of $r$
 obeying $r|d^2$,
and for any integer $1\le \ell\le{d^2\over r}$ obeying $\ell^2\equiv 1$
(mod ${4d^2\over r}$); \smallskip

\item{(b)} $\D(d_1,\ell_1|d_2,\ell_2)$ for any divisors $d_i$ of $r$ obeying
$r|d_i^2$, as well as the additional property that $2d_1|r$ iff $2d_2|r$,
and for any integers $1\le \ell_i\le {d_i^2\over r}$ obeying
$\ell_i^2\equiv 1$ (mod ${4d_i^2\over r}$);

\smallskip\item{(c)} when $r$ is a perfect square and $16|r$, there are 8 
other modular invariants: $C_i\,\D^i\,C_j${,} $C_i\,\D^{ii}$, and
$\D^{iii}\,C_j$.      \smallskip

Take $C_i=I$ in (a) unless $2d|r$. The number of these grows asymptotically
with the square of the number of divisors of $r$. 
The following are simple-current invariants:
$\D(r,1)=I$ and $\D(r,1|r,1)=M[J_v]$ (for all $r$), and $\D(r,r-1)=M[J_s]$
and $\D(r,r-1|r,r-1)=M[J_v]\,M[J_s]$ (when ${r\over 2}$ is odd), and
$\D({r\over 2},1)=M[J_s]$ and $\D({r\over 2},1|{r\over 2},1)=M[J_v]\,
M[J_s]$ (when $4|r$).

For example, for $4\le r\le 16$, respectively, there are precisely $16, 3, 7, 3, 8, 7$,
and 7 modular invariants. Of these, $0, 0, 1, 0, 0, 4$, and 1
are exceptional. The seven modular invariants for $D_{6,2}$
are: the identity ${\cal D}(6,1)=I$; 
$$\eqalign{{\cal D}(6,5)=&\,|\chi_{000000}|^2\!+\!|\chi_{000020}|^2\!+\!
|\chi_{000002}|^2
\!+\!|\chi_{010000}|^2\!+\!|\chi_{001000}|^2\!+\!|\chi_{000100}|^2\!+\!|\chi_{000010}|^2\cr
&+\chi_{100000}\,{\chi_{000011}^*}+\chi_{000011}\,{\chi_{100000}^*}
+|\chi_{000001}|^2+|\chi_{100010}|^2+|\chi_{100001}|^2\cr
{\cal D}(6,1|6,1)=&\,|\chi_{000000}+\chi_{200000}|^2+|\chi_{000020}
+\chi_{000002}|^2+2|\chi_{100000}|^2+2|\chi_{010000}|^2\cr&+2|\chi_{001000}|^2
+2|\chi_{000100}|^2+2|\chi_{000011}|^2\cr
{\cal D}(6,1|6,5)=&\,|\chi_{000000}+\chi_{200000}|^2+|\chi_{000020}+
\chi_{000002}|^2+|\chi_{100000}+\chi_{000011}|^2\cr
&+2|\chi_{010000}|^2+2|\chi_{001000}|^2+2|\chi_{000100}|^2\cr
{\cal D}(6,5|6,5)=&\,|\chi_{000000}+\chi_{200000}|^2+
|\chi_{000020}+\chi_{000002}|^2+2|\chi_{010000}|^2+2|\chi_{001000}|^2
\cr&+2|\chi_{000100}|^2
+2\chi_{100000}\,{\chi_{000011}^*}+2\chi_{000011}\,{\chi_{100000}^*}}$$
as well as the conjugates $C_1$ and $C_1\,{\cal D}(6,5)$.

Recall the matrices $I_m$, $1_{mm'}$, $0_{mm'}$, and $M^{(m|i,j)}$ from \S5.2.
Let $C_{mn}$ and $C'_{mn}$ be the $m\times n$ checkerboard matrices,
i.e.\ their $(i,j)$th entries are 1 if $i+j$ is even/odd respectively (all other
entries are 0). Write $I^s_m$ for the $m\times m$ {\it skew-identity}:
$$I^s_m=\left(\matrix{0&\cdots&0&1\cr \vdots&&1&0\cr 0&\adots&&\vdots\cr
1&0&\cdots&0}\right)\eqno(5.2)$$
The circle $A_3^{(1)}$ should also be interpreted here as the graph `$D_3^{(1)}$',
with adjacency matrix
$$\left(\matrix{0&0&1&1\cr 0&0&1&1\cr 1&1&0&0\cr 1&1&0&0}\right)$$

The NIM-reps of $\widehat{{\rm so}}(2r)={D}_r^{(1)}$ level 2 are as follows.

\smallskip\noindent{{\bf (i)}} {\it Provided $r$ is a perfect square:} Choose any
odd divisor $m\ge 3$ of $\sqrt{r}$ (the case of even divisors will be treated
shortly). Define  $\N(m)_{\la^i}=M^{(m|i,-i)}$ and $\N(m)_J=I_m$
for any of the simple-currents $J$. For all of the four spinors $\si$,
put $\N(m)_\si={\sqrt{r}\over m}\cdot 1_{mm}$.

If $r$ is even,  choose any even divisor $m$ of
$2\sqrt{r}$, as well as either simple-current $J'=J_s$ or $J'=J_c$.
Define $\N(m,J')_{\la^i}$ and $\N(m,J')_J$ as for $\N(odd)$. Write $\si=\L_r$
or $\si=\L_{r-1}$ depending on whether or not $J'=J_s$. Then
$\N(m,J')_{\si}=\N(m,J')_{J_v\si}={2\sqrt{r}\over m}C_{mm}$ and
$\N(m,J')_{C_1\si}=\N(m,J')_{J_vC_1\si}={2\sqrt{r}\over m}C'_{mm}$.

These are $m$-dimensional NIM-reps. The fusion graph of $\L_1$ is the circle
$A_{m-1}^{(1)}$. When $m$ is odd, the exponents are $\E=\{0,2\langle\la^{2r/m}\rangle\}$,
and when $m$ is even the exponents are $\{0,J',2\langle\la^{2r/m}\rangle\}$.
(We write $\langle\la^d\rangle$ for $\{\la^d,\la^{2d},\ldots,\la^{r-d}\}$
when $d|r$, and for $\{\la^d,\la^{2d},\ldots,\la^{r-d/2}\}$ when otherwise
$d|2r$; the coefficient `$2$' means all those primaries appear with multiplicity
2.)

\smallskip\noindent{{\bf (ii)}} {\it Provided $r$ is a perfect square:} Choose any
divisor $m$ of $\sqrt{r}$. Let $\N^0(m)_{\L_1}$ be the adjacency matrix of
${}^0\!A^0_m$. How to get the other $\N^0(m)_{\la^i}$ will be explained shortly.
Define $\N^0(m)_{J_v}=I_m$, and for any spinor $\si$ put $\N^0(m)_\si=
{\sqrt{r}\over m}\cdot 1_{mm}$. If $r$ is even, then $\N^0(m)_{J_s}=\N^0(m)_{J_c}
=I_m$, otherwise for odd $m$ $\N^0(m)_{J_s}=\N^0(m)_{J_c}$ will be the
unique order-2 symmetry of ${}^0\!A^0_m$, namely the skew-identity $I^s_m$.

$\N^0(m)$ is an $m$-dimensional NIM-rep.
The fusion graph of $\L_1$ will be ${}^0\!A^0_m$. The exponents are
$\{0,\langle\la^{r/m}\rangle\}$.
The simplest example is quantum-dimension $\la\mapsto S_{\la 0}/S_{00}$.

\smallskip\noindent{{\bf (iii)}} {\it Provided $r$ is even and a perfect square:} Choose any
$m\ge 5$ so that $\sqrt{r}/(m-3)$ is an odd integer, and choose either
simple-current $J'=J_s$ or $J'=J_c$. Define $\si=\L_r$ or $\si=\L_{r-1}$,
as in {\bf (i)}. Put $\N'(m,J')_{\L_1}$ to be the adjacency matrix of
$D_{m-1}^{(1)}$. We'll discuss how to obtain the matrices $\N'(m,J')_{\la^i}$
shortly. Put $\N'(m,J')_{J'}=I_m$ and $\N'(m,J')_{J_v}=\N'(m,J')_{J_vJ'}=
I_2^s\oplus I_{m-4}\oplus I_2^s$. Define
$$\eqalign{\N'(m,J')_\si=&\,\left(\matrix{X&a Y_m&X\cr a Y_m^t&2a
C_{m-4,m-4}&aY_m^t\cr X&aY_m&X'}\right)\cr
\N'(m,J')_{C_1\si}=&\,\N'(m,J')_{J_vC_1\si}=a\left(\matrix{0_{22}&Y_m'&0_{22}\cr
Y_m'{}^t&2 C'_{m-4,m-4}&Y_m'{}^t\cr 0_{22}&Y_m'&0_{22}}\right)}$$
where $X=\left(\matrix{x&x'\cr x'&x}\right)$, $X'=\left(\matrix{x'&x\cr x&x'}
\right)$, $Y_m=\left(\matrix{0&1&0&\cdots\cr
0&1&0&\cdots}\right)$, $Y_m'=\left(\matrix{1&0&1&\cdots\cr 1&0&1&\cdots}\right)$,
$a=\sqrt{r}/(m-3)$, $x=(a+1)/2$, $x'=(a-1)/2$. That is, $Y_m$ is the $2\times m$
matrix whose $(i,j)$ entry is 0 or 1
provided $j$ is odd or even, respectively, and $Y_m'=1_{2,m}-Y_m$.
The matrix for $\N'(m,J')_{J_v\si}$
is obtained from that of $\N'(m,J')_\si$ by interchanging the submatrices
 $X$ and $X'$.

This NIM-rep is $m$-dimensional. The fusion graph for $\L_1$ is $D_{m-1}^{(1)}$.
The exponents are $\{0,J',\si,J_v\si,\langle \la^{r/(m-3)}\rangle\}$.

\smallskip\noindent{{\bf (iv)}} {\it Valid for any $r$:} Choose any $m,m'\ge 1$
such that $\sqrt{r/mm'}\in\Z$. Define $\N^{00}(m,m')_{\mu}=\N^0(m)_{\mu}
\oplus\N^0(m')_{\mu}$ (see {\bf (ii)} above) for any $\mu=\la^i$, and any
simple-current $\mu$. For any spinor $\si$, put $\N^{00}(m,m')_\si=\left(
\matrix{0&A\cr A^t&0}\right)$ where $A=\sqrt{r/mm'}\cdot 1_{mm'}$.

Next, when in addition $m'$ is odd and $\ge 3$, define another NIM-rep
by $\N^{0}(m,m')_{\mu}
=\N^0(m)_{\mu}\oplus\N(m')_{\mu}$ (see also {\bf (i)} above) for
any $\mu=\la^i$, and any simple-current $\mu$. For any spinor $\si$, put $\N^{0}(m,m')_\si=
\N^{00}(m,m')_\si$.

Next, when both $m$ and $m'$ are odd and $\ge 3$, define $\N(m,m')_{\mu}
=\N(m)_{\mu}\oplus\N(m')_{\mu}$  for
any $\mu=\la^i$, and any simple-current $\mu$. For any spinor $\si$, put $\N(m,m')_\si=
\N^{00}(m,m')_\si$.

Finally, if $m$ and $m'$ are both even, we can weaken the condition
$\sqrt{r/mm'}\in\Z$ to $\sqrt{4r/mm'}\in\Z$. Define $\N(m,m')_{\mu}
=\N(m)_{\mu}\oplus\N(m')_{\mu}$  for
any $\mu=\la^i$.          The simple-currents are
$\N(m,m')_{J_v}=I_{m+m'}$ and $\N(m,m')_{J_s}=\N(m,m')_{J_c}=M^{(m|r)}\oplus
M^{(m'|r)}$. For $r$ even put
$$\eqalign{\N(m,m')_{\L_r}=&\, \N(m,m')_{\L_1+\L_{r-1}}=a\left(\matrix{0_{mm}&C_{mm'}\cr
C_{m'm}&0_{m'm'}}\right)\cr
\N(m,m')_{\L_{r-1}}=&\, \N(m,m')_{\L_1+\L_{r}}=a\left(\matrix{0_{mm}&C'_{mm'}\cr
C'_{m'm}&0_{m'm'}}\right)}$$
while for $r$ odd put
$$\eqalign{\N(m,m')_{\L_r}=&\, \N(m,m')_{\L_1+\L_{r-1}}=a\left(\matrix{0_{mm}&C_{mm'}\cr
C'_{m'm}&0_{m'm'}}\right)\cr
\N(m,m')_{\L_{r-1}}=&\, \N(m,m')_{\L_1+\L_{r}}=a\left(\matrix{0_{mm}&C'_{mm'}\cr
C_{m'm}&0_{m'm'}}\right)}$$
In both cases,  $a=2\sqrt{r/mm'}$.

All of these NIM-reps are $(m+m')$-dimensional. Their exponents are, respectively,
$$\eqalign{\E^{00}=&\,\{0,J_v,\langle\la^{r/m}
\rangle,\langle\la^{r/m'}\rangle\}\cr \E^0=&\,\{0,J_v,\langle\la^{r/m}
\rangle,2\langle\la^{2r/m'}\rangle\}\cr \E=&\,\{0,J_v,2\langle\la^{2r/m}
\rangle,2\langle\la^{2r/m'}\rangle\}\cr \E=&\,
\{0,J_v,J_s,J_c,2\langle\la^{2r/m}\rangle,2\langle\la^{2r/m'}\rangle\}}$$
The fusion graph of $\L_1$ is ${}^0\!A^0_m\cup{}^0\!A^0_{m'}$,
${}^0\!A^0_m\cup A^{(1)}_{m'-1}$, $A^{(1)}_{m-1}\cup A^{(1)}_{m'-1}$,
$A^{(1)}_{m-1}\cup A^{(1)}_{m'-1}$, resp.

For instance, the conjugation $C_1$  corresponds to $\N^{00}(r,1)$ and
$M[J_v]$ to $\N(2r,2)$.

\smallskip\noindent{{\bf (v)}} {\it Valid for any $r$:}
Choose any $m,m'\ge 4$ such that $a:=\sqrt{r/(m-3)(m'-3)}$ is an odd integer
--- when $r$ is even we require in addition that $m$ be odd and $m'$ even.
Let $\N''(m,m')_{\L_1}$ be the adjacency matrix of the graph $D_{m-1}^{(1)}
\cup D_{m'-1}^{(1)}$. We'll discuss shortly how to obtain the other matrices
$\N''(m,m')_{\la^i}$. Put
$$\N''(m,m')_{J_v}=I_2^s\oplus I_{m-4}\oplus I_2^s\oplus I_2^s
\oplus I_{m'-4}\oplus I_2^s$$
For $r$ even, $\N''(m,m')_{J_s}=I_m\oplus I^s_{m'}$ (see (5.2)), while for $r$ odd
$\N''(m,m')_{J_s}=I'_m\oplus I'_{m'}$ where $I'_m$ is the order-4 symmetry of the
Dynkin diagram of $D_{m-1}^{(1)}$, i.e.\ the $m\times m$
matrix
$$I'_m:=\left(\matrix{0_{2,m-2}&I_2\cr I^s_{m-2}&0_{m-2,2}}\right)$$
For $r$ even, put $\N''(m,m')_{\L_r}=\left(\matrix{0_{mm}&D\cr D^t&0_{m'm'}}
\right)$ and $\N''(m,m')_{\L_{r-1}}=\left(\matrix{0_{mm}& E\cr E^t&
0_{m'm'}}\right)$ where
$$D=\left(\matrix{X&aY_{m'}&0_{22}\cr aY_m^t&2aC_{m-4,m'-4}&aY_m^t\cr X&aY_{m'}&0_{22}}\right)
\qquad E=\left(\matrix{0_{22}&aY_{m'}'&X\cr aY_m'{}^t&2aC'_{m-4,m'-4}&aY_m'{}^t
\cr 0_{22}&aY_{m'}'&X}\right)$$
For $r$ odd, put
$\N''(m,m')_{\L_r}=\left(\matrix{0_{mm}&P\cr Q^t&0_{m'm'}}\right)=
(\N''(m,m')_{\L_{r-1}})^t$ where
$$P=\left(\matrix{X&aY_{m'}&0_{22}\cr aY_m^t&2aC_{m-4,m'-4}&aY_m^t\cr 0_{22}
&aY_{m'}&X'}\right)
\qquad Q=\left(\matrix{0_{22}&aY_{m'}'&X\cr aY_{m}'{}^t&2aC'_{m-4,m'-4}&aY_m'{}^t\cr
X&aY_{m'}'&0_{22}}\right)$$
Here, $x=(a+1)/2$, $x'=(a-1)/2$, 
and the matrices $X,Y,\ldots$ are as in {\bf (iii)}.
The matrices $\N''(m,m')_{J_c}$, $\N''(m,m')_{J_v\L_{r-1}}$ and $\N''(m,m')_{J_v\L_r}$ are obtained
by the obvious matrix products.

These are $(m+m')$-dimensional NIM-reps, with exponents 
$$\E=\{0,J_v,J_s,J_c,
\langle\la^{r/(m-3)}\rangle,\langle\la^{r/(m'-3)}\rangle,\L_r,\L_{r-1},
\L_1+\L_r,\L_1+\L_{r-1}\}$$ 
The fusion graph of $\L_1$ is $D_{m-1}^{(1)}\cup D^{(1)}_{m'-1}$ (recall that
$D_3^{(1)}=A_3^{(1)}$).

The regular NIM-rep, corresponding to fusion matrices, is $\N''(r+3,4)$.

\smallskip\noindent{{\bf (vi)}} {\it Valid whenever $4|r$:}
Choose any odd $m,m'\ge 5$ such that $a:=\sqrt{r/(m-3)(m'-3)}$ is an odd integer,
and pick either $\si\in\{\L_r,\L_{r-1}\}$. Put $\N''(m,m',\si)_{J'}=I_{m+m'}$,
where $J'$ denotes $J_s$ or $J_c$ when $\si=\L_r$ or $\L_{r-1}$, respectively.
Put $\N''(m,m',\si)_{\si}=\left(\matrix{0_{mm}&D\cr D^t&0_{m'm'}}
\right)$ and $\N''(m,m',\si)_{C_1\si}=\N''(m,m',\si)_{J_vC_1\si}=
\left(\matrix{0_{mm}& E\cr E^t&0_{m'm'}}\right)$ where now
$$D=\left(\matrix{X&aY_{m'}&X\cr aY_m^t&2aC_{m-4,m'-4}&aY_m^t\cr X&aY_{m'}&X'}\right)
\qquad E=\left(\matrix{0_{22}&aY_{m'}'&0_{22}\cr aY_m'{}^t&2aC'_{m-4,m'-4}&aY_m'{}^t
\cr 0_{22}&aY_{m'}'&0_{22}}\right)$$
Then $\N''(m,m',\si)_{v}$ and the
 other matrices are as in {\bf (v)}.

This is an $(m+m')$-dimensional NIM-rep, with exponents 
$$\E=\{0,J_v,J_s,J_c,
\langle\la^{r/(m-3)}\rangle,\langle\la^{r/(m'-3)}\rangle,\si,\si,J_v\si,J_v\si
\}$$ 
The fusion graph of $\L_1$ is $D_{m-1}^{(1)}\cup D^{(1)}_{m'-1}$.
The simple-current invariant $M[J_s]$ corresponds to $\N''(5,{r\over 2}+3,
\L_r)$.

\medskip For any NIM-rep, the matrices for ${\la^i}$ are obtained recursively
from (A.6a),(A.6b). For the NIM-reps $\N^0$ and $\N^{00}$ based on the ${}^0\!A^0_m$
diagram, these are most easily
found by using the explicit formula $M^{(2m|i,-i)}$ for the $A^{(1)}_{2m-1}$
graph, and then folding the result in the obvious way. For the NIM-reps
$\N'$ and $\N''$ based on the diagram $D^{(1)}_{m-1}$, the matrices for
$\la^i$ will be a union of graphs from Figure 2.
For $i$ odd there will be a total of $(1+{\rm gcd}(i,m-3))/2$
components, all bipartite. When $i$ is even, and the exact power of 2
dividing $i$ also divides $m-3$,
then there will be ${\rm gcd}(i,m-3)/2$ bipartite components, together with
one graph of type ${}^0\!A^0$. Otherwise, when the power of 2 dividing
$i$ exceeds that of $m-3$, there will be $1+{\rm gcd}(i,m-3)$ components,
none of them bipartite, and a total of two loops.

For instance, there are precisely eight NIM-reps for $D_{6,2}$: namely,
$\N^{00}(6,1)$, $\N^0(6,1)$, $\N^{00}(2,3)$, $\N^0(2,3)$, $\N(2,12)$,
$\N(4,6)$, $\N''(9,4)$,
$\N''(5,6)$, of dimensions 7,7,5,5,14,10 respectively. Only $\N^0(6,1)$ and
$\N^0(2,3)$ fail to have a corresponding modular invariant. Only the modular
invariant ${\cal D}(6,1|6,5)$ is NIM-less.

More generally, the exceptionals $\D^i$ and $C_1\D^i C_1$ correspond to $\N'(\sqrt{r}+3,J_s)$
and $\N'(\sqrt{r}+3,J_c)$, respectively, while both $C_1\D^i$ and $\D^iC_1$
correspond to $\N^0(\sqrt{r})$. Both $\D^{ii}$ and $\D^{iii}$ correspond
to $\N(2\sqrt{r},J_s)$, while both $C_1\D^{ii}$ and $\D^{iii}C_1$ correspond
to $\N(2\sqrt{r},J_c)$.

Given parameters $d,\ell$, define $m$ as follows:
if $4|(\ell\mp 1)$, put $m={\rm gcd}(d,{(\ell\pm 1)r\over 2d})$. When $2d|r$,
 the modular invariant ${\cal D}(d,\ell)$ corresponds
to $\N''(m+3,{r\over m}+3,\L_r)$ and  $C_1\,{\cal D}(d,\ell)\,C_1$ to 
$\N''(m+3,{r\over m}+3,\L_r)$. Otherwise, 
 ${\cal D}(d,\ell)$ corresponds to  $\N''(m+3,{r\over m}+3)$.
In both cases, both  $C_1\,{\cal D}(d,\ell)$
and  ${\cal D}(d,\ell)\,C_1$  correspond to $\N^{00}(m,r/m)$.
The modular invariant ${\cal D}(d_1,\ell_1|d_2,\ell_2)$ is NIMmed iff
both $d_1=d_2$ and $\ell_1=\ell_2$, in which case the corresponding NIM-rep
is $\N(2m,{2r\over m})$.

Triality for $\widehat{{\rm so}}(8)$ introduces some additional so$(8)_2$
modular invariants, but all are NIMmed. In particular, the two order-3
conjugations both correspond to $\N^0(2)$, while the other two additional
conjugations correspond to $\N'(5,J_s)$ and $\N'(5,J_c)$. The additional
modular invariants arising from conjugations of ${\cal D}(2,1)=M[J_s]$
correspond to $\N(4,J_s)$ and $\N(4,J_c)$.

The current algebra so$(2r)_2=D_{r,2}$ will have NIM-less modular invariants, unless either $r$ is
prime, or $r=4$ or 8.

\bigskip\noindent{{\bf 6. The simplest NIM-less modular invariants}}\medskip

There is modular data canonically associated to finite groups [15].
The $S$ and $T$ matrices, and a list of modular invariants, is given in
[34] for the symmetric group $S_3$.
For convenience label the
primary fields here $0,1,\ldots,7$ as in [34]. Then:
$$S={1\over 6}\left(\matrix{1&1&2&2&2&2&3&3\cr 1&1&2&2&2&2&-3&-3\cr
2&2&4&-2&-2&-2&0& 0\cr 2&2&-2&4&-2&-2&0&0\cr
2&2&-2&-2&-2&4&0&0\cr 2&2&-2&-2&4&-2&0&0\cr 3&-3&0&0&0&0&3&-3\cr 3&-3&0&0&0&0&-3&3}\right)$$
Some of its modular invariants  don't have an associated
NIM-rep. For example consider
$${\cal Z}=(ch_0+ch_1+ch_2+ch_3)(ch_0+ch_2+ch_6)^*$$
Its exponents(=spin-0 primaries) are the
primaries `0' and `2', so we're looking for a 2-dimensional NIM-rep.
Let's consider the existence of the NIM-rep matrix $\N_3=\left(\matrix{a& b\cr c&d}\right)$,
corresponding to primary `3'.
Like all the $S_3$ primaries, `3' is self-conjugate, so $b=c$. So we're looking for
a $2\times 2$ symmetric $\Z_{\ge}$-matrix, with eigenvalues 
$S_{30}/S_{00}=2$ and $S_{32}/S_{02}=-1$.
One way to  see such a matrix can't exist is to consider its
trace and determinant: Tr$(\N_3)=1=a+d$ (so either $a$ or $d$ vanishes), and
det$(\N_3)=-2=ad-b^2=-b^2$, i.e. $b=\sqrt{2}\not\in\Z$.

So no NIM-rep can correspond to that modular invariant ${\cal Z}$.
More generally, this probably accounts for the abundance of modular
invariants arising for finite groups [34].

The simplest example of a NIM-less WZW modular invariant 
is $\B(3,1|9,1)$ for $\widehat{{\rm so}}(9)$ level 2, given explicitly in
\S5.2. It has exponents $\{0, 2\Lambda_1,
\Lambda_1,\Lambda_2,\Lambda_3,\Lambda_3,\Lambda_3,2\Lambda_4\}$, all
with multiplicity 1. Here's a simple argument that it is NIM-less:

Let $A=\N_{\Lambda_1}$ be the matrix for the first-fundamental
weight. It is an $8\times 8$ $\Z_{\ge}$-matrix.
Since charge-conjugation is trivial here,
we know $A=A^t$. The
quantum-dimension of $\L_1$ is $2$, so this must be the
maximal eigenvalue $r(A)$ of $A$.
The traces of a NIM-rep matrix $\N_\mu$ can be obtained in terms of the
exponents ${\cal E}$ by (3.8).
Using this we see that Tr$(A)=1$.
We want to show that no such matrix can
exist, and also respect the fusion 
$$\Lambda_1\stimes\Lambda_1=0\splus(2\Lambda_1)\splus\Lambda_2$$

Looking at the exponents, we see that all eigenvalues of both the vacuum 0
and simple-current $2\L_1$ are +1, and thus $\N_0=\N_{2\L_1}=I$. Nonnegativity
of $\N_{\L_2}$ thus requires $2\le (A^2)_{ii}=\sum_j(A_{ij})^2$ for
all $i$ and so
each row sum $\sum_jA_{ij}\ge 2$. But a standard fact of Perron-Frobenius
theory is that for any nonnegative matrix $B$, the minimum row-sum can equal
the maximum eigenvalue $r(B)$ iff all row-sums equal $r(B)$. Thus all row-sums
equal 2, and $\sum_{i,j}A_{ij}$ is even. However, since $A$ is symmetric,
$\sum_{i,j}A_{ij}\equiv {\rm Tr}(A)=1$ (mod 2).

This contradiction means that  no such matrix $\N_{\Lambda_1}$ can exist,
and so we can't have a NIM-rep for this modular invariant.
As was proved last section, most of the modular invariants for
$\widehat{{\rm so}}(n)$ level 2 are likewise NIM-less.

%
%
%

\bigskip\centerline{{\bf 7. How to make your own NIM-rep classifications}}\bigskip

In this short section we explain how to put some of the ideas of \S3.3 together, in
order to obtain NIM-rep classifications (of a sort) for any choice of modular
data. For definiteness, consider $\widehat{{\rm sl}}(3)$ level $k=8$
(our methods though are completely general). It has
$45$ weights$=$primaries. We'll find all
possible sets of exponents. 

At first glance this seems challenging, since the generator $\N_{\L_1}$
will have largest eigenvalue about $2.6825$, significantly beyond
any known matrix or graph classification. However, the Galois symmetry enormously simplifies this task, making it
essentially do-able by hand. In particular, Thm.3(iv) says that the
exponent ${\cal E}({\N})$ is a union of Galois
orbits. $P_+$ here has only four orbits with respect to its Galois group
$(\Z/33\Z)^\times$. They are 
$$\eqalign{{\cal O}_0=&\{(0,0),(1,1),(2,2),(3,3),(4,4)\}\cr
{\cal O}_1=&\{(0,3),(0,6),(1,4),(1,7),(2,5),(3,0),(4,1),(5,2),(6,0),(7,1)\}}$$
and ${\cal O}_2=J{\cal O}_0\cup J^2{\cal O}_0$ and ${\cal O}_3=J{\cal
O}_1\cup J^2{\cal O}_1$, using obvious notation, where the action of
the simple-current $J$ here is given by $J(a,b)=(8-a-b,a)$. It suffices then to 
determine the four multiplicities $m_0,m_1,m_2,m_3$. However, the
multiplicity of the vacuum must be 1 if the NIM-rep is to be
indecomposable, so we must have $m_0=1$. A simple-current can only have
multiplicities 0 or 1 (Thm.3(iii)), so $m_2=0,1$.

Consider first $m_2=1$, i.e.\ $J\in{\cal E}(\N)$. Then by Thm.3(iii), all
$J$-orbits must have constant multiplicity, i.e.\ $m_1=m_3$.
Now the trace (3.8) for $\la=(0,6)$ gives us
$$8m_0-6m_1+16m_2-12m_3\ge 0$$
that is, $24\ge 18m_1$, i.e.\ $m_1=0,1$. So two possible exponents are
$(m_0,m_1,m_2,m_3)=(1,0,1,0)$ and $(1,1,1,1)$.

Now consider $m_2=0$. Then the same trace inequality now becomes $8\ge
6m_1+12m_3$, i.e.\ $m_3=0$ and $m_1=0,1$. So the two remaining possible
exponents are $(m_0,m_1,m_2,m_3)=(1,0,0,0)$ and $(1,1,0,0)$.

Each of these four possible exponent multi-sets are in fact realised by each of
the four ${{\rm sl}}(3)_8$ modular invariants. For example, charge-conjugation
corresponds to $(1,0,0,0)={\cal O}_0$. Corresponding NIM-reps are given in [7].

Apparently no other NIM-reps are known for $\widehat{{\rm sl}}(3)$ level 8,
but that may be simply because no one has looked really hard (e.g.\
we give at the end of \S3.2 a NIM-rep for sl$(3)_3$ which seems to be new).
The Perron-Frobenius eigenvalue of $\N_{\L_1}$ here is
large enough to conceivably allow more than one realisation
for a given exponent.

Incidentally, the same method severely constrains some off-diagonal
entries of any ${{\rm sl}}(3)_8$ modular invariant. For instance, we likewise
get four possibilities for each of the multi-sets $\{m_\mu^\pi=M_{\mu,\pi\mu}\}_{\mu\in P_+}$
where $\pi$ is any of the four fusion-automorphisms of ${{\rm sl}}(3)_8$. The fusion-automorphisms
for any current algebra were classified in [29]; for sl$(n)_k$ they are
given by $\la\mapsto C^jJ^{a\sum_{i=1}^{n-1}i\la_i}\la$, where $j=0,1$ and gcd$(ak+1,n)=1$.

\bigskip\centerline{{\bf 8. Final remarks: speculations and questions}}\medskip

To get the main thrust of the paper with a minimum of effort, read \S\S6,7 and this
conclusion. Our main results are the sl$(n)_2$ and so$(n)_2$ NIM-rep
classifications, as well as Thm.3 and its comparison to Thm.1.

\smallskip\noindent{{\bf (1)}} We've found infinitely many NIM-reps lacking a corresponding
modular invariant (this is very typical behaviour). We've found infinitely
many modular invariants lacking a NIM-rep (e.g.\ $\widehat{{\rm so}}(n)$ level 2).
We've found infinitely many pairs of distinct modular invariants which
correspond to identical NIM-reps (e.g.\ $\widehat{{\rm so}}(8n)$ level 1, or triality
and its inverse for $\widehat{{\rm so}}(8)$ at any level) --- this refutes
a hope expressed in \S2.4 of [33].
There are also different NIM-reps corresponding to identical modular invariants
(e.g.\ the $\widehat{{\rm sl}}(3)$ level 9 NIM-reps called ${\cal
E}^{(12)}_i$, $i=1,2,3$, in [7]).

Incidentally, {\it different modular invariants can correspond to identical
RCFTs!}\footnote{$^*$}{{\little On this simple but (to me) surprising point, as well as paragraph (2) below, I've
benefitted from conversations with M.\ Gaberdiel. The referee informs
me that `it is already known for years', but not to me!}}
A simple example  is WZW $\widehat{{\rm so}}(16)$ level 1, where
the four distinct modular invariants $C_1^i\,M[J_s]\,C_1^j$ ($i,j=0,1$)
all correspond to the WZW $\widehat{E}_8$ level 1 theory
(see \S4.2 if this notation seems obscure) --- there is, after all, only one $c=8$ holomorphic theory!
 More precisely, the partition functions ${\cal Z}$
of these seemingly different modular invariants are indeed different functions
of the modular parameter $q=e^{2\pi\i\tau}$ and the left- and right-moving
Cartan angles $\vec{z}_L,\vec{z}_R\in{\Bbb C}^8$ --- though their $q$-dependence
is the same, their $\vec{z}$-dependence differs by a change-of-basis.
Different so(16)'s sit inside $E_8$, and they yield different decompositions
${\cal H}=\oplus_{\la,\mu}M_{\la\mu}{\cal H}_\la\otimes\overline{{\cal H}}_\mu$
of the state space in terms of so$(16)_1$ modules, and hence express the
same $E_{8,1}$ theory by {\it distinct} so$(16)_1$ modular invariants!

It's long been known that different RCFTs can have identical partition
functions ${\cal Z}$, but this always seems to be because ${\cal Z}$ isn't
taken with full variable dependence. For instance the $q$-functions of the
two holomorphic $c=16$ theories are identical, but can be distinguished
when their Cartan angles are considered. Or a more interesting example: the simple-current modular
invariant $M[J]$ for sl$(3)_3$ is indistinguishable from its charge-conjugate,
even when all sl$(3)$ Cartan angles are included; by contrast we get 6 different
modular invariants (all restricting to $M[J]$)
 when the Cartan angles of so$(8)_1$ (the maximally extended chiral
algebra here) are considered. Incidentally, other hints that this sl$(3)_3$
modular invariant is `degenerate' come from NIM-reps [7] and
twisted partition functions ${\cal Z}_{g,g'}$ [33].

But what this new observation
tells us is that there    can be
different ways to introduce the Cartan angles into a given RCFT, and they result in
different `full-variable' partition functions. Potentially, a similar
problem can arise any time the modular invariant is written in terms of
characters of a nonmaximal chiral algebra.

\smallskip\noindent{{\bf (2)}} What does it mean when a modular invariant is NIM-less? The simplest
guess is that it is nonphysical (i.e.\ can't be realised as the 1-loop partition
function in a consistent RCFT). In fact, Verstegen argued in [39] that the
so$(9)_2$ modular invariant ${\cal B}(3,1|9,1)$ is nonphysical, by saying that no chiral extension with
 modular data could
 be found in which ${\cal B}(3,1|9,1)$ would be diagonal. A similar claim
is made in [40], regarding the ${{\rm so}}(15)_2$
 modular invariant we call ${\cal B}(15,1|15,4)$. It is tempting to conjecture
 that any NIM-less modular invariant for ${{\rm so}}(n)_2$ will have a similar
 problem: its `maximal chiral extension', if it exists in some form,
 won't have healthy $S$ and $T$ matrices. It should be emphasised
though that the requirement that a consistent RCFT have a compatible
NIM-rep is not as solid as for instance the requirement that its torus
partition function be modular invariant.

\smallskip\noindent{{\bf (3)}} The dual  question is: What about the NIM-reps
that fail to correspond to a modular invariant? Most notable among these are
the tadpoles $T_n$ of $\widehat{{\rm sl}}(2)$ level $2n-1$. In fact these correspond
to the {\it sub}modular invariant $M_{\la\mu}=\delta_{\mu,J^{\la_1}\la}$
where $J$ is the simple-current, taking $\la=(\la_0,\la_1)$ to $(\la_1,\la_0)$.
This $M$ is not a true modular invariant (e.g.\ 
it commutes with $T^4$ but not $T$); because it's invariant under a (small-index)
subgroup
of the modular group, we call it a submodular invariant.
Similar remarks apply to the NIM-reps for ${{\rm sl}}(n)_1$ which don't
have a corresponding modular invariant (see \S4.2).

So a natural question is: can an RCFT (or string theoretic) interpretation
be given to the assignment of NIM-reps to certain submodular functions?

\smallskip\noindent{{\bf (4)}}
It is also natural to ask: Find a {\it simple} explanation
(there are many complicated ones) for why there is no $\widehat{\rm sl}(2)$
modular invariant at level $2n-1$, corresponding to the tadpole $T_n$.
Then this could give rise to an additional NIM-rep axiom, permitting us to
automatically dismiss nonphysical ones.

An original axiom of sl$(n)_k$ fusion graphs [7,35] was that there be a
$\Z_n$-grading on the vertices of the graph, compatible with the $n$-ality
$t(\la):=\sum_j j\la_j$. This was introduced because for sl$(2)_k$ it threw
away the unwanted tadpoles and retained the A-D-E NIM-reps. This axiom has now been
dropped, because we now understand it to be too restrictive --- Thm.3(viii)
tells us that it is equivalent to demanding that the simple-current
$J$ be an exponent, which isn't always true of healthy modular invariants.

However, the most appropriate NIM-rep axiomatisation may be
in between these two extremes. As mentioned in Prop.2, for most current algebras
(including every sl$(n)_k$),
we know that all modular invariants (known and unknown) are required to
have certain simple-currents as exponents, and hence the corresponding NIM-reps
will necessarily have nontrivial gradings.

If we are interested only in NIM-reps which correspond to modular invariants
(for suitable pairing $\omega$ --- this is discussed in \S2.2), then we should dismiss from any consideration
those NIM-reps which will necessarily fail for an elementary reason. In this
view, it was correct to require for $\widehat{{\rm sl}}(2)$ that the fusion graphs
be bipartite. If we permit ourselves the freedom of choosing an appropriate
pairing $\omega$, as apparently we should [5,28], then for example we can also demand that
the NIM-reps for $\widehat{{\rm sl}}(n)$ be $\Z_n$-graded, for $n<8$
(for $\widehat{{\rm sl}}(8)$ we can only demand the
NIM-reps to be $\Z_4$-graded).

More generally, we could demand that any weight $\kappa$ satisfying
(2.12d) be an exponent of our NIM-rep.

\smallskip\noindent{{\bf (5)}} A property (hence a possible additional
axiom for NIM-reps) which any physically realised NIM-rep must obey, has been
suggested recently [41]. Namely, there must exist a vertex $1\in{\cal
B}$ such that, for all $\la\in P_+$, 
$${\rm min}_{x\in {\cal B}}\N_{\la x}^x=\N_{\la 1}^1$$ 
It would clearly be interesting to test the spurious NIM-reps obtained here
(and elsewhere) with this relation, and also to derive some
consequences in the spirit of \S3. Two quick examples are: 

\noindent\item{{\it (i)}}
$U_{1,0}={\rm min}_x U_{x,0}$, where $U_{\updownarrow,0}$ is the
common Perron-Frobenius eigenvector of all $\N_\la$; 

\noindent\item{{\it (ii)}} for any $\la\in P_+$, the
norm-squared $\sum_y (\N_{\la x}^y)^2$ of any row of $\N_\la$ will be
minimal for $x=1$. 

\noindent To get (i), consider the sum $\sum_\la S_{0\la}\N_{\la
x}^x$. To get (ii), consider the product $\N_\la\N_{C\la}$.
 
\smallskip\noindent{{\bf (6)}} Some authors (e.g.\ [6]) have  suggested
 that the study of NIM-reps may shed light on
modular invariant classifications. However our view is that, although
NIM-rep classifications are extremely pleasant in the simplest cases,
their complexity rises much quicker than that of modular invariants. For
instance we get an immediate understanding of the A-D-E in $\widehat{{\rm sl}}(2)$
NIM-reps, while the corresponding explanation is still lacking in the 
$\widehat{{\rm sl}}(2)$ modular invariant classification. On the other
 hand, it is possible
to obtain fairly easily the full modular invariant classification for e.g.\
$\widehat{E}_8$ level 380 [12] (the answer is simply $M=I$), although it
would be completely hopeless to determine its NIM-reps --- certainly we would
expect enormous numbers of them. For sl$(2)_k$ there is a single generating
primary, and its quantum-dimension is $<2$; for $E_{8,380}$ we need
8 generating primaries, and the smallest has quantum-dimension
very nearly 248.

Decades ago, it was conjectured that a graph was uniquely determined by its
eigenvalues. By now many pairs of {\it cospectral} graphs (graphs with identical
eigenvalues) are known. The simplest pair is $A_3^{(1)}\cup A_1$ and $D_4^{(1)}$.
It turns out that $5.9\%$ of all graphs with 5 vertices, are not determined by
their eigenvalues; the percentage is $6.4\%$ for 6 vertices, $10.5\%$ for 7,
$13.9\%$ for 8, and $18.6\%$ for graphs with 9 vertices. It is now conjectured
that this percentage rises to 100\% as the number of vertices increases ---
in other words, to almost every graph there would be at least one other with
exactly the same eigenvalues and multiplicities. This is already known to
be true for trees [32]. And the situation is far worse if you allow (as typically
we must) {\it directed} edges and {\it loops} --- even for 2 vertices, almost never do the
eigenvalues identify the multi-digraph. What this seems to suggest is that,
for more typical modular data, there will be several NIM-reps possessing
the same exponents.

Although it is a natural instinct of the mathematically inclined to classify,
in hindsight the resulting lists rarely seem to be of much value. What we seek are classifications
which have structure and in that way suggest new questions. Or we want to classify something
which is so interesting or useful that even if its classification were a
complicated tangle, it would still be of value. We suspect further NIM-rep
classifications will typically be not worth the trouble. Similar comments
apply to the modular invariants corresponding to the finite group modular data
of [15]. By contrast, a typical current algebra has a list of
 modular invariants which is simple and structured --- see e.g.\ the Tables
 in [18,12].

That said, the handful of NIM-rep classifications we now have {\it do} cast light
on the modular invariant $\leftrightarrow$ NIM-rep correspondence. It
would be interesting to study the NIM-reps for a `typical' current algebra
whose smallest nontrivial quantum-dimension is much larger than 2. This
would test our speculation that its number of NIM-reps would be large.
Also, it would be interesting to study the NIM-reps for the finite group
`pre-orbifold' modular data [15], say for the symmetric group $S_3$ and
the dihedral group $D_4$. This would test our speculation that most of the
remarkable numbers of modular invariants there are spurious.

So our view is that the value of NIM-reps to modular invariant classifications
is indirect: eliminating spurious modular invariants.
However this inefficacy of the NIM-rep hypothesis 
could change if someone would find a simple
property of a NIM-rep spectrum which isn't automatically obeyed by
modular invariants.

\smallskip\noindent{\bf (7)}
Why is so$(n)_2$ so special here? Because it has so many modular
invariants. One reason for this is that rank-level duality associates
so$(n)_2$ with u$(1)_{n+2}$, and $\widehat{{\rm u}}(1)$
has a relatively rich variety of modular invariants
coming from its
simple-currents. However, a better reason is that the so$(n)_{2}$ matrix $S$ formally looks like the
character table of the dihedral group and for some $r$ actually equals
the Verlinde matrix $S$ associated to the dihedral group ${D}_{n}$
twisted by an appropriate 3-cocycle [34]. Finite group modular data yields
swarms of modular invariants. The critical factor is the impotence of the Galois
parity condition (2.10b) here as most (for ${{\rm so}}(n)_2$)
or all (for finite groups) of the parities $\epsilon_\ell$ are identically +1.
This is very different from the other current algebras.

\smallskip\noindent{\bf (8)} As mentioned in {\bf (6)}, we suspect that classifying NIM-reps is
probably hopeless for all but the smallest ranks or levels. There will be
too many of them. (This is in
marked contrast to modular invariants, at least for the current algebras.)
This speculation leads to an intriguing question: Could this be hinting that there
will typically (e.g.\ current algebras of large rank and level)
be several different RCFTs for a given modular invariant?

\bigskip
\noindent{{\bf Acknowledgements}} \smallskip

It is a pleasure to thank David Evans,
J\"urgen Fuchs, Matthias Gaberdiel, Adrian Ocneanu, Philippe Ruelle, Christoph Schweigert, Mark
Walton, and Jean-Bernard Zuber for discussions and comments on early
drafts. I learned about boundary CFT from the
conferences at Warwick (1999) and Kyoto (2000) on ``Modular invariants,
subfactors, and singularity theory'', and I thank the organisers for
the invitations and financial support. This paper was largely written
at St.\ John's College, Cambridge, who I warmly thank for their
generous hospitality. This research was also supported in part by NSERC.

\bigskip \centerline{{\bf Appendix A. Proofs}}

\bigskip\noindent{{\it A.1. The $\widehat{{\rm sl}}(n)$ level 2 proof.}}\medskip

Recall the parametrisation $\la(ab)$ of $P_+$ given in \S5.1.
The $S$ entries are given by the formula
$$S_{\la(ab),\la(cd)}={2\over \sqrt{n\,\k}}\exp[\pi \i{(a+b)(c+d)\over n}]\,
\sin(\pi\,{(b-a+1)(d-c+1)\over \k})\eqno(A.1)$$
where we require $0\le a\le b<n$ and $0\le c\le d<n$, and put $\k=n+2$.

For  $\widehat{{\rm sl}}(n)$ level 2 and any $1\le \ell<n-1$, we have the
 fusion product
$$\L_1\stimes \L_\ell=\L_{\ell+1}\splus(\L_1+\L_\ell)\eqno(A.2)$$

Let $\N$ be any NIM-rep. Write  $n=2^hm$ where $m$ is odd.
 Write $P=\N_J$ and $A=\N_\la$, where
 $\la:=J^{(m-1)/2}\L_1=\la({m-1\over 2},{m+1\over 2})$; $P$ is a
 permutation matrix corresponding to a permutation $\pi$ of the
 vertices, and  $A$ will correspond to a multi-digraph ${\cal G}$. As
 together $\la$ and $J$ are fusion-generators, it suffices 
to find both $\pi$ and ${\cal G}$. The point is that $A^t=P^{-m}A$ and
that  $r({\cal G})=2\cos(\pi/\k)<2$, so
Lemma A of Appendix A.4 applies: we find that the components of ${\cal
 G}$ are digraphs corresponding to the diagrams of Figure 3.
Those diagrams are explained in \S5.1.

The eigenvalues of these digraphs are given in Table 3. The
$m_i$ in the first six rows come from Table 1. The multiplicity of 0
 for $C_n(k)$ is $k$ (for $n$ even) and $2k$ (for $n$ odd --- 
the extra $k$ coming from $2m-1=n$).
We obtained these eigenvalues by twisting by roots of 1 the
eigenvectors for Figure 1.

\vfill\eject\centerline {{\bf Table 3. Eigenvalues of Graphs in Figure 3}}
$$\vbox{\tabskip=0pt\offinterlineskip
  \def\tablerule{\noalign{\hrule}}
  \halign to 5.4in{
    \strut#&\vrule#\tabskip=0em plus1em &    
    \hfil#&\vrule#&\hfil#&\vrule#&\hfil#&\vrule#&    
    \hfil#&\vrule#\tabskip=0pt\cr\tablerule         
&&\omit\hidewidth Graph \hidewidth&&\omit\hidewidth $\#$ vertices\hidewidth&&
\omit\hidewidth eigenvalues \hidewidth&&\omit\hidewidth range\hidewidth&\cr\tablerule
&& $A_n(k)$, $n\ge 1$ &&\hfill $kn$\hfill && \hfill$2\,\exp[\pi\i\ell/k]\,\cos(\pi\,m_i/(n+1))$\hfill&&
\hfill$0\le \ell<k$, $1\le i\le n$\hfill& \cr
&& $D_n(k)$, $n\ge 4$ && \hfill$kn$\hfill && \hfill$2\,\exp[\pi\i\ell/k]\,\cos(\pi\,m_i/(2n-2))$\hfill&&
\hfill$0\le \ell<k$, $1\le i\le n$\hfill& \cr
&&\hfill  $E_6(k)$\hfill  && \hfill$6k$\hfill && \hfill$2\,\exp[\pi\i\ell/k]\,\cos(\pi\,m_i/12)$\hfill&&
\hfill$0\le \ell<k$, $1\le i\le 6$\hfill& \cr
&&\hfill  $E_7(k)$\hfill  && \hfill$7k$\hfill&&\hfill$2\,\exp[\pi\i\ell/k]\,\cos(\pi\,m_i/18)$\hfill&&
\hfill$0\le \ell<k$, $1\le i\le 7$\hfill& \cr
&&\hfill  $E_8(k)$\hfill  &&\hfill$8k$\hfill&&\hfill$2\,\exp[\pi\i\ell/k]\,\cos(\pi\,m_i/30)$\hfill&&
\hfill$0\le \ell<k$, $1\le i\le 8$\hfill& \cr
&&\hfill $T_n$, $n\ge 1$\hfill &&\hfill$n$\hfill&&\hfill$2\,\cos(\pi\,m_i/(2n+1))$
\hfill&&\hfill $1\le i\le n$\hfill& \cr
&& $B_n(k)$, $n\ge 3$ &&\hfill$(2n-1)k$\hfill&&\hfill$2\,\exp[\pi\i\ell/k]\,\cos(\pi\,(2m-1)/2n)$,
\hfill&&\hfill$1\le m\le n$,\hfill&\cr
&&&&&&\hfill$2\,\exp[\pi\i\,(2\ell+1)/2k]\,\cos(\pi\,m'/n)$\hfill&&\hfill$0\le \ell<k$,
$1\le m'<n$\hfill& \cr
&& $C_n(k)$, $n\ge 2$ &&\hfill$(n+1)k$\hfill&&\hfill$2\,\exp[\pi\i\ell/k]\,\cos(\pi\,(2m-1)/2n)$,\hfill
&&\hfill$0\le \ell<k$, $1\le m\le n$\hfill& \cr &&&&&&\hfill 0 (mult $k$)
\hfill&&&\cr
&& \hfill $F_4(k)$\hfill  &&\hfill$6k$\hfill&&\hfill$2\,\exp[\pi\i\ell/k]\,\cos(\pi\,m/12)$,
\hfill&&\hfill$m\in\{1,5,7,11\},$\hfill& \cr
&&&&&&\hfill$\pm\exp[\pi\i\,(2\ell+1)/2k]$\hfill&&\hfill$0\le \ell<k$\hfill&\cr
\tablerule\noalign{\smallskip}
 }} $$

Write $2^a d$ ($d$ odd) for the order of $P$. So $a\le h$ and $d$ divides
$m$. The permutation $\pi$ must interchange all components of ${\cal
G}$, since our NIM-rep $\N$ is indecomposable. 
Since $P$ and $A$ commute, we
get $A_{\pi i,\pi j}=A_{ij}$ --- i.e.\ each component of ${\cal G}$ is equivalent.

Now, $P^m$ permutes vertices
within each component, so so must $P^d$, and the number of components must then
divide $d$. If it's
a proper divisor ($d'$ say), then $P^{2^a d'}$ will also permute the vertices
of each component, and so would constitute an odd-order symmetry of each
component. But of the diagrams $X_s(2^\ell)$ in Figure 3,  only $D_4(2^\ell)$ has 
a nontrivial odd-order symmetry. However
$r(D_4(2^\ell))=2\cos(\pi/6)=2\cos(\pi/(n+2))$, so $D_4(2^\ell)$ corresponds to
$\widehat{{\rm sl}}(4)$, in which case $d=1=d'$. Thus for any $\N$, the number of components
$d'$ of ${\cal G}$ must exactly equal $d$.

Write $X_s(2^\ell)$ for the common name of the components of ${\cal G}$, as given in Figure 3.
Since $P^m$ has order $2^a$, and $A^t=P^{-m}A$, either the order $a=\ell+1$ (if both weights `$k$' and
`$2k$' appear in Figure 3), or $a=\ell$ (otherwise). 

Given the matrix $A$, i.e.\ $d$ copies of the digraph $X_s(2^\ell)$,
and the matrix $P^m$, i.e.\ $d$ copies of $\Pi^t$, we
can uniquely determine the permutation matrix $P$ as follows. 
The order $d$ permutation $P^{2^h}$ must permute the $d$ different
components, because otherwise the NIM-rep would be
decomposable. Ordering the components appropriately, we can require
that $P^{2^h}$ takes the $j$th component to the $(j+1)$th  one. We can
label the vertices of each component compatibly, in the sense that
$P^{2^h}$ takes vertex $(v,i)$ of one component to $(v,i)$ of the next
one. By fixing $P^m$ and $P^{2^h}$ in this way, we've determined $P$. Thus,
the whole NIM-rep $\N$ is uniquely determined by the component diagram
$X_s(2^\ell)$ and the number $d$.

Let's now run through the possibilities:

\smallskip\noindent{{\it Case 1:}} Suppose the components are $X_s(2^\ell)=
A_s(2^\ell)$. The largest eigenvalue tells us $s=n+1$. Counting the
simple-current exponents
of $A_{n+1}(2^\ell)$, we get precisely $2^{\ell+1}d$; they all have multiplicity
1 and form a subgroup of $\Z_{n}$, so $\ell<h$ and hence $\k=n+2$
must be even. Then
$$2\exp[\pi\i\,(a+b)/2^h]\,\cos[\pi\, (b-a+1)/\k]=2
\cos[\pi\,m_i/\k]$$
has no solution $a,b$ when $m_i$ is even (except for $n=2$, which
fails because $\N_J$ would have order $2^\ell=1$, even though $\L_1$
would be an exponent). This impossibility means that
$A_{n+1}(2^\ell)$ can never generate a representation of our fusion ring,
so can't appear in an sl$(n)_2$ NIM-rep.

\smallskip\noindent{{\it Case 2:}} Suppose the components are $D_s(2^\ell)$.
Then $s=(n+4)/2$ (so $n$ is even). For the same reason as in Case 1, we must have
$\ell<h$. Again, $m_i=s-1$ cannot be even, so $4|n$. The rest
is trivial.

\smallskip\noindent{{\it Case 3:}} Suppose the components are $T_s$.
Then $s=(n+1)/2$, and the rest follows.

\smallskip\noindent{{\it Case 4:}} Suppose the components are $B_s(2^\ell)$.
Then $s=(n+2)/2$ and $\ell<h$, as usual. In fact, we can fix $\ell$:
no $a,b$ can be found obeying 
$$2\exp[\pi\i\,(a+b)/2^h]\,\cos[\pi (b-a+1)/\k]=2\exp[\pi\i\,(2j-1)/2^{\ell+1}]\,
\cos[2\pi\,m'/\k]$$
for $m'=j=1$, unless $\ell={h-1}$.

\smallskip\noindent{{\it Case 5:}} Suppose the components are $C_s(2^\ell)$.
Then $s=(n+2)/2$, and everything else proceeds as in Case 2. When
$4|n$ and $\ell>n-2$, what we find
though is that $n/2^\ell d$ will divide $i+j$ for all exponents
$\la(ij)\in\E$, which would mean by (3.5) that $P=\N_J$ would have order $2^\ell d$, not $2^{\ell+1}d$
as it should here. When $n/2$ is odd, we are saved by the fixed-points.

\smallskip\noindent{{\it Case 6:}} The exceptional digraphs are all handled
in similar ways. For instance, suppose the components are $E_6(2^\ell)$.
Then $\ell=0$, and the graph eigenvalue for $m_i=4$ won't equal any
Verlinde eigenvalue $S_{\la,\la(ab)}/S_{0,\la(ab)}$. So $E_6(2^\ell)$
cannot appear here.

\bigskip \noindent{{\it A.2. The $\widehat{{\rm so}}(odd)$ level 2 proof.}}\medskip

Recall the  weights $\ga^a$ parametrised in \S5.2.
The $S$ matrix entries for so$(n)_2=B_{r,2}$, where $n=2r+1$, are [38]
$$\eqalignno{S_{J^i0,J^j0}={1\over 2}S_{J^i0,\ga^a}=&\,{(-1)^i\over\sqrt{n}}
S_{J^i0,J^j\L_r}={1\over 2\sqrt{n}} &(A.3a)\cr
S_{\L_r\L_r}=&\,S_{J\L_r,J\L_r}=-S_{\L_r,J\L_r}=0.5&(A.3b)\cr
S_{\ga^a\ga^b}=&\,{2\over \sqrt{n}}\cos{2\pi ab\over n}&(A.3c)\cr
S_{\L_r\ga^a}=&\,S_{J\L_r,\ga^a}=0&(A.3d)}$$
for each $a,b\in\{1,\ldots,r\}$, $i,j\in\{0,1\}$. 
The fusion products we need are
$$\eqalignno{\L_1\stimes\L_1=&\,0\splus(2\L_1)\splus\L_2&(A.4a)\cr
\L_1\stimes\ga^i=&\,\ga^{i-1}\splus\ga^{i+1} &(A.4b)\cr
\L_1\stimes \L_r=&\L_r\splus(\L_1+\L_r)&(A.4c)}$$
for $1<i<r$. Hence the obvious fusion-generator consists of $\L_1,$ the spinor $\L_r$, and the
simple-current $J=2\L_1$.

Let $\N$ be any indecomposable NIM-rep of ${{\rm so}}(n)_2$. Put $\N_i:=\N_{\L_i}$.
Write $m_\mu$ for the multiplicities of its exponents $\mu\in\E$.
The charge-conjugation $C$ is trivial here, so all matrices $\N_\la$ are
symmetric.
Let's try to find $\N_1$: its quantum-dimension is $S_{\L_1 0}/
S_{00}=2$. Now, 
the connected multigraphs ${\cal G}$ with maximum eigenvalue 2 are given in Figure 2.
The proof that this list is complete is given in \S A.4, and their
eigenvalues are given in Table 2.
Hence $\N_1$ will be the adjacency matrix of a disjoint union of graphs from Figure 2.

Now, there are only two $\la\in P_+$ with $S_{\L_1\la}/S_{0\la}=2$: namely
$\la=0,J$. Moreover, $m_0=1$ and $m_J=0,1$, so $\N_1$ is made up of at most two
connected graphs (Thm.3(vi)).

\smallskip\noindent{\it Case 1:}  $\N_1$ is a single connected graph ${\cal G}_0$.
Then $m_J=0$.

\smallskip\noindent{\it Case 2:} $\N_1$ has precisely two components, ${\cal
G}_1$ and ${\cal G}_2$. Then $m_J=1$, i.e.\ $J$ is an exponent.\smallskip

Now, from (A.3) any eigenvalues of the graph ${\cal G}_i$ will be either 0, or of the form $2\cos(2\pi
a/n)$, for $0\le a\le r$. In particular, $-2$ is not an allowed value, which
excludes anything bipartite (e.g.\ trees). We find that the only
possibilities for the components ${\cal G}_i$, are $A_m^{(1)}$ when $m+1$
divides $n$, $D^0_m$ when $2m-3$ divides $n$, and ${}^0\!A_1^0=(2)$ and
${}^0\!A_2^0=\left(\matrix{1&1\cr 1&1}\right)$. For later convenience, denote
${}^0\!A_1^0$ by $A_0^{(1)}$, and  ${}^0\!A_2^0$ by $D_2^0$.

By (A.3), the multiplicity of eigenvalue 0
will equal the number $m_{\L_r}+m_{\L_1+\L_r}$ of spinors
in $\E$. If $J\in\E$, then $m_{\L_r}=m_{\L_1+\L_r}$. Also, note that the
$A_{even}^{(1)}$ do not have 0 as an eigenvalue, while it is an eigenvalue
of $D_m^0$ with multiplicity 1. There are spinors in $\E$, iff $\N_J\ne I$,
in which case $\N_J$ will be an order-2 permutation matrix.
Note that $\N_J\N_{1}=\N_{1}$, while $\N_{\L_1+\L_r}=\N_J
\N_{r}$. 

Thus we get the following refinement of our cases:

\smallskip\noindent{\it Case 1(a):}  $J\not\in\E$, and no spinors are in $\E$.
$\N_1$ is the adjacency matrix of ${\cal G}_0=A_{m-1}^{(1)}$, for some $1\le m$
dividing $n$. Also, $\N_J=I$ and $\N_{r}=\N_{\L_1+\L_r}$. 

\smallskip\noindent{\it Case 1(b):}  $J\not\in\E$, but one spinor (call it $\si$)
is in $\E$; it has multiplicity $m_\si=1$.
$\N_1$ is the adjacency matrix of ${\cal G}_0=D_m^{0}$, for some $2\le m$
obeying $2m-3$ divides $n$. Also, $\N_J\ne I$ and corresponds to an order-2
symmetry of $D_m^0$.

\smallskip\noindent{\it Case 2(a):} $J\in\E$, but no spinors are in $\E$.
$\N_1$ is given by the direct sum of the adjacency matrices of ${\cal G}_1
=A_{m-1}^{(1)}$ and ${\cal G}_2=A_{m'-1}^{(1)}$, where $1\le m\le m'$ and both
$m,m'$ divide $n$. Also, $\N_J=I$ and  $\N_{r}=\N_{\L_1+\L_r}$.\smallskip

\smallskip\noindent{\it Case 2(b):} $J\in\E$, and both spinors are in $\E$
with multiplicity 1.
$\N_1$ is given by the direct sum of the adjacency matrices of ${\cal G}_1
=D_{m}^{0}$ and ${\cal G}_2=D_{m'}^{0}$, where $2\le m\le m'$ and both
$2m-3,2m'-3$ divide $n$. $\N_J\ne I$ and corresponds to an order-2 symmetry
of the graph ${\cal G}_1\cup{\cal G}_2$.\smallskip

Consider first Case 1(a).
Recall the definition of the matrices $M^{(m|i)},M^{(m|i,j)},\widetilde{M}^{(
m)}$ from \S5.2. We may put $\N_1=M^{(m|1,-1)}$. Note that
$M^{(m|i)}M^{(m|j)}=M^{(m|i+j)}$ so $M^{(m|i,-i)}M^{(m|j,-j)}=M^{(m|i+j,-i-j)}+
M^{(m|i-j,j-i)}$. From this and (A.4a),(A.4b) we obtain $\N_{\ga^i}=M^{(m|i,-i)}$.
Finally, we need the matrix $\N_r=\N_{\L_r}$. (A.4c) says $\N_1\N_{r}
=2\N_{r}$, so each column of $\N_r$ is an eigenvector of $\N_1$ with
eigenvalue 2. This eigenspace is 1-dimensional, spanned by $(1,1,\ldots,1)^t$,
so each column of $\N_r$ is constant.  Since also $\N_r^t=\N_r$, we get that
$\N_r=a\cdot 1_{mm}$.
The constant $a$ can be determined by quantum-dimension calculations.
The result is $\N=\N(m)$, given in \S5.2.

Next, turn to Case 1(b). Here we can put $\N_1=\widetilde{M}^{(m)}$.
There is only one  order-2 symmetry of the graph
$D_m^0$: $\N_J$ must interchange the two degree-1 vertices (i.e.\
the nodes $m-1$ and $m$). 

All that remains is to determine the matrix $\N_r$.
 From (A.4c) we get that its columns lie in the nullspace Null$(\N_1-I-\N_J)$, and so are of
 the form $(x,x,\ldots,x,y,z)^t$ where $y+z=x$. Now use $\N_r^t=\N_r$
 and $\N_J^{-1}\N_r\N_J=\N_r$ to get
 $$\N_r=\left(\matrix{2a&\cdots&2a&a&a\cr \vdots&&\vdots&\vdots&\vdots\cr
2a&\cdots&2a&a&a\cr a&\cdots&a&b&c\cr a&\cdots&a&c&b}\right)$$
where $a=b+c$.  Its Perron-Frobenius eigenvector is $(2,2,\ldots,2,1,1)^t$,
with eigenvalue $a\,(2m-3)/2$, and this must equal the quantum-dimension
$S_{\L_r,0}/S_{00}=\sqrt{n}$. This fixes $a$. 
The trace (3.8) tells us $b={a+1\over 2}$ (if $\si=\L_r$) or $b={a-1\over 2}$ (if $\si=J\L_{r}$).
Then $\N_{\L_1+\L_r}=\N_J\N_r$ will be the same as $\N_r$, except with
$b$ and $c$ interchanged. The result is $\N'(m,\si)$.

Incidentally, the fact that case 1 requires $n$ to be a perfect square
follows from Galois (Thm.3(iv)): when $\sqrt{n}\not\in \Z$, the Galois orbit of 0
is $\{0,J\}$.

Case 2(a) is similar to Case 1(a), so the details won't be repeated. To get that
the upper-left $m\times m$ and lower-right $m'\times m'$ blocks in $\N_r$
are 0, use Thm.3(x). The result is $\N(m,m')$.

Case 2(b) essentially reduces to two copies of the Case 1(b) argument.
To determine $\N_J$, use the nonnegativity of $\N_{2}=(\N_1)^2-I-\N_J$,
as well as the fact that $\N_J$ must be a symmetry of the graph $D_m^0\cup
D_{m'}^0$. Again we get Tr$(\N_r)=0$. The result is $\N'(m,m')$.

\bigskip \noindent{{\it A.3. The $\widehat{{\rm so}}(even)$ level 2 proof.}}\medskip

Recall the weights $\la^i$ of so$(n)_2=D_{r,2}$, where  $n=2r$. In [38] we find
that the $S$ entries are
$$\eqalignno{S_{00}={1\over \sqrt{r}}S_{0\L_r}&\,={1\over 2}
S_{0\la^a}={1\over 2\sqrt{n}}&(A.5a)\cr S_{\la^a\la^b}=&\,
{2\over \sqrt{n}}\cos(\pi{ab\over r})&(A.5b)\cr S_{\la^a\L_r}=&\,
S_{\la^a\L_{r-1}}=0&(A.5c)\cr S_{\L_r\L_r}=&\,S_{\L_{r-1}\L_{r-1}}=
{1\over 4}(1+(-\i)^r)&(A.5d)\cr S_{\L_r\L_{r-1}}=&\,{1\over 4}(1-(-\i)^r)
&(A.5e)}$$
for $a,b\in\{1,2,\ldots,r-1\}$.
The remaining entries of $S$ are given by (2.4) and $S=S^t$.

The only fusion products  we need are
$$\eqalignno{\L_1\stimes\L_1=&\,0\splus(2\L_1)\splus\L_2&(A.6a)\cr
\L_1\stimes\la^i=&\,\la^{i-1}\splus\la^{i+1}  &(A.6b)\cr
\L_1\stimes \la^{r-1}=&\,\la^{r-2}\splus(2\L_r)\splus(2\L_{r-1})&(A.6c)\cr
\L_1\stimes \L_{r-1}=&\,\L_{r}\splus(\L_1+\L_{r-1})&(A.6d)\cr
\L_1\stimes \L_r=&\,\L_{r-1}\splus(\L_1+\L_r)&(A.6e)
}$$
where $1<i<r-1$. Hence the obvious fusion-generator consists of $\L_1,$ the spinors
$\L_r$ and $\L_{r-1}$, and the
simple-currents $J_v=2\L_1$ and $J_s=2\L_r$. 

Let $\N$ be any NIM-rep, with exponent $\E$. Write $\N_i:=\N_{\L_i}$,
$\N_v:=\N_{J_v}$, $\N_s:=\N_{J_s}$ and $\N_c:=\N_{J_c}$. 
Consider the matrix $\N_1$:
since $r(\N_1)=2$ and  $\N_1^t=\N_1$, its graph is a disjoint union of the
graphs of Figure 2. Their eigenvalues are given in Table 2.

The weight $\L_1$ has $S_{\L_1\mu}/S_{0\mu}=2$ only
for the simple-currents $0$ and $J_v$, so we'll have 1 or 2 indecomposable
components, as in \S A.2. Likewise, $S_{\L_1\mu}/S_{0\mu}=-2$ iff
$\mu=J_s,J_c$, so a component of $\N_1$ will be bipartite iff either $J_s$ or $J_c$ are  exponents. 

$\N_{v}$ will be an order 1 or 2 symmetry of the fusion graph of
$\L_1$: $\N_1\N_v=\N_v\N_1=\N_1$. It stabilises each component. If the graph has a degree-1 vertex $i$, then this symmetry
must move that vertex to a different degree-1 vertex (otherwise the $(i,i)$ entry of 
$\N_{2}=\N_1^2-I-\N_{v}$ will equal $-1$). This eliminates the possibility of
having components $E_6^{(1)},E_7^{(1)},E_8^{(1)}$, and determines the
permutation $\N_{v}$ restricted to any of the other possible
components from Table 2 (except for ${}^0\!A_2^0$ and
$A_3^{(1)}$). For later convenience, we'll write
$D_2^0:={}^0\!A_2^0$,  $D_3^{(1)}:=A_3^{(1)}$ and $E_4:=A_3^{(1)}$, and
give both $D_3^{(1)}$ and $E_4$ the adjacency matrix displayed in \S5.3.
We take $\N_v$ to act trivially on both ${}^0\!A_2^0$ and $A_3^{(1)}$,
 but to switch the
vertices of $D_2^0$, switch the last two vertices of $E_4$, and to switch
vertices $1\leftrightarrow 2$ and $3\leftrightarrow 4$ of $D_3^{(1)}$.

The Galois orbits of the spinors are: for $r$ odd,
$\{\L_r,\L_{r-1},\L_1+\L_r, \L_1+\L_{r-1}\}$; for $r$ even,
$\{\L_r,\L_1+\L_{r-1}\}$ and $\{\L_{r-1},\L_1+\L_{r}\}$. The Galois
orbit of 0 is: $\{0,J_v\}$ unless $r$ is a perfect square, in which case
it's only $\{0\}$. Finding the Galois orbit of some primary $\mu$ is
easy once you know the $S$ entries: just apply (2.6) to the ratios 
$S_{\nu\mu}/S_{0\mu}$.

Note that the components $A_m^{(1)}$ and ${}^0\!A^0_m$ contribute no spinors to $\E$,
while the components $D^0_m$ ($m\ge 2$) and $E_4$ contribute exactly one spinor to
$\E$, and $D_m^{(1)}$ ($m\ge 3$) exactly two. The reason is that the number of
spinors in $\E$ is precisely the dimension of the common nullspace of $\N_1$
and $\N_2=\N_1^2-I-\N_v$.

Consider first when the graph ${\cal G}$ of $\N_{\L_1}$ is connected. 
Then we know $J_v\not\in\E$. By the Galois symmetry 
Thm.\ 3(iv), $r$ must then be a perfect square. Likewise, we know the graph
cannot be $D_m^0$ ($m\ge 2$) or $E_4$, because then there would only be one spinor in
$\E$.

The following matrices will be useful: Write 
$1^-_k$ for the $2\times k$ matrix whose $(i,j)$th entry is
$(-1)^j$. Write $(\pm 1)_{k\ell}$ for the $k\times\ell$ matrix whose
$(i,j)$th entry is $(-1)^{i+j}$. 

\medskip\noindent{\it Case 1(a):} The graph ${\cal G}$ is the circle
$A_{m-1}^{(1)}$. Then from Table
2, $m$ must divide $2r$. No spinors are in $\E$.
The set $\E$ of exponents  is now determined (up to the choice of $J'$ when $m$ is
even), and by (3.6) we see that all simple-currents
must map to $I_m$. Also, for $m$ even, $r$ must be even because
otherwise $CJ'=J_vJ'$ would also be in $\E$.

Note that $\N_r+\N_{r-1}$ is symmetric, and by (A.6d),(A.6e) obeys
$\N_1\,(\N_r+\N_{r-1})=2\,(\N_r+\N_{r-1})$. So we get
$\N_r+\N_{r-1}={2\sqrt{r}\over m} 1_{mm}$. For $m$ odd, all spinors
must map to the same matrix, by (3.6). For $m$
even, define the spinor $\si$ as in {\bf (i)} in \S5.3 and consider $\N_\si-\N_{C_1\si}$:
it also must be symmetric (since $C$ is trivial) and by (A.6d),(A.6e)
its columns will be eigenvectors of $\N_1$ with eigenvalue $-2$. The
rest follows.

\medskip\noindent{\it Case 1(b):} When the
graph is ${}^0\!A^0_m$, the argument is similar to but simpler than that of Case 1(a).
$\N_s$ is determined as follows: it is nontrivial iff $r/m$ is odd, i.e.\
iff $r$ is odd (since $m|\sqrt{r}$); it also must be a symmetry of
${}^0\!A_m^0$ (since $\N_1=\N_s\N_1\N_s^{-1}$).

\medskip\noindent{\it Case 1(c):} Suppose  the fusion graph is
$D_{m-1}^{(1)}$, $m\ge 4$. Then we know $\E$ has precisely two spinors, so by
Galois $r$ must be even (hence a multiple of 4) and the spinors are
$\si,J_v\si$ for some $\si=\L_r,\L_{r-1}$. This fixes the exponents,
apart from some choice of $J'=J_s,J_c$.

By (A.6d),(A.6e), the columns of $\N_r+\N_{r-1}$ will be 0-eigenvectors of $\N_1-I-\N_v$,
and any $\N_{spinor}$ must commute with $\N_v$. Hence
 we get
 $$\N_r+\N_{r-1}=\left(\matrix{U&a\,1_{2,m'-4}&V\cr
a\,1_{m-4,2}&2a\,1_{m-4,m'-4}& a\,1_{m-4,2}\cr V&a\,1_{2,m'-4}&W}\right)$$
where $U=\left(\matrix{u&a-u\cr a-u&u}\right)$, etc. By rearranging
appropriately
the row/column indices, we may suppose $u\ge a-u$ and $v\ge
a-v$. By computing maximal eigenvalues, we obtain $a=\sqrt{r}/(m-3)$.

By (3.6) we get $\N_{C_1\si}=\N_v\N_{C_1\si}$ and hence
$(\N_r+\N_{r-1})-\N_v\,(\N_r+\N_{r-1})= \N_\si-\N_v\N_\si$ has
eigenvalues $\pm 2\sqrt{2}$ (multiplicity 1) and 0. So the nonzero eigenvalues of
$$\left(\matrix{\Delta u&-\Delta u&\Delta v&-\Delta v\cr
-\Delta u&\Delta u&-\Delta v&\Delta v\cr
\Delta v&-\Delta v&\Delta w&-\Delta w\cr
-\Delta v&\Delta v&-\Delta w&\Delta w}\right)$$
must be $\pm 2\sqrt{2}$, where $\Delta u=2u-a$, etc. Its trace should be 0, so
$\Delta w=-\Delta u$. We obtain $\Delta u=\Delta v=1$, so $a$ and $m$ are odd
and $u,v,w$ are all determined.

Note from $\Delta u>0$ that $\N_\si$ must have nonzero diagonal entries
and hence a nonzero trace, so $S_{\si,J'}/S_{0J'}=+\sqrt{r}$ --- i.e.\ $\si$ and $J'$ are
related as in {\bf (i)} in \S5.3. Thus Tr$(\N_{C_1\si})={\rm Tr}(\N_{J_vC_1
\si})=0$, so the upper-left and lower-right $2\times 2$ blocks of $\N_{C_1\si}$
are $0_{22}$. This also implies  $\N_s=I_m$, by (3.6).

Arguing as above, we find
$$\N_\si-\N_{C_1\si}=\left(\matrix{U&a\,{1^{-}}_{m'-4}&V\cr
a\,(1^{-}_{m-4})^t&2a\,(\pm 1)_{m-4,m'-4}& a\,(1^{-}_{m-4})^t\cr V&a\,1^{-}_{m'-4}&W}\right)$$
where $(\pm 1)_{k\ell}$ and $1^-_k$ were defined earlier. This determines everything. 

\medskip
That completes the discussion of the fusion graph of $\N_1$ being
connected. The other possibility is that the fusion graph possesses
two connected components ${\cal G}_1$ and ${\cal G}_2$, and that $J_v\in\E$. 
Because $J_v\in\E$, the matrices $\N_{spinor}$ must
all be traceless. Also,
note that ${\cal G}_1$ is bipartite iff {\it either} $J_s$ or $J_c$ is in $\E$; but
by Thm.3(iii), that's true iff {\it both} $J_s$ and $J_c=J_sJ_v$ are in $\E$.
Hence ${\cal G}_1$ is bipartite iff ${\cal G}_2$ is bipartite. 
We will first eliminate the possibility that ${\cal G}_1$ is $D_m^0$ or $E_4$.

\medskip\noindent{{\it Case 2(a):}} Suppose ${\cal G}_1$ is the graph $D_m^0$, $m\ge 2$.
The total number of spinors in $\E$ must be even, by Galois, so ${\cal G}_2$
must be some $D_{m'}^0$ ($E_4$, unlike $D_m^0$, is bipartite). Then $r$ must be even (since there are only two spinors
in $\E$), and the spinors must be $\si,J_v\si$ for either $\si=\L_{r-1}$
or $\si=\L_r$. The matrix $\N_v=I_{m-2}\oplus I_2^s
\oplus I_{m'-2}\oplus I_2^s$,
and the exponents $\E=\{0,J_v,\langle\la^{2r/(2m-3)}\rangle,
\langle\la^{2r/(2m'-3)}\rangle,\si,J_v\si\}$, are now determined.

By the usual argument (see Case 1(c)) and using the fact that Tr$(\N_{spinor})=0$,
we get $\N_r+\N_{r-1}=\left(\matrix{0_{mm}&B\cr B^t&0_{m'm'}}\right)$ where
$B$ is the $m\times m'$ matrix
$$\left(\matrix{2a&\cdots&2a&a&a\cr\vdots&&\vdots&\vdots&\vdots\cr
2a&\cdots&2a&a&a\cr a&\cdots&a&b&c\cr a&\cdots&a&c&b}\right)$$
where $b+c=a=2\sqrt{r/(2m-3)(2m'-3)}$.
Hence 4 divides $r$. We find that the eigenvalues of that $(m+m')\times(m+m')$
matrix are $\pm 2\sqrt{r},\pm(b-c),$ and 0 (multiplicity $m+m'-4$).
However, the eigenvalue $S_{\L_r\mu}/S_{0\mu}+S_{\L_{r-1}\mu}/S_{0\mu}$
corresponding to exponent $\mu=\si$ will equal $\sqrt{2}$.
This forces $b-c=\pm\sqrt{2}$, which
contradicts integrality.

\medskip\noindent{{\it Case 2(b):}} Suppose ${\cal G}_1$ is the graph $E_4$;
then so must be ${\cal G}_2$. By the usual arguments we get that $r$ is even
and $\E=\{0,J_v,J_s,J_c,\L_{r/2},\L_{r/2},\si,J_v\si\}$ for some
$\si\in\{\L_{r-1},\L_r\}$. Now proceed as in Case 2(a).

\medskip\noindent{{\it Case 2(c):}} Consider next ${\cal G}_1$ being nonbipartite
(i.e.\ of type $A_{even}^{(1)}$ or ${}^0\!A_m^0$). Then so must be ${\cal G}_2$.
There are no spinors in $\E$, and neither $J_s,J_c$ are in $\E$. The exponents $\E$
are thus determined, and we find from (3.6) that all spinors map to the identical matrix,
which is easy to find by Tr$(\N_{spinor})=0$ and the method of Case 1(a).

\medskip\noindent{{\it Case 2(d):}} Now consider ${\cal G}_1=A_{m-1}^{(1)}$,
when $m$ is even. Suppose for contradiction that ${\cal G}_2=D_{m'-1}^{(1)}$.
Then there are only two spinors in $\E$ --- say $\si,J_v\si$ for $\si\in
\{\L_{r-1},\L_r\}$ --- so $r$ must be even. We know $\N_v$ is as in Case 2(a).
In the usual way (eigenvectors of $\N_1-I-\N_v$, etc), we find that
$\N_r+\N_{r-1}=\left(\matrix{0_{mm}&D\cr D^t&0_{m'm'}}\right)$ where
$D$ is the $m\times m'$ matrix
$$D=\left(\matrix{a&a&2a&\cdots&2a&a&a\cr a&a&2a&\cdots&2a&a&a\cr
\vdots&\vdots&\vdots&&\vdots&\vdots&\vdots\cr a&a&2a&\cdots&2a&a&a}\right)$$
Choose $\si'$ to be the spinor $\L_r,\L_{r-1}$
for which $S_{\si,\si'}\ne 0$ (so $\si'=C_1^{r/2}\si$). Then we find
by (3.6) that $\N_{\si'}\ne \N_{J_v\si'}$ while
$\N_{C_1\si'}=\N_{J_vC_1\si'}$. So $0\ne (I-\N_{v})\,(\N_r+\N_{r-1})=0$.

That contradiction means ${\cal G}_2$ must be $A_{m'-1}^{(1)}$ for some even
$m'$. 
Then we know the exponents,  and we get that $\N_r=\N_{J_v\L_r}\ne
\N_{r-1}=\N_{J_v\L_{r-1}}$. We find $\N_r\pm\N_{r-1}$ as in Case 1(c);
flipping if necessary the order of the vertices $1,2,\ldots,m$ yields the
precise placement of $C$'s and $C'$'s given in {\bf (iv)}.

\medskip\noindent{{\it Case 2(e):}} Finally, consider the case where $\N_1$
is $D_{m-1}^{(1)}\cup D_{m'-1}^{(1)}$. The exponents consist
of $0,J_v,J_s,J_c, \langle\la^{r/(m-3)}\rangle,\langle\la^{r/(m'-3)}
\rangle$, and four spinors. These spinors must be closed under Galois, so when
$r$ is odd all four distinct spinors must appear, each with multiplicity 1.
When $r$ is even, this can also happen, but another possibility is that the spinors are $\si,J_v\si$,
for $\si\in\{\L_r,\L_{r-1}\}$, each appearing with multiplicity 2. 

  We can compute  $\L_r\pm\L_{r-1}$ in the usual way, and thus obtain
  $\N_r$ and $\N_{r-1}$ (the answer depends on $r$ being odd or even, and
  also depends on some parameters).
  The exponents tell us the eigenvalues of $\N_{\si'}-\N_{J_v\si'}$ for
  either choice of $\si'=\L_r,\L_{r-1}$, and this then fixes the values of
the various parameters.

Incidentally, in reading off eigenvalues from the matrices arising here, it
is helpful to recall facts such as the sum of the squares of the eigenvalues of
a symmetric matrix $D$, equals the sum of the squares of the entries
of $D$.

\bigskip\noindent{{\it A.4. The matrix classifications.}}\medskip

Let ${\cal G}$ be any multi-digraph. Write $r({\cal G})$ for its largest
eigenvalue. We will prove first that the only connected {\it multigraphs} (multiple edges
and loops are allowed, but no directed edges)
with $r({\cal G})=2$, are listed in Figure 2.

Incidentally, to find the eigenvalues and eigenvectors of the graphs ${\cal G}=A_n^{(1)}$,
$D_n^{(1)}$, $E_6^{(1)}$, $E_7^{(1)}$, $E_8^{(1)}$, use the fact that they're the McKay
graphs for cyclic, dihedral, and $S_4,A_5,S_5$ groups. To find the
eigenvalues and eigenvectors of ${}^0\!A_n^0$ and $D_n^0$, use the fact
that they're $\Z_2$-foldings of $A_{2n}^{(1)}$ and $D_{2n}^{(1)}$.

Let ${\cal G}$ be any connected multigraph with $r=2$. To prove that it must lie in
Figure 2, we simply
use the following fact over and over:

\smallskip\noindent{\bf (PF1)} [30] If $A,B$ are nonnegative matrices, and entry-wise
$A_{ab}\le B_{ab}$ $\forall a,b$, then $r(A)\le r(B)$. If in addition $A,B$
are symmetric and indecomposable, then $r(A)<r(B)$ unless $A=B$.
\smallskip

For example, the Fact tells us that if ${\cal G}$ has any multiple edges,
then it must be $A_1^{(1)}$. If ${\cal G}$ has at least 2 loops, then it
must be one of the ${}^0\!A_n^0$. If ${\cal G}$ has a vertex with at least
4 edges leaving it, then it must be $D_4^{(1)}$. Etc.

\smallskip
The other matrix classification we need is much more difficult: finding all
indecomposable $\Z_\ge$-matrices $A$ with largest eigenvalue $<2$, and which obey $A^t=\Pi A
=A\Pi$ for some permutation matrix $\Pi=(\delta_{b,\pi a})$. By replacing
$A$ with some $\Pi^k A$, we can (and in \S A.1 we do) assume $\Pi$ has order a power of 2.
We will write $n_a$ for the smallest positive power of this permutation
$\pi$ which fixes $a$ --- so each $n_a$ will likewise be a power of 2.

\medskip\noindent{{\bf Lemma A.}} Let $A$ be an indecomposable matrix with nonegative integer
entries and with $r(A)<2$, and obeying the relations $A^t=\Pi A=A\Pi$  for some permutation matrix
$\Pi$ whose order is a power of 2. Then up to equivalence, 
$(A,\Pi)$ corresponds to a diagram in Figure 3.\medskip

In \S5.1 we explain how to obtain the digraph $A$ and permutation
$\Pi$, given a diagram in Figure 3. Note e.g.\ that both $D_n(1)$ and
$C_{n-1}(1)$ have the same digraph $A$, but different permutations $\Pi$.

The equations $A=\Pi A\Pi^t$ and $A^t=\Pi A$ tell us, for all $a,b,\ell$,
$$A_{ab}=A_{\pi^\ell a,\pi^\ell b}=A_{\pi^\ell b,\pi^{\ell+1}a}\eqno(A.7)$$
Note also (from Perron-Frobenius theory [30]) that $(A^tA)_{aa}\le
r(A^tA)=r(A)^2<4$, so:

\smallskip\noindent{{\bf (PF2)}} all entries of $A$ are 0's and 1's;

\noindent{{\bf (PF3)}} there are at most three 1's in any row or column.

\medskip\noindent{{\bf Claim 1.}} Let $A_{ab}\ne 0$. Then either:

\noindent{(i)} $n_a=n_b$, and $A_{\pi^ia,\pi^jb}\ne 0$ iff $i\equiv j$ (mod $n_a$),
in which case $A_{\pi^ia,\pi^jb}=A_{ab}=1$.

\noindent{(ii)} $n_a=2n_b$, and $A_{\pi^i a,\pi^jb}\ne 0$ iff $i\equiv j$ (mod $n_b$),
in which case $A_{\pi^ia,\pi^jb}=A_{ab}=1$.

\noindent{(iii)} $n_b=2n_a$, and $A_{\pi^ia,\pi^jb}\ne 0$ iff $i\equiv j$ (mod $n_a)$,
in which case $A_{\pi^ia,\pi^jb}=A_{ab}=1$.

\medskip To see this, first write $n$ for the maximum of $n_a$
and $n_b$; then (A.7) and (PF3) tell us that $n/n_a$ and $n/n_b$ either
equal 1 or 2 (since $n$ must be a power of 2). The more refined statements
in (i)--(iii) arise by restricting to the submatrix of $A$ consisting of the
rows and columns lying in the $\pi$-orbits of $a$ and $b$, and using (PF1).\qquad QED

\medskip It is convenient to assign to  $A$ a diagram in which
each $\pi$-orbit $\langle \pi\rangle a$ is associated a node. Each node is
labeled with its size $n_a$, and we connect nodes $\langle\pi\rangle a$ and
$\langle \pi\rangle b$ if $A_{\pi^ia,
\pi^jb}\ne 0$ for some $i,j$. We want to show that this diagram lies in
Figure 3.

The next Claim takes care of the possibility that our diagram has a loop.

\medskip\noindent{{\bf Claim 2.}} Suppose $A_{a,\pi^\ell a}\ne 0$ for some $\ell$. Then
$A$ is a tadpole $T_m$ whose loop is at $a$, and whose permutation $\Pi=I$.

\medskip \noindent{\it Proof.} Assume first that $n_a>1$, i.e.\ that $\pi a\ne a$. Then (A.7) tells us
$A_{a,\pi^{1-\ell}a}=A_{a,\pi^\ell a}= 1$. But $1-\ell\not\equiv \ell$
(mod $n_a$), since $n_{a}>1$ is a power of 2, which contradicts 
Claim 1(i).

So $\pi a=a$ and $A_{aa}=1$. Suppose $A_{ab}\ne 0$ for some $b\ne a$.
If there were any other entry $A_{ac}\ne 0$ (e.g.\ if $n_b=2$), then
$A$ would have a submatrix of type $D_3^0$, contradicting $r(A)<2$.
Continuing in this way, we obtain that $A$ is the adjacency matrix for the
tadpole, and that $\pi$ fixes everything.\qquad QED\medskip

So it suffices to consider $A_{a,\pi^ia}=0$ for all $a$ and $i$
(i.e.\ no loops in the diagram). First we'll show that the diagram must be a tree.

If instead it contains a cycle (i.e.\  a subdiagram of shape $A_m^{(1)}$ for
some $m$), then every vertex corresponding to a node in that cycle will have a row-sum at least equal
to 2, and so $r$ for that subdiagram will be at least 2. 

To see that our diagram can consist only of weights $k$ and $2k$, suppose
for contradiction that
it has  a subdiagram consisting of a node with weight $k$,
followed by any number (say $i$) of weight-$2k$ nodes, followed by a weight-$4k$ node.
The corresponding subgraph has an eigenvector consisting of $k+2ik$ 2's and $4k$ 1's,
with eigenvalue 2.

Thus our diagram will be of two types: either all nodes are of the same size
$k$; or all the nodes are labelled
by $k$ or $2k$, for some $k$. 
That the diagram must be one of the ones listed in Figure 3, now follows
from routine arguments.

\bigskip\noindent{{\it A.5. Proof of Theorem 3. }}\medskip

We conclude the Appendix with the remaining proofs of Thm.3.

A useful fact is that, whenever there is an integer $i$ and primaries
$\mu,\nu$ such that 
$$S_{\la\mu}\,S_{0\mu}^i=S_{\la\nu}\,S_{0\nu}^i\eqno(A.8a)$$
holds for all $\la\in P_+$, then $\mu=\nu$. To see this, hit both
sides with $S_{\la\mu}^*$ and sum over $\la\in P_+$.
We also use the fact, evident from Verlinde's formula (2.2), that 
the Verlinde ratios form a 1-dimensional representation of the fusion
ring: for any $\mu\in P_+$,
$${S_{\la\mu}\over S_{0\mu}}\,{S_{\la'\mu}\over S_{0\mu}}=\sum_{\nu\in P_+}
N_{\la\la'}^\nu\,{S_{\nu\mu}\over S_{0\mu}}\eqno(A.8b)$$

\medskip\noindent{{\it Proof of (iii)}.} Let $J$ be a simple-current of
order $n$ in $\E$. Write $P_i$ for all $\la\in P_+$ with
$Q_J(\la)\equiv i/n$ (mod 1).          
Consider any $\mu\in\E$. We want to show $J\mu\in \E$. To do this, we
will use this fact from Perron-Frobenius theory [30]: 
\smallskip

\noindent{{\bf (PF4)}} When an irreducible nonnegative matrix $A$ has eigenvalue
$e^{2\pi\i/D}\,r(A)$, then the eigenvalues of $A$ are invariant under
rotation by $2\pi/D$, i.e.\ if $s$ is an eigenvalue of $A$ with
multiplicity $m$, then so is $e^{2\pi\i/D}s$. \smallskip

 The problem is that this only applies to individual
matrices, and we need it simultaneously for all $\N_\la$.
Write $d(\mu)$ for the largest divisor $d$ of $n$ such that
$\mu=J^{n/d}\mu$. So $d=1$ if $\mu$ is not a fixed-point of $J$. Note
that $S_{\la\mu}=0$ when $\la\in P_i$ and $d(\mu)$ does not divide
$i$ ({\it Proof:} apply (2.4) to $S_{\la,J^{n/d}\mu}=S_{\la\mu}$). In
Claim 1 we establish a converse of that simple fact.

\smallskip\noindent{{\bf Claim 1.}} Choose any $\mu\in P_+$. Then
$S_{\la\mu}\ne 0$ for some $\la\in P_m$, iff $d(\mu)$ divides $m$.

\smallskip\noindent{{\it Proof.}} Call a number $m$ `$\mu$-nice' if $S_{\la\mu}\ne 0$
for some $\la\in P_m$. The Galois automorphism $\si=\si_\ell$ in (2.6)
obeys $\si_\ell(P_m)=P_{\ell m}$, because of the calculation 
$$\eps_\si(\la)\,e^{2\pi\i\,\ell m/n}S_{\si\la,0}=\si_\ell(e^{2\pi\i\,
m/n}S_{\la 0})=\si_\ell S_{\la,J}=\eps_\si(\la)\,S_{\si \la,J}$$
Thus by (2.6), $m$ is `$\mu$-nice' iff the greatest common divisor gcd$(n,m)$ is `$\mu$-nice'. By considering
various products of the form
$(S_{\la\mu}/S_{0\mu})^a(S_{\la'\mu}/S_{0\mu})^b$ for $\la\in P_m,\la'\in P_{m'}$,
repeatedly using (A.8b) to write them as sums of various 
$S_{\la''\mu}/S_{0\mu}$ for $\la''\in P_{am+bm'}$, we
get that $m$ and $m'$ are both `$\mu$-nice' iff gcd$(n,m,m')$ is
`$\mu$-nice'. Continuing in this way, we ultimately obtain some
divisor $D$ of $n$ such that $m$ is `$\mu$-nice' iff $D$ divides $m$ (we're
just using the fact that $\Z$ is a Principal Ideal Domain). The point is that
then $S_{\la,J^{n/D}\mu}=S_{\la\mu}$ $\forall\la\in P_+$, so by (A.8a) we
get that $\mu=J^{n/D}\mu$. \qquad QED to Claim 1\smallskip

\noindent{{\bf Claim 2.}} Choose any two primaries $\mu,\mu'$, and write $d=d(\mu)$. If
$S_{\la\mu}/S_{0\mu}=S_{\la\mu'}/S_{0\mu'}$ for all $\la\in P_d$, then
$\mu=\mu'$. \smallskip

\noindent{{\it Proof.}} Let $d'=d(\mu')$. Then $d'$ divides $d$, by
Claim 1. Choose any $\la\in P_{d'}$ so that $S_{\la\mu'}\ne 0$. Then 
use (A.8b) repeatedly to obtain
 $$0\ne({S_{\la\mu'}\over S_{0\mu'}})^{d'/d}=\sum_{\nu\in P_d}n_\nu\,
{S_{\nu\mu'}\over S_{0\mu'}}=\sum_{\nu\in P_d}n_\nu\,{S_{\nu\mu}\over
S_{0\mu}}=({S_{\la\mu}\over S_{0\mu}})^{d'/d}$$
for certain numbers $n_\nu$.
Thus by Claim 1, $d$ must also divide $d'$, and hence $d=d'$.

Now choose any $\la\in P_m$. We want to show that $S_{\la\mu}/S_{0\mu}=
S_{\la\mu'}/S_{0\mu'}$. Assume that $d$
divides $m$ (otherwise both ratios trivially vanish). By Claim 1 there exists a primary $\nu\in P_d$ such that
$S_{\nu\mu}\ne 0$. Then again by (A.8b)
$${S_{\la\mu}\over S_{0\mu}} \,({S_{\nu\mu}\over S_{0\mu}})^{n+1-m/d}=
\sum_{\gamma\in P_d}n_\ga{S_{\ga\mu}\over S_{0\mu}}=
\sum_{\gamma\in P_d}n_\ga{S_{\ga\mu'}\over S_{0\mu'}}=
{S_{\la\mu'}\over S_{0\mu'}} \,({S_{\nu\mu'}\over S_{0\mu'}})^{n+1-m/d}$$
Hence $S_{\la\mu}/S_{0\mu}=S_{\la\mu'}/S_{0\mu'}$
$\forall \la\in P_+$, and so $\mu=\mu'$ by (A.8a).\qquad QED to
Claim 2\smallskip

Recall that for us both $J$ and $\mu$ are in $\E$, and we want to show
that the multiplicity $m_{J\mu}$ equals $m_\mu$.  Write $d=d(\mu)$. Then
Claim 2 says that we can find a nonnegative linear combination
$N'=\sum_{\la\in P_d} a_\la N_\la\ge 0$ (in fact almost every
nonnegative linear combination will do) of fusion matrices for which
the eigenvalue $\sum_{\la\in P_d} a_\la S_{\la\mu}/S_{0\mu}$ has
multiplicity 1. For such a choice of coefficients $a_\la$, the
fact (PF4) tells us that the eigenvalue $e^{2\pi\i
d/n}\sum_{\la\in P_d} a_\la S_{\la\mu}/S_{0\mu}$ of $N'$ also will have 
multiplicity 1, which is to say that the only primary  $\mu'$ obeying
$$e^{2\pi\i d/n}\sum_{\la\in P_d} a_\la {S_{\la\mu}\over S_{0\la}}=
\sum_{\la\in P_d} a_\la{S_{\la\mu'}\over S_{\la\mu'}}$$
is $\mu'=J\mu$. Applying this to the eigenvalues of $\N':=\sum_{\la\in P_d}
\N_\la$, we get that $J\mu$ is indeed in $\E$, with multiplicity $m_\mu$.

Taking $\mu=J^{-1}$, we see that $m_J=1$, and we also get from this
that the $J\in\E$ form a  group. \qquad QED to (iii)\medskip

\noindent{{\it Proof of (viii).}} We will first show that we get an
$\N_1$-grading associated to any simple-current $J\in\E$.  Fix a
simple-current $J\in\E$ and vertex $1\in\B$. Define as in \S3.3
$g_\la=Q_J(\la)$, and $g(y)=Q_J(\la)$ when $\N_{\la 1}^y\ne 0$ (recall
that for $\N$ irreducible, $\sum_{\la\in P_+}\N_{\la 1}^y>0$).

First note that $g(y)$ is well-defined (mod 1). For if also $\N_{\mu
1}^y\ne 0$, then $(\N_{\la}\N_{C\mu})_{11}\ne 0$, i.e.\ Tr($\N_\nu)\ne
0$ for some primary $\nu\in P_+$ with $Q_J(\nu)\equiv Q_J(\la)-Q_J(\mu)$ (mod
1). But then Thm.3(x) requires that $Q_J(\la)\equiv Q_J(\mu) $ (mod 1)
as desired.

Now, suppose $\N_{\la y}^z\ne 0$, for some $\la\in P_+$ and $y,z\in
\B$. Choose $\mu\in P_+$ so that $\N_{\mu 1}^y\ne 0$. Then
$(\N_\mu\N_\la)_{1z}\ne 0$ and thus $\N_{\nu 1}^z\ne 0$ for some
primary $\nu\in P_+$ with $Q_J(\nu)\equiv Q_J(\mu)+Q_J(\la)$ (mod 1),
i.e.\ $g(z)\equiv g(y)+g_\la$ (mod 1), as desired. Thus this gives us
an $\N_1$-grading.

Suppose conversely that we have an $\N_1$-grading. Suppose $\la,\mu,\nu$
are three primaries in $P_+$ with fusion coefficient $N_{\la \mu}^\nu\ne 0$.
Choose any vertices $x,y\in\B$ such that $(\N_\nu)_{xy}\ne 0$. Then
there must exist a vertex $w\in\B$ such that $(\N_\la)_{xw}\ne 0$ and
$(\N_\mu)_{wy}\ne 0$. Comparing $g_\la+g(x)\equiv g(w)$,
$g_\mu+g(z)\equiv g(y)$, and $g_\nu+g(x)\equiv g(y)$, all taken (mod
1), we find that $g_\la+g_\mu\equiv g_\nu$ (mod 1), i.e.\ $\la\mapsto
g_\la$ is a $\Q$-grading of the fusion ring. Note that the map
$\la\mapsto e^{2\pi\i g_\la}S_{\la 0}/S_{00}$ defines a 1-dimensional
representation of the fusion ring, and thus 
$$e^{2\pi\i g_\la}\,{S_{\la 0}\over S_{00}}={S_{\la 0'}\over
S_{00'}}\qquad \forall \la\in P_+$$ 
for some $0'\in P_+$. Taking the norm-squared of both sides and
summing over $\la\in P_+$, we get $S_{00}^{-2}=S_{00'}^{-2}$, i.e.\
$0'$ is a simple-current $J$, and $g_\la=Q_J(\la)$. Let $n$ be its order.

Now let $\vec{v}$ be the simultaneous Perron-Frobenius eigenvector, for
which the matrix $\N_\la$ has eigenvalue $S_{\la 0}/S_{00}$. Note
that 
$$\eqalign{\sum_{y\in\B}\N_{\la x}^y\, (e^{2\pi\i\, g(y)}\,\vec{v}_{y})=&\,e^{2\pi\i\,(g_\la+
g(x))}\sum_{y\in\B}\N_{\la x}^y
\vec{v}_{y}=e^{2\pi\i\,(g_\la+g(x))}\,{S_{\la 0}\over S_{00}}\,\vec{v}_{x}\cr =&\,{S_{\la
J}\over S_{0J}}\,e^{2\pi\i g(x)}\,\vec{v}_{x}}$$
and thus $J\in\E$. This is
precisely the $\N_1$-grading arising from the simple-current $J$,
constructed in \S3.3.\qquad QED to Thm.3(viii)

\bigskip\centerline{{\bf References}}
\medskip

\item{[1]} J.\ Cardy, {Nucl.\ Phys.} {B324} (1989) 581.

\item{[2]} I.\ Affleck, J.\ Phys.\ A33 (2000) 6473.

\item{[3]} G.\ Pradisi, A.\ Sagnotti, Y.\ S.\ Stanev, Phys.\ Lett.\
B381 (1996) 97.

\item{[4]} A.\ Sagnotti, Y.\ S.\ Stanev, Nucl.\ Phys.\ Proc.\ Suppl.\
55B (1997) 200.

\item{[5]} J.\ Fuchs, C.\ Schweigert, {Nucl.\ Phys.} {B530} (1998)
99.

\item{[6]} R.\ E.\ Behrend, P.\ A.\ Pearce, V.\ B.\ Petkova, J.-B.\
Zuber, 
{Nucl.\ Phys.} {B579} (2000) 707.

\item{[7]} {P.\ Di Francesco}, {J.-B.\ Zuber}, Nucl.\ Phys. {B338} (1990) 602.

\item{[8]} J.-B.\ Zuber, In: {\it Proc.\ Intern.\ Congress of Math.\ Phys.},
ed by D.\ Iagolnitzer, International Press, 1995, p.674.

\item{[9]} {D.\ E.\ Evans}, {Y.\ Kawahigashi}, {\it Quantum symmetries on
operator algebras}, Oxford University Press, 1998.

\item{[10]} J.\ B\"ockenhauer, D.E.\ Evans, ``Subfactors and modular invariants'',
math.OA/0008056.

\item{[11]} A.\ Ocneanu, talks (Toronto June 2000, and Kyoto November 2000).

\item{[12]} {T.\ Gannon}, ``The modular invariants for the exceptional
algebras'', in preparation;

\item{} {T.\ Gannon}, ``The modular invariants for the classical
algebras'', work in progress.

\item{[13]} V.B.\ Petkova, J.-B.\ Zuber, Nucl.\ Phys.\ B463 (1996) 161.

\item{[14]} J.\ Fuchs, C.\ Schweigert, Nucl.\ Phys.\ B558 (1999) 419;

\item{} J.\ Fuchs, C.\ Schweigert, Nucl.\ Phys.\ B568 (2000) 543.

\item{[15]} R.\ Dijkgraaf, C.\ Vafa, E.\ Verlinde, H.\ Verlinde, Commun.\
Math.\ Phys.\ 123 (1989) 485.

\item{[16]} P.\ Di Francesco, J.-B.\ Zuber, In: {\it Recent Developments in
Conformal Field Theory}, World Scientific, 1990, p.179.

\item{[17]} {Ph.\ Di Francesco, P.\ Mathieu,} {D.\
S\'en\'echal}, {\it Conformal Field Theory}, Springer, New York, 1997.

\item{[18]} {T.\ Gannon}, ``Modular data: the algebraic combinatorics of
conformal field theory'', math.QA/0103044.

\item{[19]} E.\ Verlinde, Nucl.\ Phys.\ B300 (1988) 360.

\item{[20]} {V.\ G.\ Kac,} {\it Infinite Dimensional Lie algebras},
3rd edition, Cambridge University Press, Cambridge, 1990.

\item{[21]} T.\ Gannon, ``Algorithms for affine Kac-Moody algebras'', to appear in
Lett.\ Math.\ Phys.


\item{[22]} A.N.\ Schellekens, S.\ Yankielowicz, 
{Phys.\ Lett.} {B227} (1989) 387.

\item{[23]} {A.\ Coste}, {T.\ Gannon}, 
Phys.\ Lett.\ {B323} (1994) 316.

\item{[24]} T.\ Gannon, P.\ Ruelle, M.A.\ Walton, Commun.\ Math.\ Phys.\ 179 (1996) 121.

\item{[25]} {T.\ Gannon}, {M.A.\ Walton}, {Commun.\ Math.\
Phys.} {206} (1999) 1.

\item{[26]} {T.\ Gannon}, 
Nucl.\ Phys.\ {B396} (1993) 708.

\item{[27]} A.\ Cappelli, C.\ Itzykson, J.-B.\ Zuber, Commun.\ Math.\ Phys.\
113 (1987) 1.

\item{[28]} A.\ Recknagel, V.\ Schomerus, Nucl.\ Phys.\ B531 (1998) 185.

\item{[29]} T.\ Gannon, ``The automorphisms of affine fusion rings'',
math.QA/0002044, to appear in Adv.\ Math.

\item{[30]} A.\ Berman, R.\ J.\ Plemmons, {\it Nonnegative Matrices in
the Mathematical Sciences}, Academic Press, New York, 1979.

\item{[31]} {F.\ M.\ Goodman, P.\ de la Harpe}, {V.\ F.\
R.\ Jones}, {\it Coxeter graphs and towers of algebras}, Springer-Verlag,
New York, 1989.

\item{[32]} D.\ Cvetkovi\'c, M.\ Doob, H.\ Sachs, {\it Spectra of
Graphs: Theory and Applications}, Academic Press, New York, 1980.

\item{[33]} S.\ Li\'enart, P.\ Ruelle, O.\ Verhoeven, Nucl.\ Phys.\
B592 (2001) 479.

\item{[34]} {A.\ Coste}, {T.\ Gannon}, {P.\
Ruelle}, {Nucl.\ Phys.} {B581} (2000) 679.

\item{[35]} J.-B.\ Zuber, in: {\it Topological Field Theory and Related
Topics}, Prog.\ Math 160, Birkh\"auser, Boston, 1998, ed by M.\ Kashiwara
et al, p.453.

\item{[36]} P.\ Degiovanni, {Commun.\ Math.\ Phys.}\ {127} (1990) 71.

\item{[37]} T.\ Gannon, {Lett.\ Math.\ Phys.} {39} (1997) 289.

\item{[38]} T.\ Gannon, {Canad.\ J.\ Math.} {52} (2000) 503.

\item{[39]} {D.\ Verstegen}, 
Nucl.\ Phys.\ {B346} (1990) 349.


\item{[40]} J.\ Fuchs, A.N.\ Schellekens, C.\ Schweigert,  Nucl.\ Phys.\
B437 (1995) 667.

\item{[41]} J.\ Fuchs, C.\ Schweigert, 
math.CT/0106050.


%


\bigskip\bigskip\epsfysize=2in \centerline{\epsffile{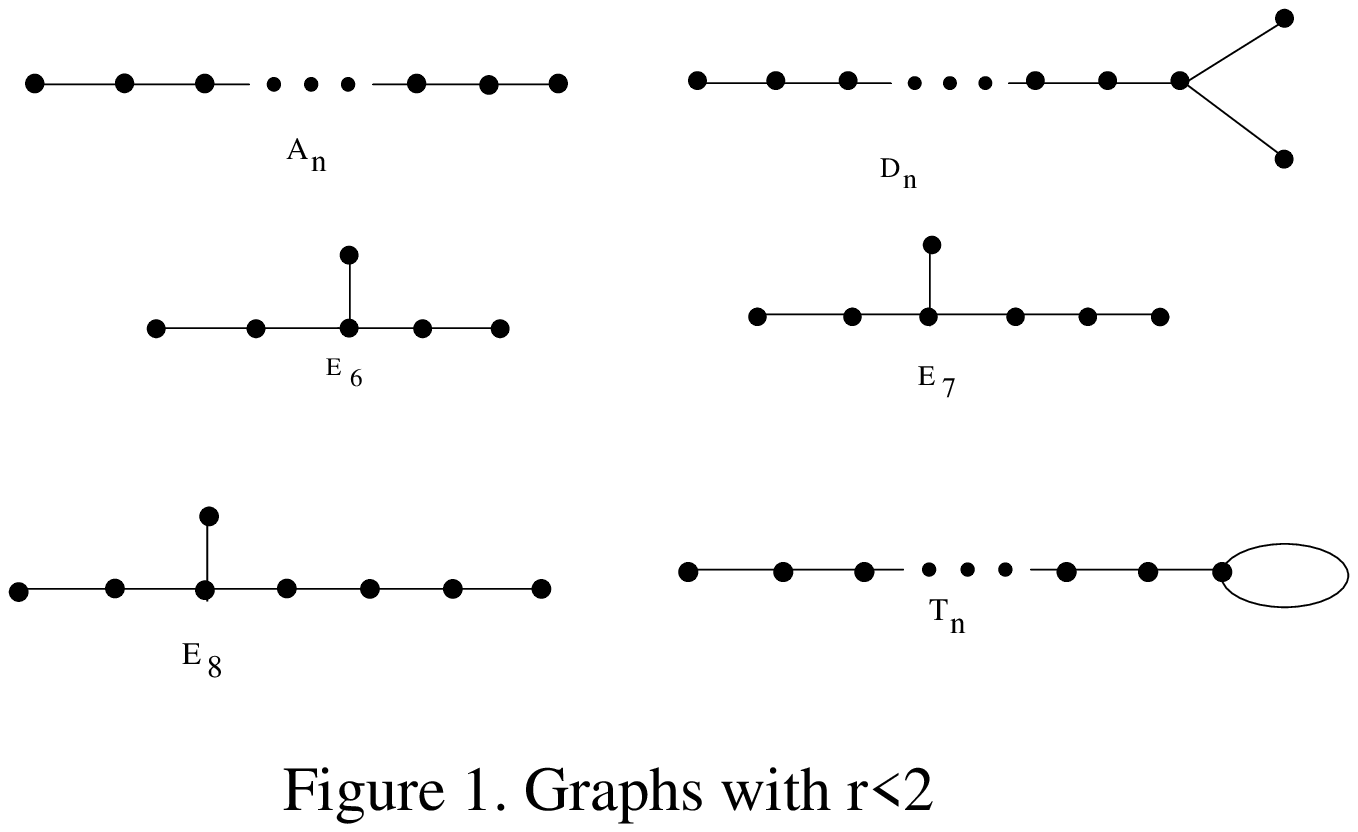}}

\bigskip\epsfysize=3in \centerline{\epsffile{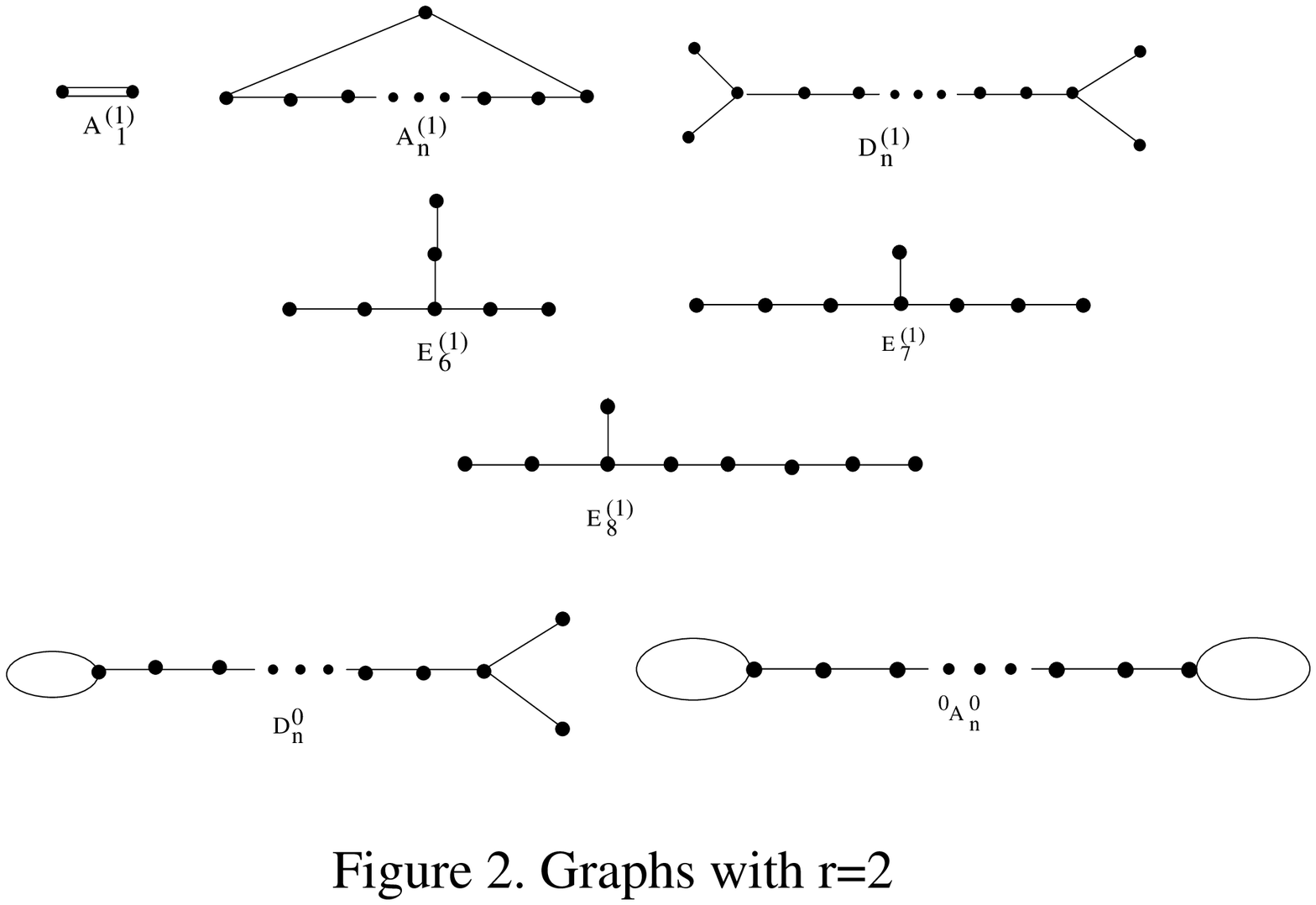}}
\bigskip
\epsfysize=4in \centerline{\epsffile{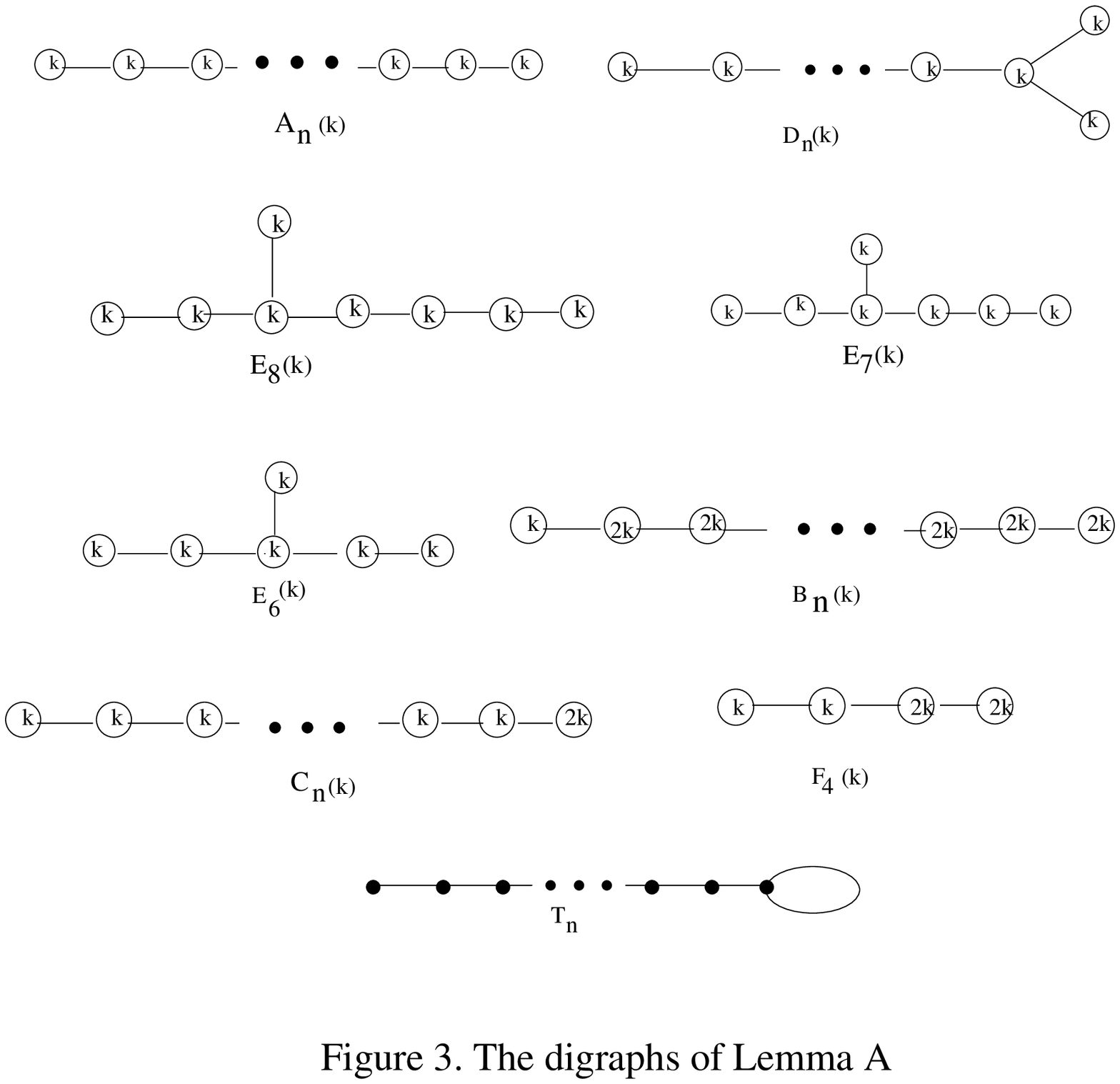}}\bigskip

\end

Suppose we have  a symmetry $\pi$ of all graphs ${\cal N}_\la$, i.e.\
$$({\cal N}_{\la})_{\pi a,\pi b}=({\cal N}_{\la})_{a,b}\qquad \forall
\la\in P_+\eqno(3.8)$$
for all indices $a,b$. Then we can fold all the graphs by it, by restricting
to orbits. More precisely, let $G$ be any group of symmetries $\pi$ as in
(3.8), and let $[a]$ be the $G$-orbit of $a$. Given any $G$-orbits $[a],[b]$,
and any $\la\in P_+$, define matrices ${\cal N}^G_{\la}$ by
$$({\cal N}^G_{\la})_{[a],[b]}=\sum_{b'\in[b]}({\cal N}_{\la})_{a,b'}$$
Then ${\cal N}^G$ is a nonnegative integer representation of the fusion ring,
and ${\cal N}^G_{0}=I$. It will be a NIM-rep iff all $G$-orbits have the
same size (e.g.\ if the action of $G$ is fixed-point-free).

This can be generalised using the idea of ...
Suppose we partition ${\cal B}$ into different sets ${\cal O}_i$. 
Choose any two such sets ${\cal O}_i,{\cal O}_j$. 
Suppose that for any vertices $x,x'\in{\cal O}_i$, the sums 
$$\sum_{y\in{\cal O}_j}\N_{\la x}^y=\sum_{z\in{\cal O}_j}\N_{\la
x'}^y\qquad \forall \la\in P_+\eqno(3.9a)$$
If this happens for all such $i,j,x,x',\la$, then we call our
partition {\it nice}. 

You can think of (3.9a) graphically as: colour the
vertices $x\in {\cal O}_i$ with colour `$i$'; then in any fixed graph $N_\la$,
(3.9a) says that $x$ sees as many nodes coloured `$j$' as $x'$ does. 
This generalises the symmetry example: in that case, the partition
${\cal O}_i$ consists of the orbits.

Then collapse each matrix $\N_\la$ into a much smaller one,
$\N_\la^{{\cal O}}$: give it one
row/column for each set ${\cal O}_i$, and put $(\N_\la^{{\cal
O}})_i^j$ equal to the value of the sum in (3.9a). Again, the result is
a nonnegative integer representation of the fusion ring, obeying
$\N_0=I$, and whose exponents are a subset of those of the parent
NIM-rep $\N$. However, $\N^{{\cal O}}$ will be a NIM-rep iff the
transpose condition (3.2c) is obeyed, iff for all $a\in{\cal O}_i$ and
$b\in{\cal O}_j$, 
$$\sum_{c\in{\cal O}_i}\N_{\la c}^b=\sum_{d\in{\cal O}_j}\N_{\la
a}^d\eqno(3.5b)$$ 

\noindent{{\bf Definition 4.}}\quad {\it By a} {fusion-homomorphism} {\it we mean a
map $\pi:P_+\rightarrow P_+'$ such that
$$N'{}^{\nu'}_{\pi\la,\pi\mu}=\sum_{\nu\in\pi^{-1}\nu'}N_{\la\mu}^\nu$$
By a} fusion-automorphism {\it we mean any permutation $\pi$ of $P_+$ such that}
$$N_{\la,\mu}^\nu=N_{\pi \la,\pi \mu}^{\pi \nu}\qquad \forall \la,\mu,\nu\in \Phi$$

Simple-currents are a major source of fusion-endomorphisms.
Let $J$ be any simple-current of order $n$. Choose any  $m\in\{0,1,
\ldots,n-1\}$. Define $\pi[m]:P_+\rightarrow P_+$ by $\la\mapsto J^{nmQ_J(\la)}\la$.
Then $\pi[m]$ is a fusion-homomorphism. It is a fusion-automorphism iff
$ {\rm gcd}(nm\,Q_J(J)+1,n)=1$.
 When the group of simple-currents is not cyclic,
this construction can be generalised in a natural way, and the resulting
fusion-symmetry will be parametrised by a collection of numbers $a_{ij}$.
We will see this next section.

We will call these {\it simple-current endomorphisms}. 

For any current algebra $\widehat{\frak{g}}$ and any sufficiently high level,
its fusion-automorphisms (classified in [G4]) consist entirely of
simple-current automorphisms and conjugations (i.e.\ symmetries of the
unextended Dynkin diagram).

-----------

Let $A=\N_{\Lambda_4}$, the matrix for the spinor. $A$ has
nonnegative integer entries. Since
charge-conjugation is trivial here, we know $A^t=A$. We already
learned $A$ is traceless, so it has only 0's on the diagonal. The
quantum dimension of the spinor is $\sqrt{9}=3$, so this must be the
maximal eigenvalue of $A$. We want to show that no such matrix can
exist, and also respect the fusion 
$$\Lambda_4\stimes\Lambda_4=0\splus\Lambda_1\splus\Lambda_2\splus\Lambda_3
\splus(2\Lambda_4)$$

From the fusion we get that $A^2$ has trace 8. But $A$ is symmetric, so
trace$(A^2)=\sum_{i,j}(A_{ij})^2$. Thus the nonzero entries of $A$
consist of either: (i) two 2's; (ii) one 2 and four 1's; or (iii) eight 1's. 

Now just run through the possibilities. We know $A$ has no nonzero diagonal
entries, and the nonzero entries are symmetrically placed about the diagonal,
so in case (i) the only possible matrix (ignoring the six identically
0 rows and columns) is the block 
$$\left(\matrix{0&2\cr 2&0}\right)\ ,$$
which however has largest eigenvalue 2. Case (ii) is impossible, since
the `2' cannot be placed on the diagonal. For case (iii),  
we can short-cut the possibilities by using the fact [BP] that the only
way for a symmetric indecomposable matrix to have a largest eigenvalue
of 3  is either to have all row sums equal to 3 (imposible here), or to have 
at least one row sum bigger than 3. So that means we have the matrix
(ignoring the three identically 0 rows/columns)
$$\left(\matrix{0&1&1&1&1\cr 1&0&0&0&0\cr 1&0&0&0&0\cr 1&0&0&0&0\cr
1&0&0&0&0}\right) $$
This is the adjacency matrix for the $\widehat{{\rm so}}(8)$ diagram, and so
also has largest eigenvalue $2\ne 3$.
